\documentclass{jfm}

\usepackage{natbib}

\usepackage{amsmath}
\usepackage{graphicx}
\usepackage{caption}
\usepackage{subcaption}
\usepackage{float}
\usepackage{multirow}
\usepackage{bm}
\usepackage{amsfonts}
\usepackage{xcolor}
\usepackage{soul}
\usepackage{color}
\usepackage{tikz}

\newcommand{\solidrule}[1]{\raisebox{1.4pt}{\tikz{\draw[-,#1,solid,line width = 1.4pt](0,4mm) -- (4mm,4mm);}}}

\begin{document}

\shorttitle{Connecting resonance to swimming performance} 
\shortauthor{A. Goza, D. Floryan, and C. Rowley} 

\title{Connections between resonance and nonlinearity in swimming performance of a flexible heaving plate}

\author
 {
Andres Goza\aff{1}\footnote{\textit{Current address}: Department of Aerospace Engineering, University of Illinois at Urbana-Champaign, Urbana, IL 61801, USA}
  \corresp{\email{ajgoza@gmail.com}},
  Daniel Floryan\aff{1},
  \and 
  Clarence Rowley\aff{1}
  }

\affiliation
{
\aff{1}
Department of Mechanical and Aerospace Engineering, Princeton University, Princeton, NJ 08540, USA

}

\maketitle

\begin{abstract}

We investigate the role of resonance in finite-amplitude swimming of a flexible flat plate in a viscous fluid. The role of resonance in performance remains unclear for two reasons: i) a lack of definition of resonance for the fully-coupled fluid-structure interaction system in a viscous flow, and ii) the presence of nonlinear effects, which makes it difficult to disentangle resonant and non-resonant mechanisms in finite-amplitude swimming. We address point i) and provide an unambiguous definition for system resonance by computing global linear stability modes of the fully-coupled fluid-structure interaction system that account for the viscous fluid, the plate, and the coupling between them. We then resolve point ii) by considering high-fidelity nonlinear simulations of systematically increased amplitude. By comparing the results for different amplitudes with one another and with the linear stability modes, we separate linear and/or resonant effects from nonlinear and/or non-resonant effects. Resonant behavior is observed over a wide range of plate stiffnesses, with peaks in trailing-edge motion and thrust occurring near the resonant frequency defined by the global linear analysis. The peaks broaden and weaken with increasing heave amplitude, consistent with an increased damping effect from the fluid. At the same time, non-resonant mechanisms are present at large heave amplitudes. The input power exhibits qualitatively different dynamics at large heave amplitudes compared to smaller heave amplitudes, where resonance dominates. Moreover, leading-edge separation is present for stiff plates at large heave amplitudes, which can drastically alter the performance characteristics from what one would expect through linear predictions.

\end{abstract}

\section{Introduction}

The dual aims of understanding fish swimming and developing agile, efficient underwater vehicles have propelled research on the canonical problem of flow past a flapping flexible plate. Experimental \citep{Ramananarivo2011, Alben2012,Dewey2013,Quinn2014,Quinn2015}, computational \citep{Vanella2009,Hua2013,Zhu2014,Zhang2017}, and theoretical work \citep{Alben2008,Ramananarivo2011,Alben2012,Floryan2018} on this problem has revealed propulsive benefits to flexibility over a range of Reynolds numbers (and for inviscid flows), plate parameters, and amplitudes and frequencies associated with the plate's kinematics. Many studies considered actuation frequencies and material parameters for which the plate exhibits `first-mode' flapping; \emph{i.e.}, with the plate shape similar to that of the first mode of a clamped Euler-Bernoulli beam. However, propulsive benefits were shown to persist for parameters that led to `higher mode' shapes \citep{Alben2008,Alben2012,Quinn2014,Floryan2018}. Because this canonical problem involves the passive motion of deformable structures, a prevailing question is the extent to which performance is connected to structural resonance. This question is the focus of the current work.

There is a lack of consensus in the literature about the role of resonance in propulsive performance: some investigations have found actuation at or near a resonant frequency to provide benefits to thrust and/or efficiency  \citep{Alben2008,Alben2012,Dewey2013,Quinn2014,Zhu2014,Floryan2018}, whereas others have observed optimality through off-resonant actuation \citep{Vanella2009,Ramananarivo2011,Zhu2014}. Connecting resonance to swimming performance is challenging for a number of reasons. First, the most relevant definition of resonance is not agreed upon in the literature for viscous flows. For an inviscid flow, \citet{Michelin2009} computed eigenvalues of the fully coupled fluid-structure interaction (FSI) system to provide a meaningful definition of resonance. Such computations have thus far not been performed for viscous flows, and resonance has instead been defined using either the scaling for an Euler-Bernoulli beam in a vacuum \citep{Vanella2009,Hua2013} or the plate response to finite-amplitude actuation \citep{Quinn2014,Zhang2017}. These approaches leave open questions about the role of resonance, as the former neglects fluid effects and the latter provides the possibility that nonlinearities are playing a role in the observed performance peaks.

This latter effect of nonlinearity provides the second challenge to connecting resonance to performance, as the vast majority of studies have been performed at finite amplitudes. A variety of nonlinear mechanisms that impact performance have been identified. \citet{Michelin2009} used inviscid theory and computation to show that while thrust peaks aligned well with the eigenvalues of the fully-coupled FSI system, input power and efficiency exhibited off-resonant peaks (though this may be due in part to the fact that the authors did not incorporate instantaneously negative values of input power in their calculations). At Reynolds numbers $O(10^3-10^4)$, \citet{Ramananarivo2011} experimentally identified superharmonic resonance for large actuation amplitudes that the authors attributed to nonlinear fluid damping. \citet{Zhu2014} demonstrated through high-fidelity computations at Reynolds numbers of $O(10^2-10^3)$ that wake asymmetry can occur for certain parameters, and observed that this effect was largely deleterious for efficiency. Using experimental data at $O(10^4)$, \citet{Moored2014} used a local linearization\footnote{This study is included in the collection of nonlinear results because the analysis was performed about a mean flow that was generated through geometrical nonlinearities in the structure as well as flow nonlinearities such as flow separation, vortex interactions, etc.} of the wake behind the plate to argue that optimal efficiency was caused by selecting parameters that were closest to the wake's natural frequency. \citet{Quinn2015} showed through experiments at Reynolds numbers of $O(10^3-10^4)$ that sufficiently large swimming motions led to flow separation that adversely affected efficiency.

Given these nonlinear mechanisms, a remaining question is how to identify the role of resonance in performance of finite-amplitude swimming. \citet{Floryan2018} clarified the role of resonant behavior in an inviscid setting by comparing eigenvalues of the fully-coupled FSI system to linear (infinitesimal amplitude) dynamics. The authors showed that actuating at the resonant frequency corresponds to peaks in thrust and input power of the linear system. The authors additionally showed that the relative growth in magnitude of these quantities at resonance is comparable, resulting in no gains in efficiency.

The goal of the present work is to address the two aforementioned challenges in the context of a viscous fluid by considering the two-dimensional flow-plate system at a Reynolds number of $240$. First, global linear stability modes of the fully coupled FSI system are investigated to unambiguously identify the natural frequencies of the system. Next, to separate the role of resonance from nonlinearities, high-fidelity nonlinear simulations of increasing amplitude are studied and compared to the linear stability modes. Resonant and non-resonant mechanisms driving performance are characterized and, where appropriate, connections are drawn to previous studies performed at other Reynolds numbers and swimming parameters.

We note that although this study is performed in the context of swimming, nothing precludes the approach and methods employed here from being applied to scenarios relevant to flexible-wing flight, where the physical mechanisms are distinct due to a greater plate inertia with respect to the flow.

\section{Problem setup and parameters considered}

\begin{figure}
\centering
	  \includegraphics[scale = 0.6,trim={0cm 0cm 0cm 0cm},clip]{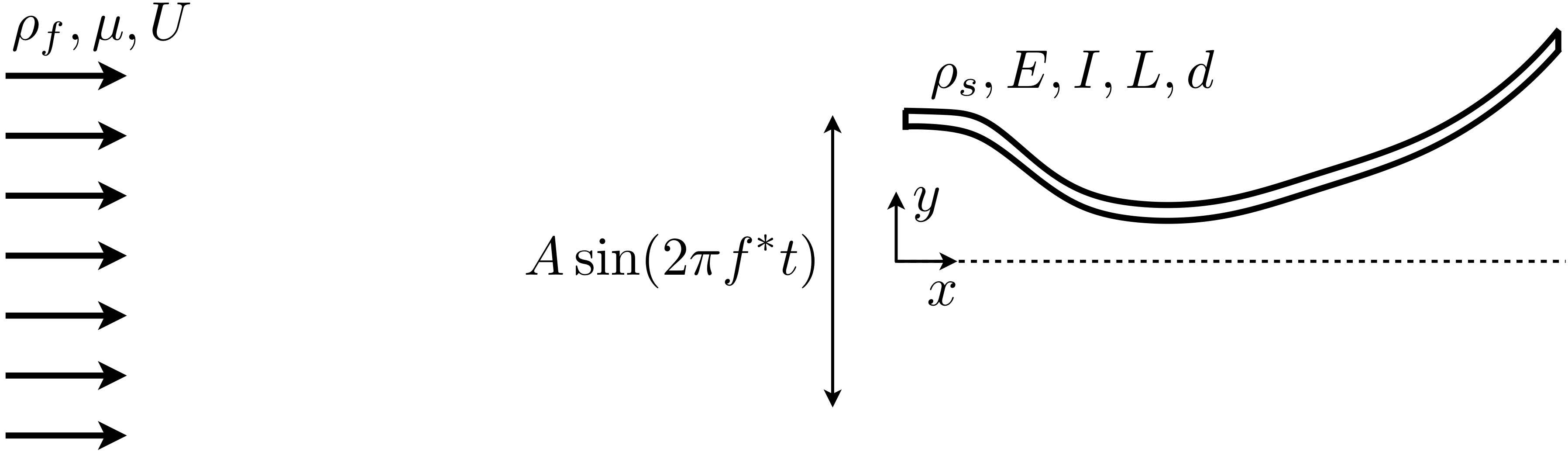}
\caption{A schematic of the problem setup and relevant dimensional quantities.}
\label{fig:problem_schem}
\end{figure}

We consider here uniform flow past a passively deformable plate that is heaved sinusoidally at its leading edge. The fluid is modeled via the two-dimensional Navier-Stokes equations, and the plate is treated as a clamped geometrically nonlinear Euler-Bernoulli beam. The important dimensionless parameters are the Reynolds number ($Re$), dimensionless mass ratio ($M$) and stiffness ratio ($S$), and dimensionless heave amplitude ($h_0$) and frequency ($f$), defined as

\begin{equation}
Re = \frac{\rho_f U L}{\mu}, \; M = \frac{\rho_s d}{\rho_f L}, \; S = \frac{EI}{\rho_f U^2 L^3}, \; h_0 = \frac{A}{L}, \; f = \frac{f^*L}{U}
\label{eqn:params}
\end{equation}
where $\rho_f$ ($\rho_s$) is the 2D fluid (structure) density, $U$ is the freestream fluid velocity, $L$ is the length of the plate, $\mu$ is the dynamic fluid viscosity, $d$ is the plate thickness, $EI$ is the flexural rigidity composed from Young's elasticity modulus ($E$) and the second moment of area ($I$), $A$ is the dimensional heave amplitude, and $f^*$ is the dimensional frequency related to the dimensional angular frequency ($\omega^*$) via $\omega^* = 2\pi f^*$. See figure~\ref{fig:problem_schem} for a schematic of the dimensional variables defining the FSI system.

In terms of these dimensionless variables, the heave motion is written as
\begin{equation}
h_{LE}(t) = h_0 \sin( \omega t )
\label{eqn:hLE}
\end{equation}
where $\omega = 2\pi f$ is the dimensionless angular frequency and $t$ is the dimensionless time defined using the convective time scale $L/U$.

In this study, the Reynolds number and mass ratio are fixed at $Re = 240$ and $M = 0.01$, respectively, whereas the stiffness, actuation frequency, and heave amplitude are varied. The moderate Reynolds number is representative of swimming of small fish and larvae \citep{Webb1988}, and simultaneously enables a comparison with higher-Reynolds-number studies and allows for a relatively large sweep of $S$ and $f$. The small mass ratio is a representative value for fish, which are neutrally buoyant and much thinner than they are long. The heave amplitudes are chosen to enable a systematic comparison of resonant and nonlinear effects in performance. Table \ref{tab:parms_summ} summarizes the range of parameter values considered in this work.

\begin{table}
\centering
\begin{tabular}{ c c c c c }
$Re$ & $M$ & $S$ & $h_0$ & $f$   \\ \hline
 240 & 0.01 & \{0.02, 0.2, 2, 20\} & \{0.001, 0.01, 0.1\} & 1--3.2
 \end{tabular}
 \caption{Range of parameter values considered for the nonlinear simulations. The frequency is varied in increments of 0.1.  }
\label{tab:parms_summ}
\end{table}

In this article, the response of the plate will often be summarized using the transverse amplitude of the plate's trailing-edge, $h_{TE}(t)$. Quantities relevant to swimming performance are the dimensionless thrust, lift, and input power, defined here as
\begin{equation}
C_T(t) = -\frac{F_x}{\frac{1}{2} \rho_f U^2 L}, \; C_L(t) = \frac{F_y}{\frac{1}{2} \rho_f U^2 L}, \; C_P(t) = - C_L(t) v_{LE}(t) \label{eqn:perf_nonlin}
\end{equation}
where $F_x$ and $F_y$ are the dimensional drag and lift defined as the horizontal and vertical forces integrated over the plate, respectively, and $v_{LE} = dh_{LE}/dt$ is the heave velocity. Performance is measured with respect to the temporal mean of these quantities---written as $\overline{C_T}$, $\overline{C_L}$, and $\overline{C_P}$, respectively---and in terms of some notion of efficiency ($\eta$) defined here as the Froude efficiency: $\eta = \overline{C_T}/\overline{C_P}$.

\section{Numerical methods: global linear stability modes and nonlinear simulations}
\label{sec:methods}

The nonlinear fully-coupled FSI equations are written symbolically here in a spatially discrete, temporally continuous setting as (see \citet{Goza2017} for details)
\begin{equation}
\bm{M\dot{y}} = \bm{r}(\bm{y})  \label{eqn:full_FSI}
\end{equation}
where $\bm{y}$ represents the full state of the system (\emph{i.e.}, all flow and structure variables required to characterize the full system dynamics). Equation (\ref{eqn:full_FSI}) contains the Navier-Stokes equations modeling the viscous flow, the geometrically nonlinear structural equations, and the no-slip boundary condition that couples the fluid and structure. Note the distinction between the matrix $\bm{M}$ and the mass ratio $M$. The matrix $\bm{M}$ is singular, which reflects the differential-algebraic nature of the governing equations (due to the no-slip condition on the plate).

The global linear system is defined from the nonlinear system (\ref{eqn:full_FSI}) by linearizing about a base state, $\bm{y}^b$. Writing $\bm{y} = \bm{y}^b + \bm{y}^p$ in (\ref{eqn:full_FSI}), where $\bm{y}^p$ is a small perturbation from the base state, and retaining only linear terms in $\bm{y}^p$ leads to the linear system
\begin{equation}
\bm{M \dot{y}}^p = \bm{A y}^p \label{eqn:lin_FSI}
\end{equation}
where $\bm{A = } d \bm{r}/ d\bm{y} |_{\bm{y}^b}$. The specific form of the various terms are provided in  \citet{Goza2018JFM}.

The global eigenvalues ($\lambda_i$) that give the natural frequency of the fully-coupled system and the corresponding modes ($\bm{\phi}_i$) satisfy the generalized eigenvalue problem $\lambda_i\bm{ M \phi}_i = \bm{A \phi}_i$, $i = 1, \dots,n$ (with $n$ defined as the state dimension). These eigenvalues and modes are computed by an implicitly restarted Arnoldi algorithm \citep{Lehoucq1998}, and are converged to within $||\lambda_i\bm{ M \phi}_i - \bm{A \phi}_i||_2 < 10^{-8}$ for all results shown below.

The nonlinear simulations are performed by solving (\ref{eqn:full_FSI}) using the immersed-boundary algorithm of \citet{Goza2017}. The method treats the fluid equations using a discrete streamfunction formulation \citep{Colonius2008} and the plate (modeled as a geometrically nonlinear Euler-Bernoulli beam) with a corotational finite element formulation \citep{Crisfield1991}. The method is strongly coupled; \emph{i.e.}, the nonlinear coupling between the structure and the fluid is enforced at each time instance, resulting in a stable algorithm in the presence of large displacements and rotations. This fluid-structure coupling is enforced by the stresses on the immersed surface, and immersed-boundary methods are well known to produce spurious computations of these stresses. These unphysical stress computations were remedied by \citet{Goza2016}, and the techniques described therein are incorporated into the FSI algorithm of \citet{Goza2017} to ensure appropriate treatment of the fluid-structure coupling. The FSI solver has been validated on several flapping flag problems for flags in both the conventional configuration (pinned at the leading edge) and the inverted configuration (clamped at the trailing edge) \citep{Goza2017}.

The simulation parameters used in this work are described in appendix \ref{app:grid_properties}. To facilitate studies of steady swimming, for all nonlinear simulations the heave amplitude was ramped up over a period of twenty convective time units via $ h^{ramp}_{LE}(t) = \alpha(t) h_{LE}(t) $, where $\alpha(t)$ is a scaling factor defined by
\begin{equation}
 \alpha(t) =
    \begin{cases}
      	0 & t \leq 5 \\
         6\left(\frac{t - 5}{15}\right)^5 -15\left(\frac{t - 5}{15}\right)^4 + 10 \left(\frac{t - 5}{15}\right)^3  & 5\leq t < 20 \\
          1 & t \ge 20
       \end{cases}
\end{equation}
The zero amplitude used for $0 \leq t \leq5$ is to allow the impulsive start from the fluid to decay. A shifted and scaled smoothstep function \citep{Ebert2003} is used over $5 \leq t \leq 20$ because of its zero first and second derivative at $t = 5, 20$. Analysis from the nonlinear simulations was performed after $t = 20$, using data from a minimum of 10 heave periods.

\section{Resonance for viscous plates: global modes of the FSI system}
\label{sec:GMresults}

\begin{figure}
\centering
	\begin{subfigure}[b]{0.30\textwidth}
		\hspace*{-6mm}
        		\includegraphics[scale = 0.37,trim={0cm 1.7cm 2.72cm 0cm},clip]{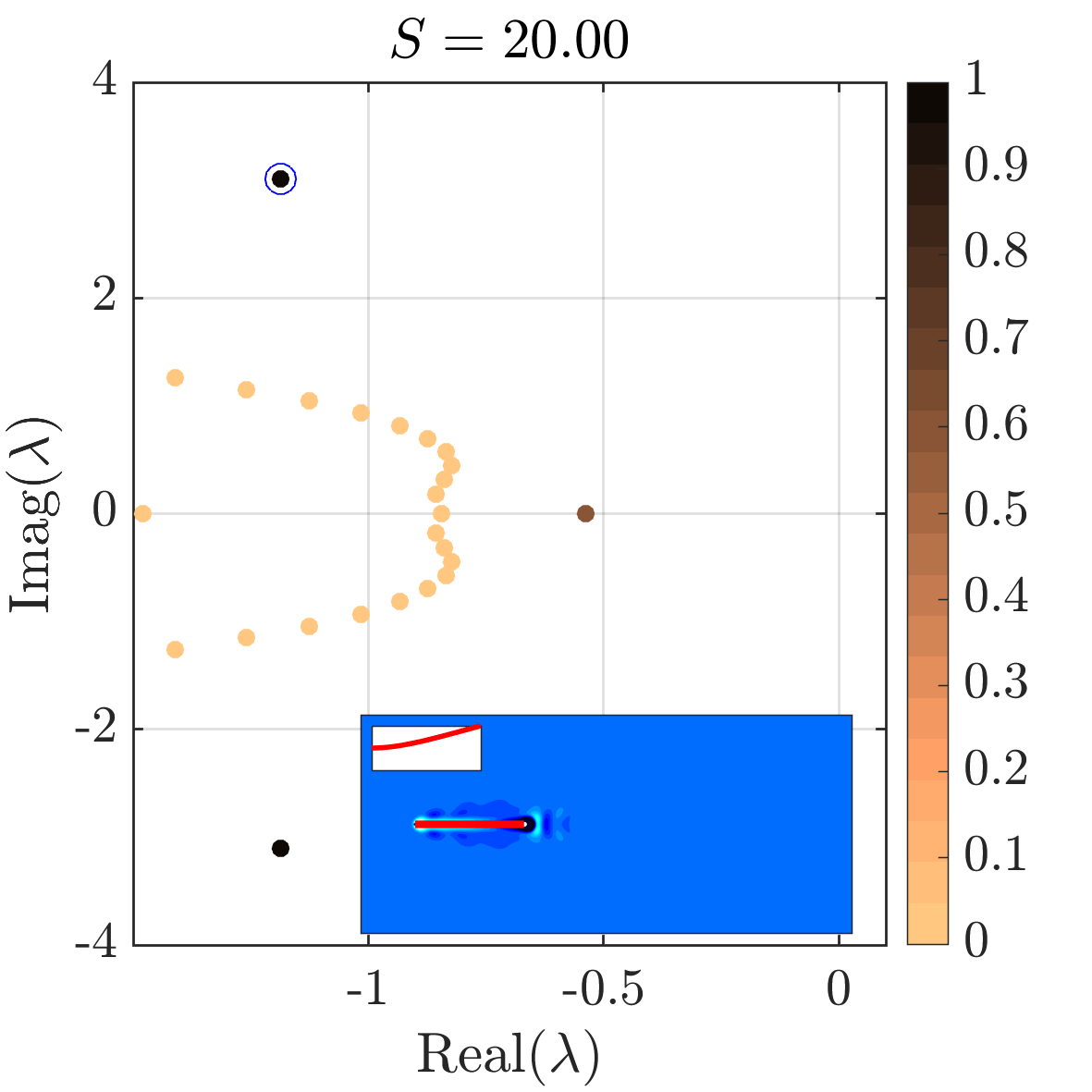}
	\end{subfigure}
	\begin{subfigure}[b]{0.30\textwidth}
		\hspace*{-2.1mm}
        		\includegraphics[scale = 0.37,trim={1.6cm 1.7cm 2.72cm 0cm},clip]{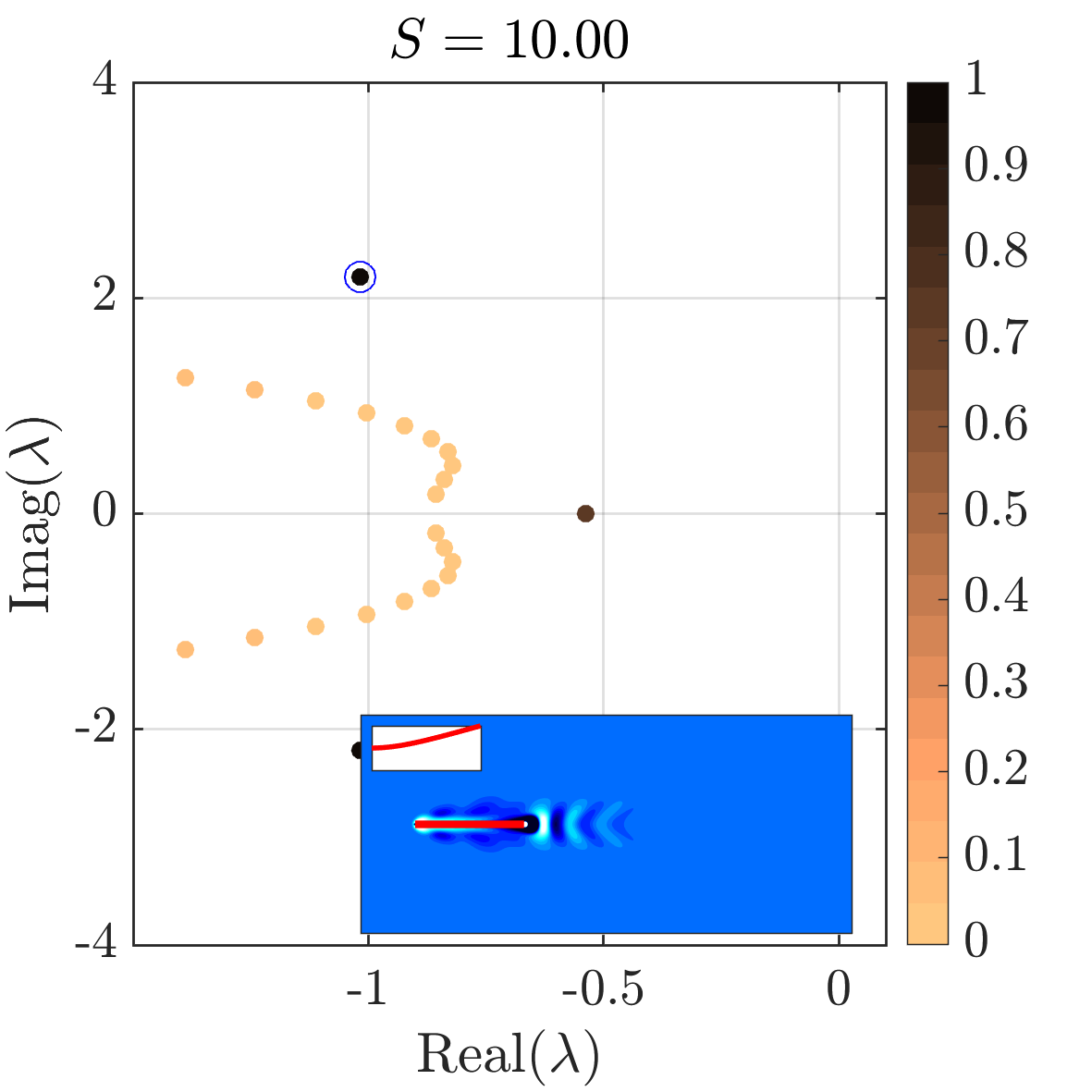}
	\end{subfigure}
	\begin{subfigure}[b]{0.30\textwidth}
		\hspace*{-3.5mm}
        		\includegraphics[scale = 0.37,trim={1.6cm 1.7cm 0cm 0cm},clip]{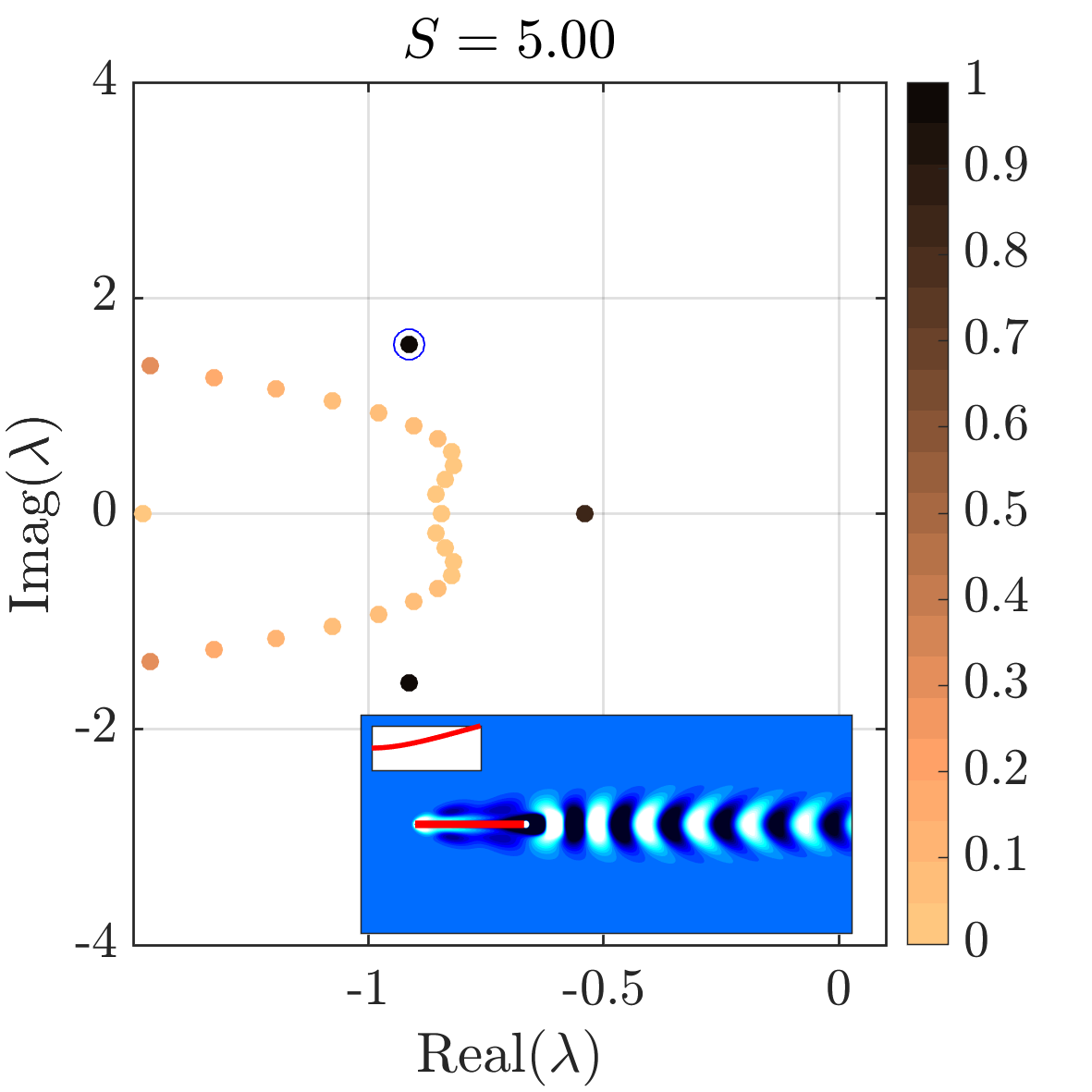}
	\end{subfigure}

	\begin{subfigure}[b]{0.30\textwidth}
		\hspace*{-6mm}
        		\includegraphics[scale = 0.37,trim={0cm 1.7cm 2.72cm 0cm},clip]{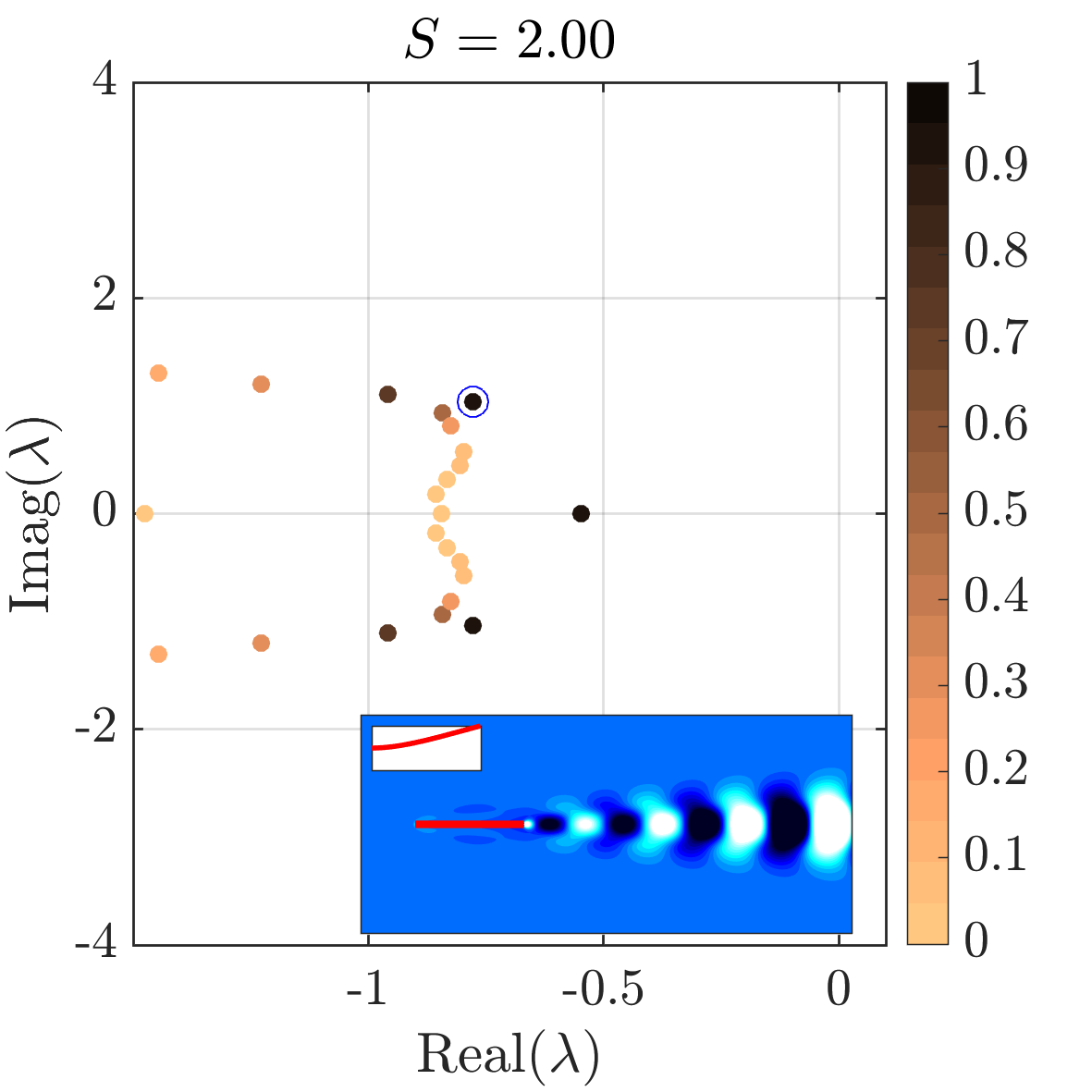}
	\end{subfigure}
	\begin{subfigure}[b]{0.30\textwidth}
		\hspace*{-2.1mm}
        		\includegraphics[scale = 0.37,trim={1.6cm 1.7cm 2.72cm 0cm},clip]{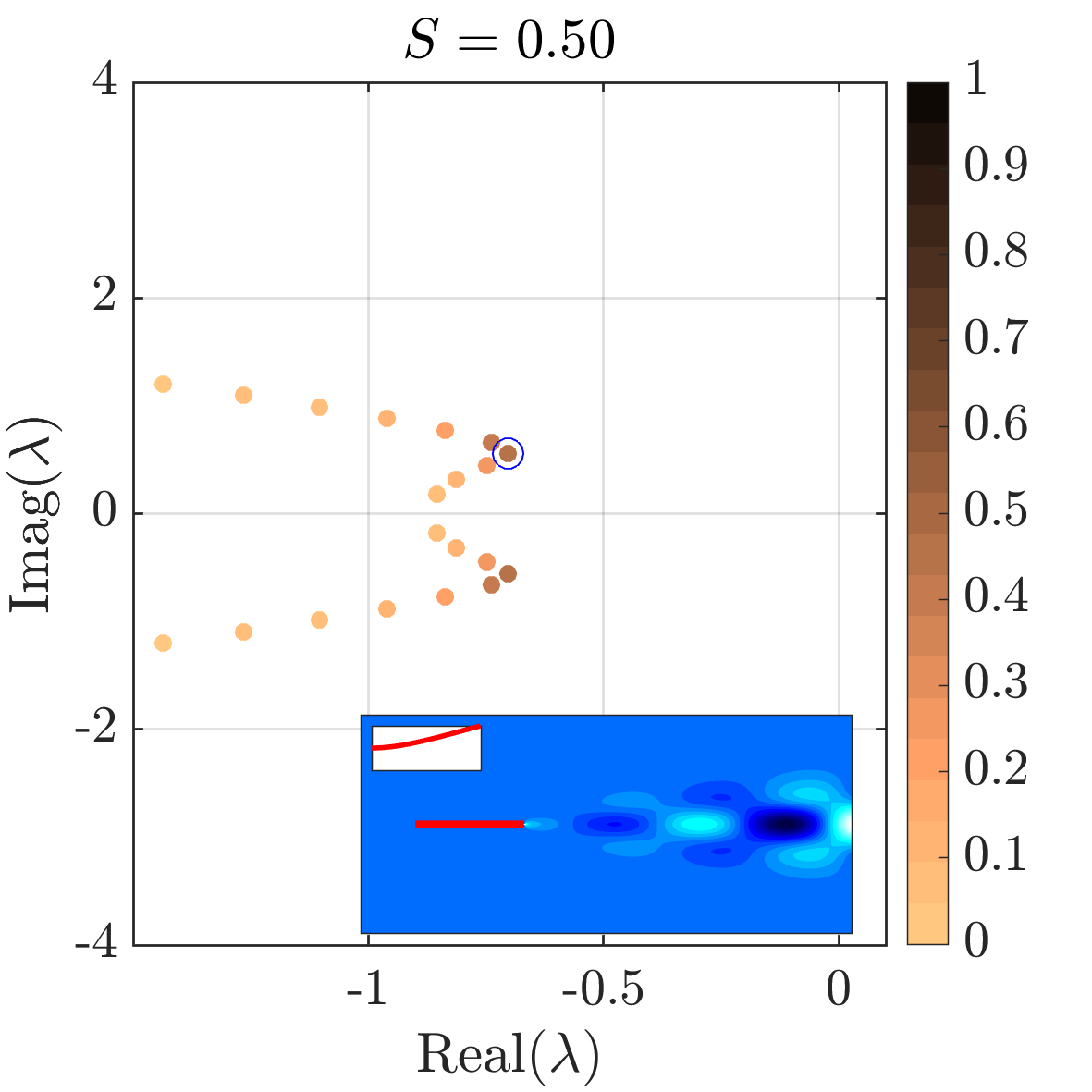}
	\end{subfigure}
	\begin{subfigure}[b]{0.30\textwidth}
		\hspace*{-3.5mm}
        		\includegraphics[scale = 0.37,trim={1.6cm 1.7cm 0cm 0cm},clip]{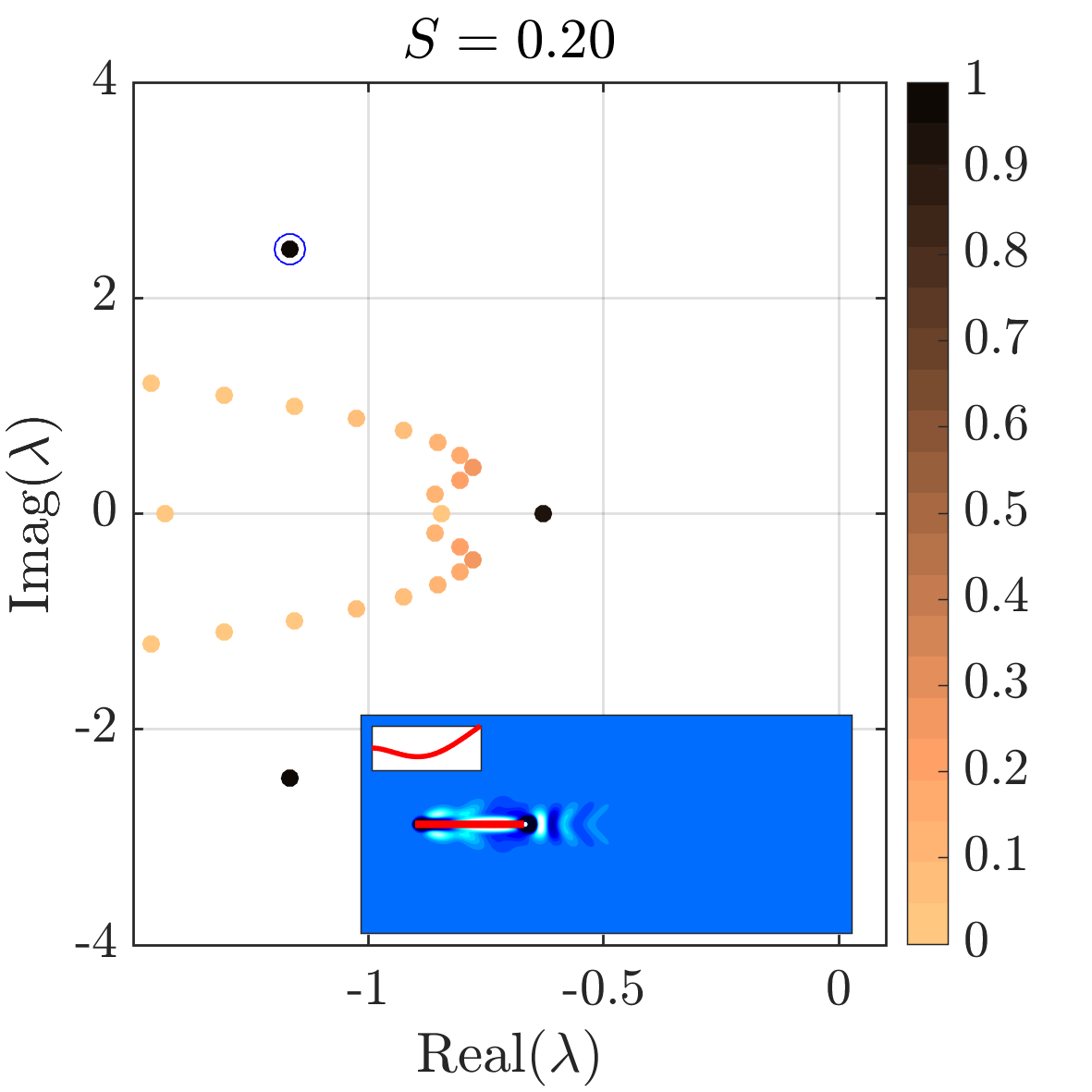}
	\end{subfigure}

	\begin{subfigure}[b]{0.30\textwidth}
		\hspace*{-6mm}
        		\includegraphics[scale = 0.37,trim={0cm 0cm 2.72cm 0cm},clip]{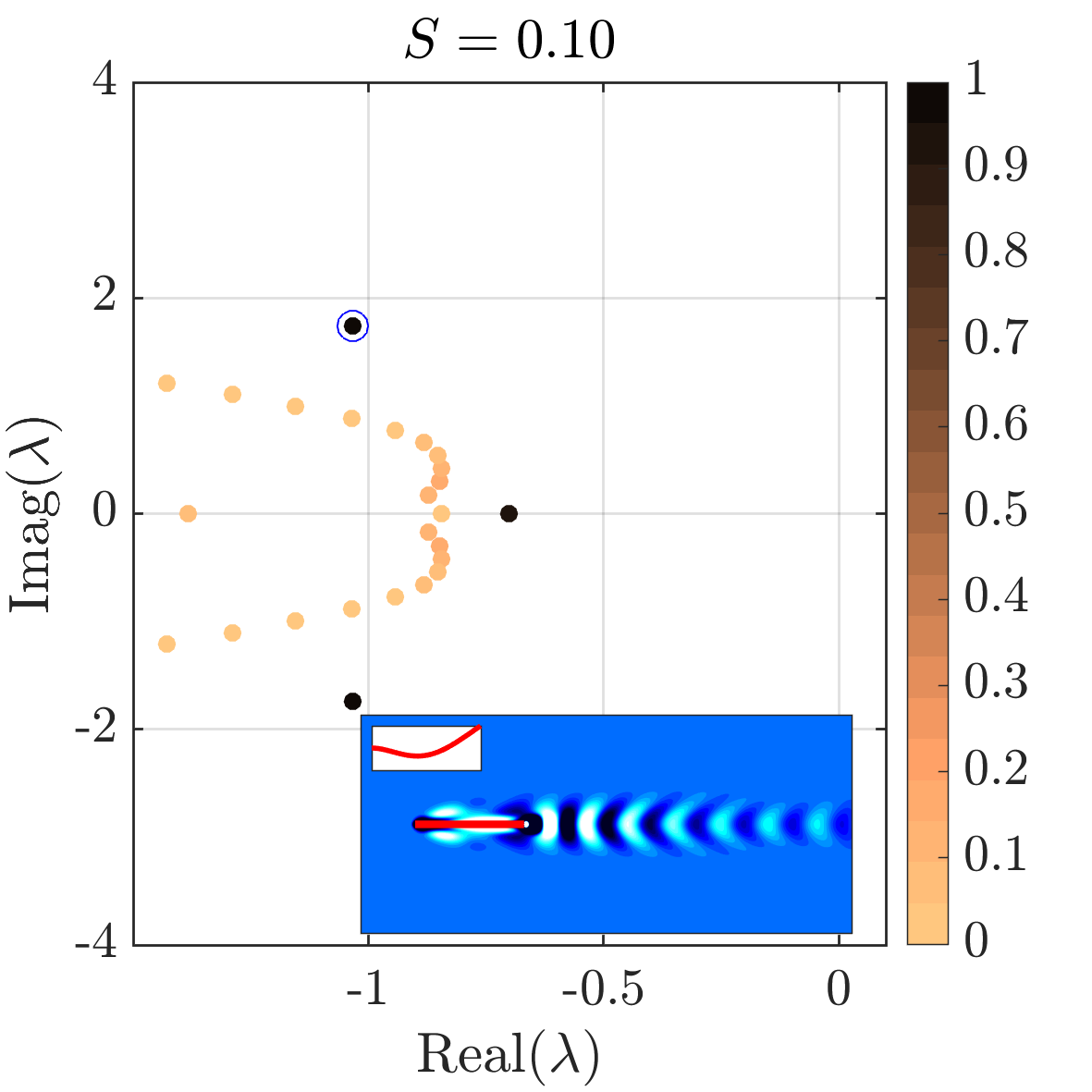}
	\end{subfigure}
	\begin{subfigure}[b]{0.30\textwidth}
		\hspace*{-2.1mm}
        		\includegraphics[scale = 0.37,trim={1.6cm 0cm 2.72cm 0cm},clip]{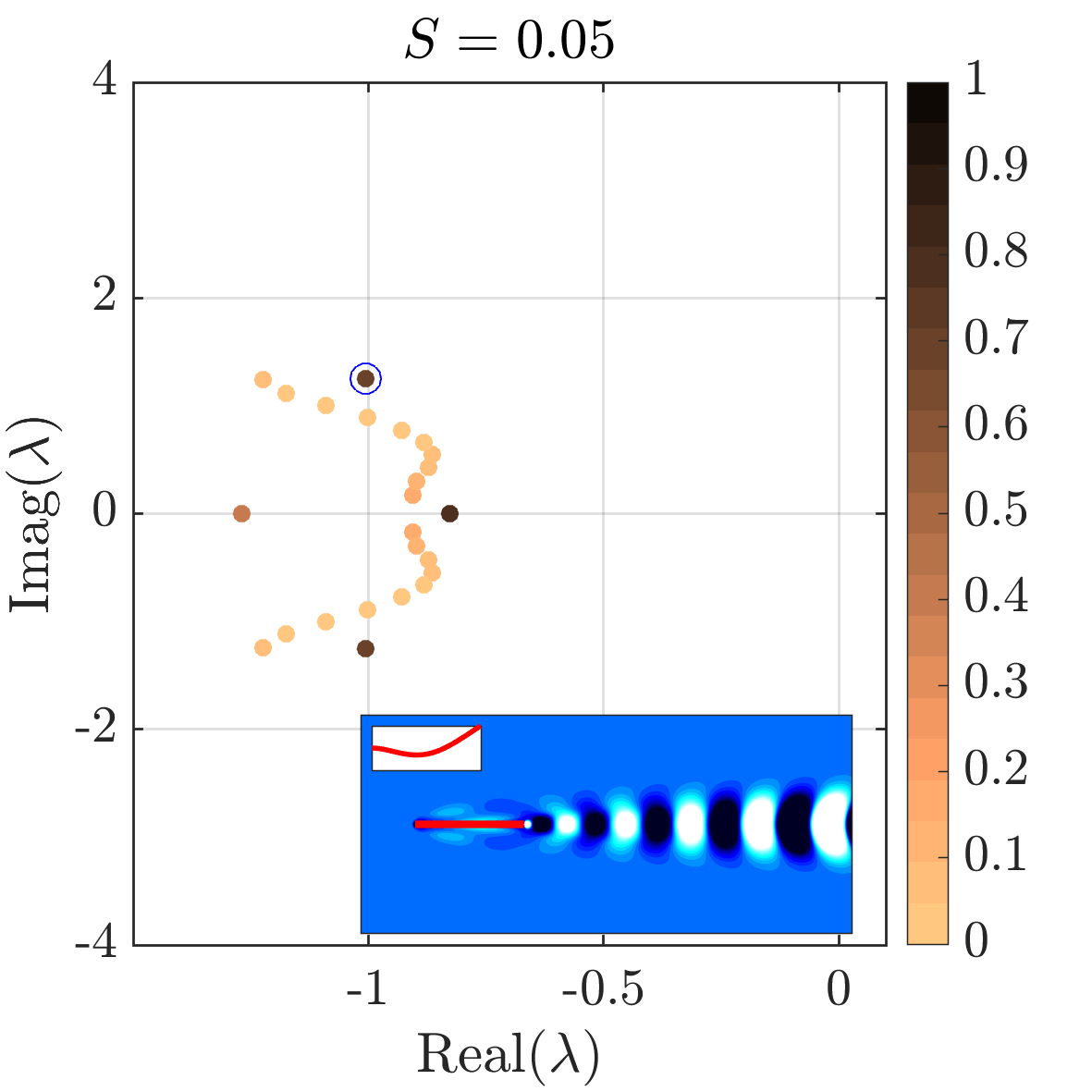}
	\end{subfigure}
	\begin{subfigure}[b]{0.30\textwidth}
		\hspace*{-3.5mm}
        		\includegraphics[scale = 0.37,trim={1.6cm 0cm 0cm 0cm},clip]{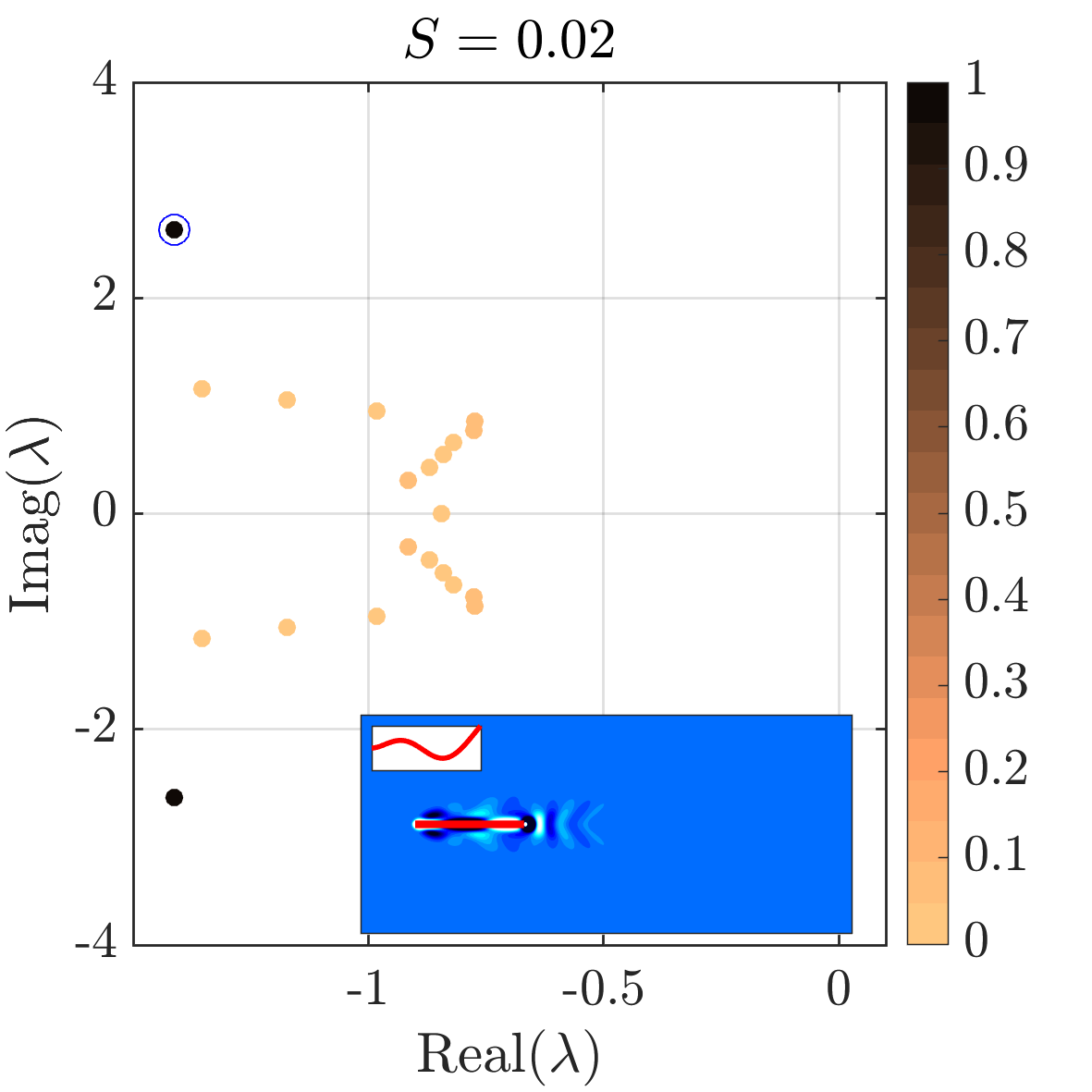}
	\end{subfigure}
	\caption{The main figures provide the first several eigenvalues with largest growth rate for nine different stiffnesses over the frequencies $[-4, 4]$. The markers are color-coded by the ratio, $E_r$ (see main text for details). For each stiffness, the eigenvalue with corresponding eigenvector of largest $E_r$ value is circled. The inserts illustrate the response of the eigenvector associated with the circled eigenvalue at the time for which the trailing-edge displacement is maximal (the sub-inserts convey zoomed-in versions of the plate shape). The eigenvectors are scaled to have unit 2-norm. Contours are of vorticity, plotted in 20 increments of $-0.05$ to 0.05, with white indicating positive and black indicating negative vorticity.}
	\label{fig:evals_flex}
\end{figure}

To provide a framework by which to define resonance in the simulations, we first present the eigenvalues of the fully coupled plate-fluid system in figure~\ref{fig:evals_flex} for several different stiffnesses. The eigenvalues are color-coded by an energy ratio, $E_r$, defined here as
\begin{equation}
	E_r = \frac{\text{plate energy}}{\text{plate energy + flow kinetic energy}}
\label{eqn:energy_metric}
\end{equation}
In (\ref{eqn:energy_metric}), the plate energy is defined as the sum of plate's bending strain energy and kinetic energy. All energies are computed from the eigenvector associated with the eigenvalue being considered. Note that because the denominator of the energy ratio is the total system energy, $E_r\le1$. 

Also provided for each stiffness in figure~\ref{fig:evals_flex} is the eigenvector associated with the largest value of $E_r$. The eigenvector is plotted using the fact that the time response associated with a given eigenvalue-eigenvector pair $(\lambda_j,\bm{\phi}_j)$ is given by the real part of $\bm{\phi}_je^{\lambda_j t}$. The inserts contain plots of this response at the time $t$ for which the trailing-edge displacement is maximal. Note that the there is no amplitude scale in the inserts, as the amplitude is irrelevant in this linear setting.

\begin{figure}
\centering
        \includegraphics[scale = 0.35,trim={1.6cm 6.5cm 2.6cm 4.3cm},clip]{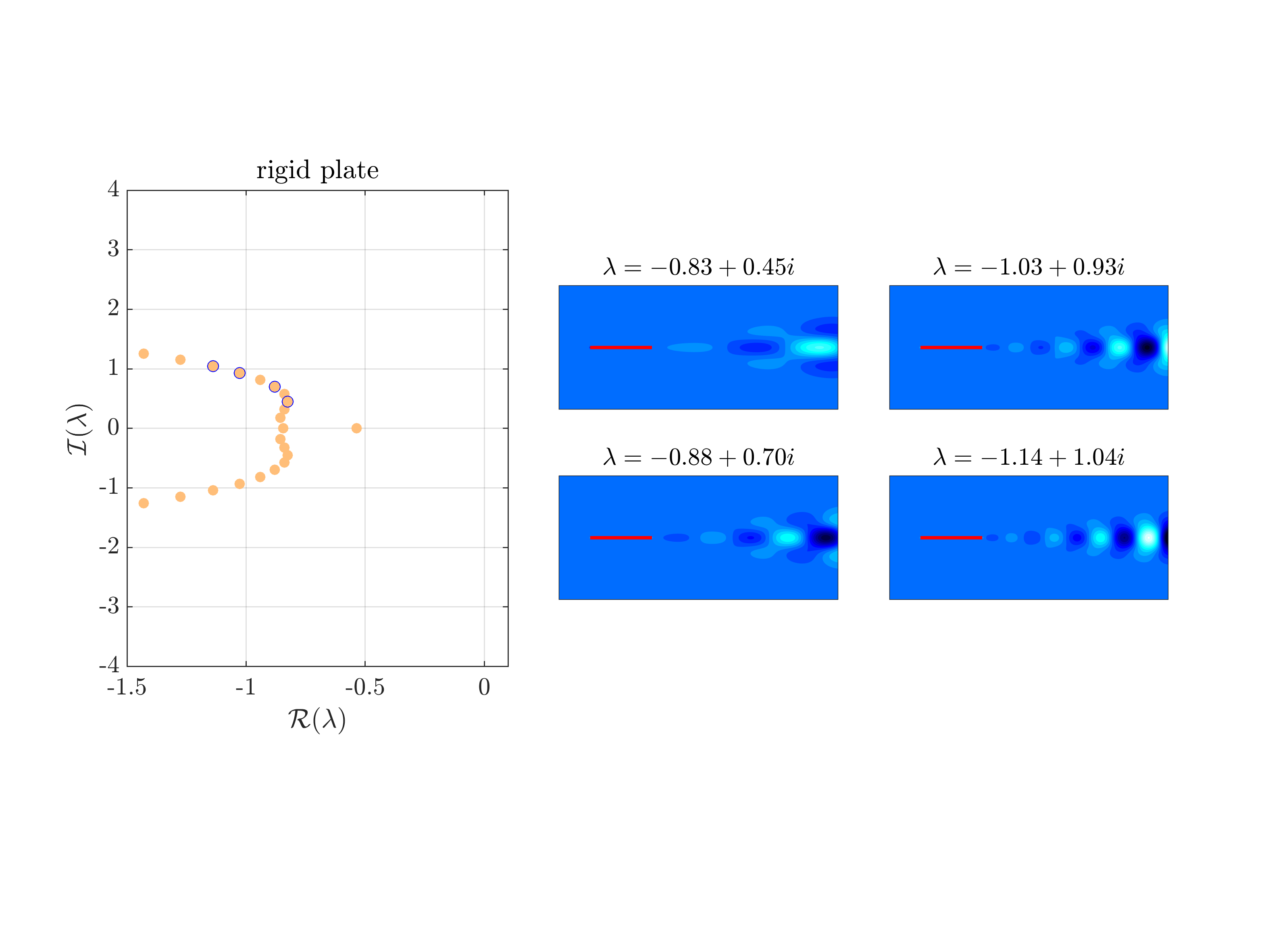}
	\caption{Left: eigenvalues associated with flow past a rigid stationary plate; \emph{i.e.}, without fluid-structure interaction. Note that all eigenvalues in the figure have corresponding eigenvectors with zero value of $E_r$ since the plate is rigid. Middle and right: eigenvectors associated with the four circled eigenvalues plotted at a time instance during the response given by the real part of $\bm{\phi}_je^{\lambda_j t}$. The eigenvectors are scaled to have unit 2-norm, and contours are plotted in 20 increments from $-0.05$ to 0.05.}
	\label{fig:evals_rigid}
\end{figure}

To explain the plots in figure~\ref{fig:evals_flex}, we first consider the specific stiffness $S = 20$ in figure~\ref{fig:evals_flex}. The top-left plot of the figure shows that there is one complex eigenvalue pair associated with a large value of $E_r$\footnote{There is also a zero frequency mode of non-negligible energy that coincides with divergence of the plate in a given transverse direction, depending on the nature of the perturbation. This mode is not expected to be significant in swimming, which inherently involves flapping behavior of non-zero frequency.}, and the insert demonstrates that the associated eigenvector has a plate shape reflective of the first mode of an Euler-Bernoulli beam. There is minimal response in the fluid for this eigenvector (also shown in the insert). Despite the similarity of plate shape to the first mode of an Euler-Bernoulli beam, note that the frequency is smaller by two orders of magnitude than for that of an Euler-Bernoulli beam in a vacuum. This discrepancy attests to the significant role of added mass and viscosity in defining resonance for plates of low $M$ values. In addition to this structure-driven mode, there are several flow-driven modes that correspond to a small value of $E_r$. Indeed, these modes are similar to those associated with flow past a \emph{rigid, stationary} body; \emph{cf.}, figure~\ref{fig:evals_rigid}.

As the stiffness is decreased from $S=20$, the structure-driven mode decreases in frequency and there is an increasingly pronounced wake response as the eigenvalue approaches those of the flow-driven modes. For $S=2$ and $S=0.5$, the proximity of the structure- and flow-driven modes leads to multiple eigenvectors with nontrivial value of $E_r$. By $S=0.2$, the mode reminiscent of the first mode of an Euler-Bernoulli beam is associated with a much smaller value of $E_r$. Moreover, a new structure-driven mode appears that is similar to the second mode of an Euler-Bernoulli beam (though again, the natural frequency is different from that of an Euler-Bernoulli beam in a vacuum by two orders of magnitude). As occurred previously, the mode is initially associated with isolated wake structures, though as the stiffness is decreased the eigenvalue approaches those of the flow-driven modes and the mode's wake becomes progressively expansive. Finally, for $S = 0.02$, there is no signature of a mode reminiscent of the second mode of an Euler-Bernoulli beam with a large value of $E_r$. Instead, a new structure-driven mode appears that resembles the third mode of an Euler-Bernoulli beam and has a wake structure localized near the plate. One expects that the pattern would persist for lower stiffnesses, with the eigenvalue approaching those of the flow-driven modes (corresponding to a more extensive wake signature), and eventually disappearing and giving way to a higher-order structure-driven mode\footnote{Though for very low stiffnesses, one would expect flutter modes to appear and affect the trends described in the main text. The appearance of these flutter modes was, for example, seen by \citet{Floryan2018} for inviscid flows. }.

The remainder of this article will investigate the extent to which this resonant behavior plays a role in finite-amplitude swimming. The nonlinear results will show that for a given stiffness, the `resonant response' that leads to performance peaks coincides with actuating at a frequency near the natural frequency associated with the mode for which $E_r$ is largest. Actuating at frequencies corresponding to the other eigenvalues is demonstrated to have a negligible effect on performance. At the same time, the results will show that there are parameters for which nonlinear mechanisms obscure this resonant behavior. An example of such a nonlinear process is the vortex shedding and interaction that is a key feature of finite-amplitude swimming. This effect is notably absent in several of the modes depicted in figure~\ref{fig:evals_flex}, where vorticity is isolated near the plate. 


\section{Connections between nonlinearity and resonance}
\label{sec:Nonlin_results}

We now compare results from nonlinear high-fidelity simulations for $S = 0.02, 0.2, 2, 20$ to the resonant behavior provided by the linear stability modes described in section \ref{sec:GMresults}. Results are first presented at relatively large amplitudes relevant to fish swimming, $h_0 = 0.1$. Connections between this large-amplitude behavior and resonance are then drawn by considering nonlinear simulations of increasing amplitude.

\subsection{General observations for large-amplitude motions}

\begin{figure}
\centering
	\begin{subfigure}[b]{0.45\textwidth}
		\hspace*{2.9mm}
        		\includegraphics[scale=0.35,trim={0cm 1.9cm 0cm 0cm},clip]{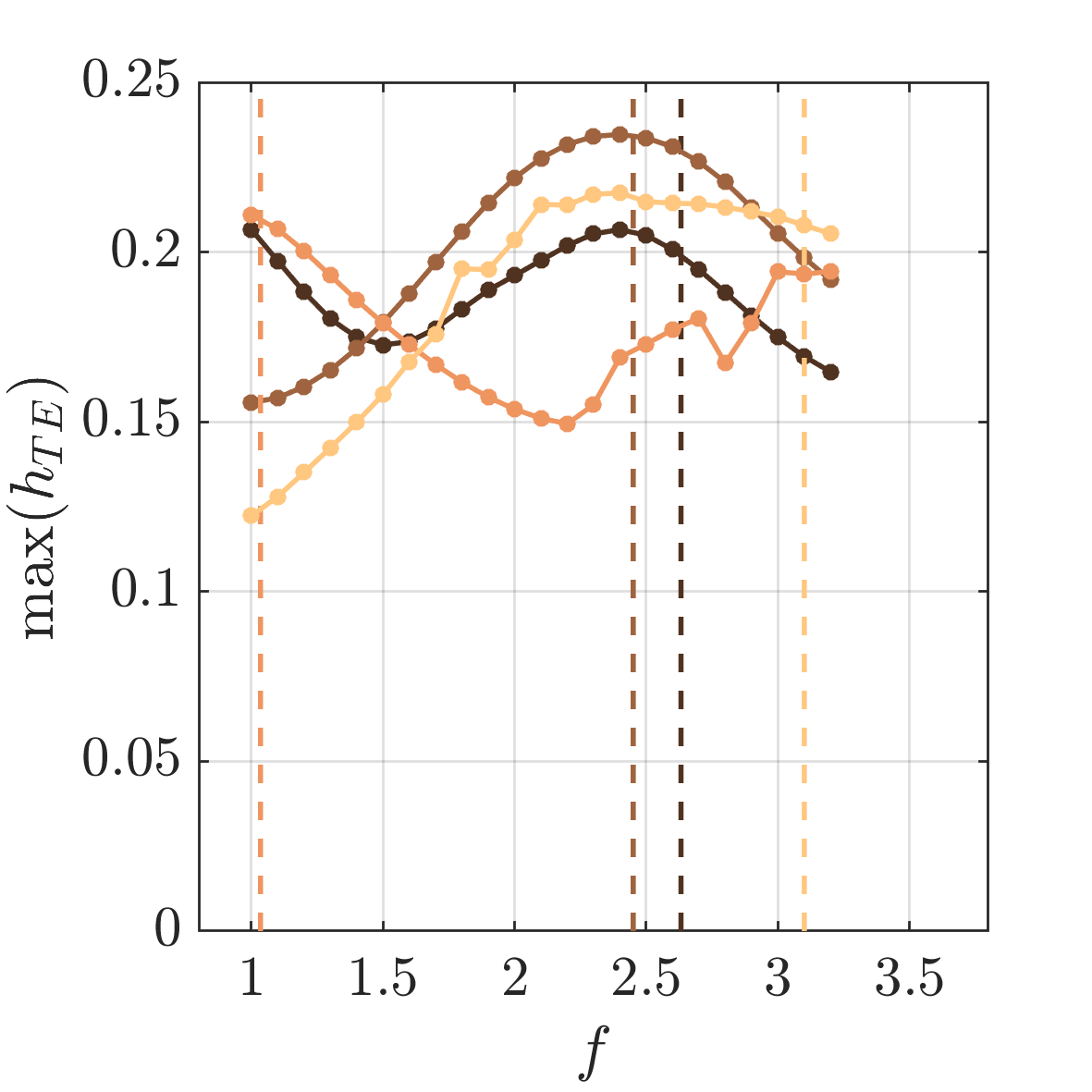}
	\end{subfigure}
	\begin{subfigure}[b]{0.45\textwidth}
        		\includegraphics[scale=0.35,trim={0cm 1.9cm 0cm 0cm},clip]{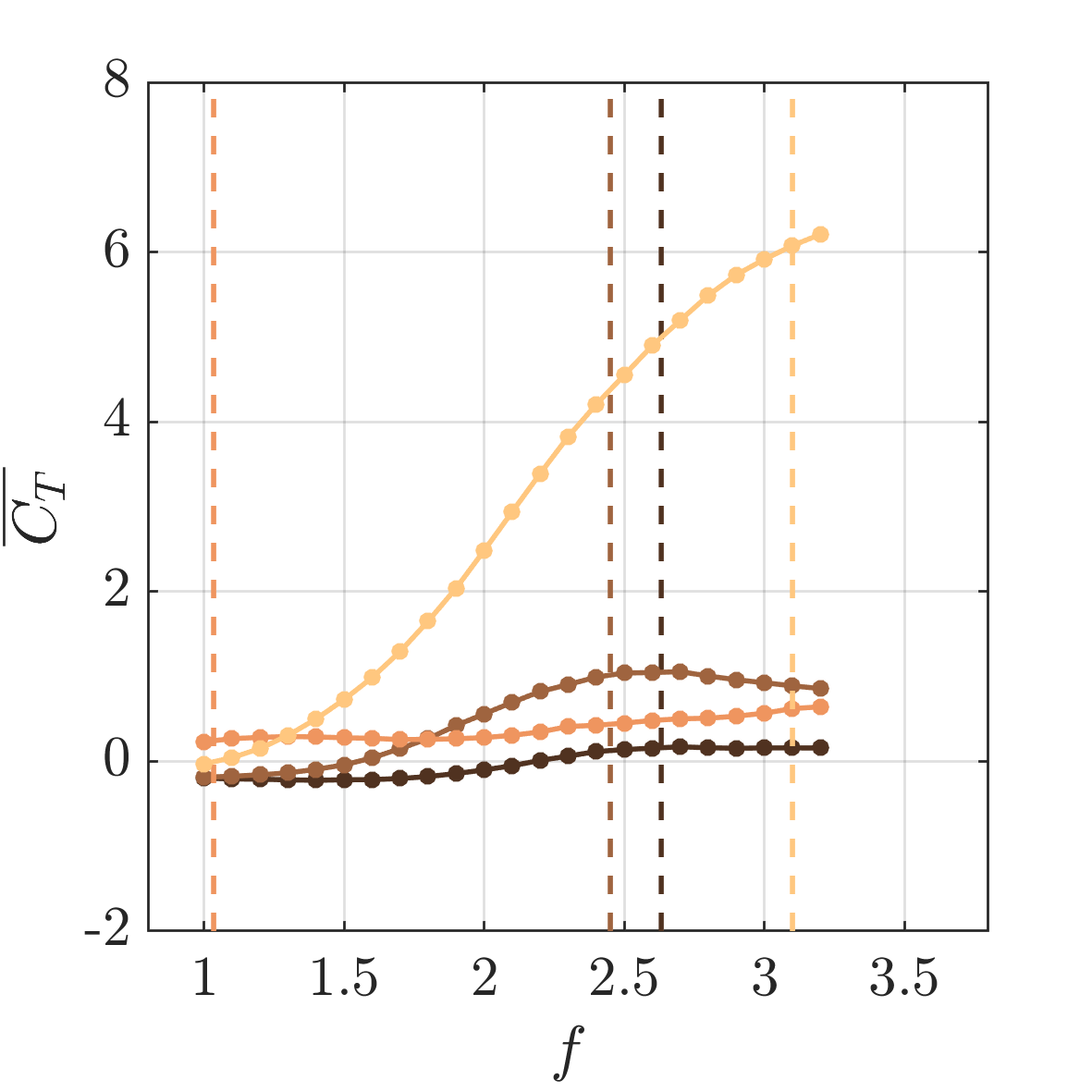}
	\end{subfigure}
	
    	\begin{subfigure}[b]{0.45\textwidth}
		\hspace*{5.5mm}
        		\includegraphics[height = 5.0cm, width = 5.0cm,trim={0cm 0cm 0cm 0cm},clip]{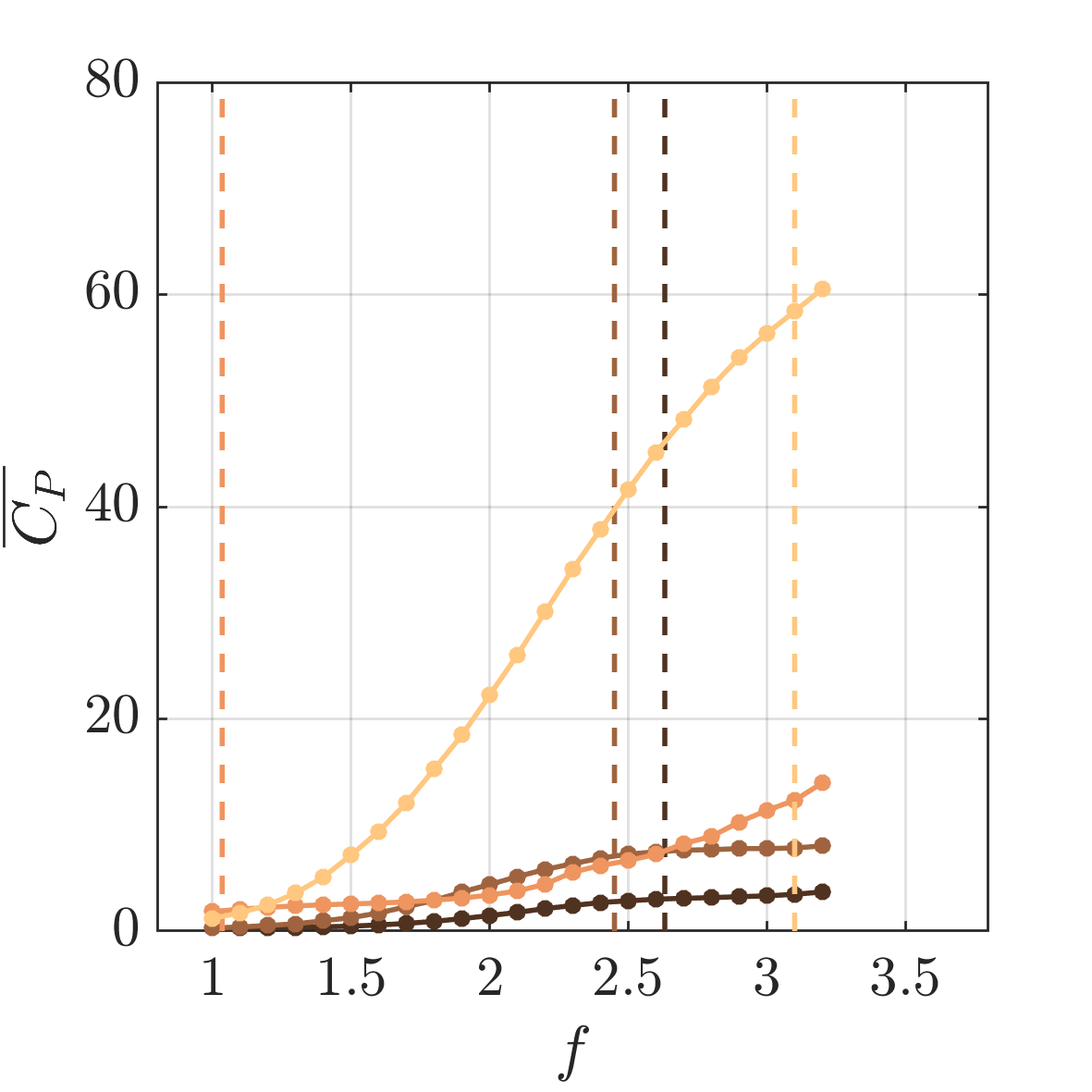}
	\end{subfigure}
	\begin{subfigure}[b]{0.45\textwidth}
		\hspace*{-3.5mm}
        		\includegraphics[height = 5.0cm, width = 5.5cm,trim={0cm 0cm 0cm 0cm},clip]{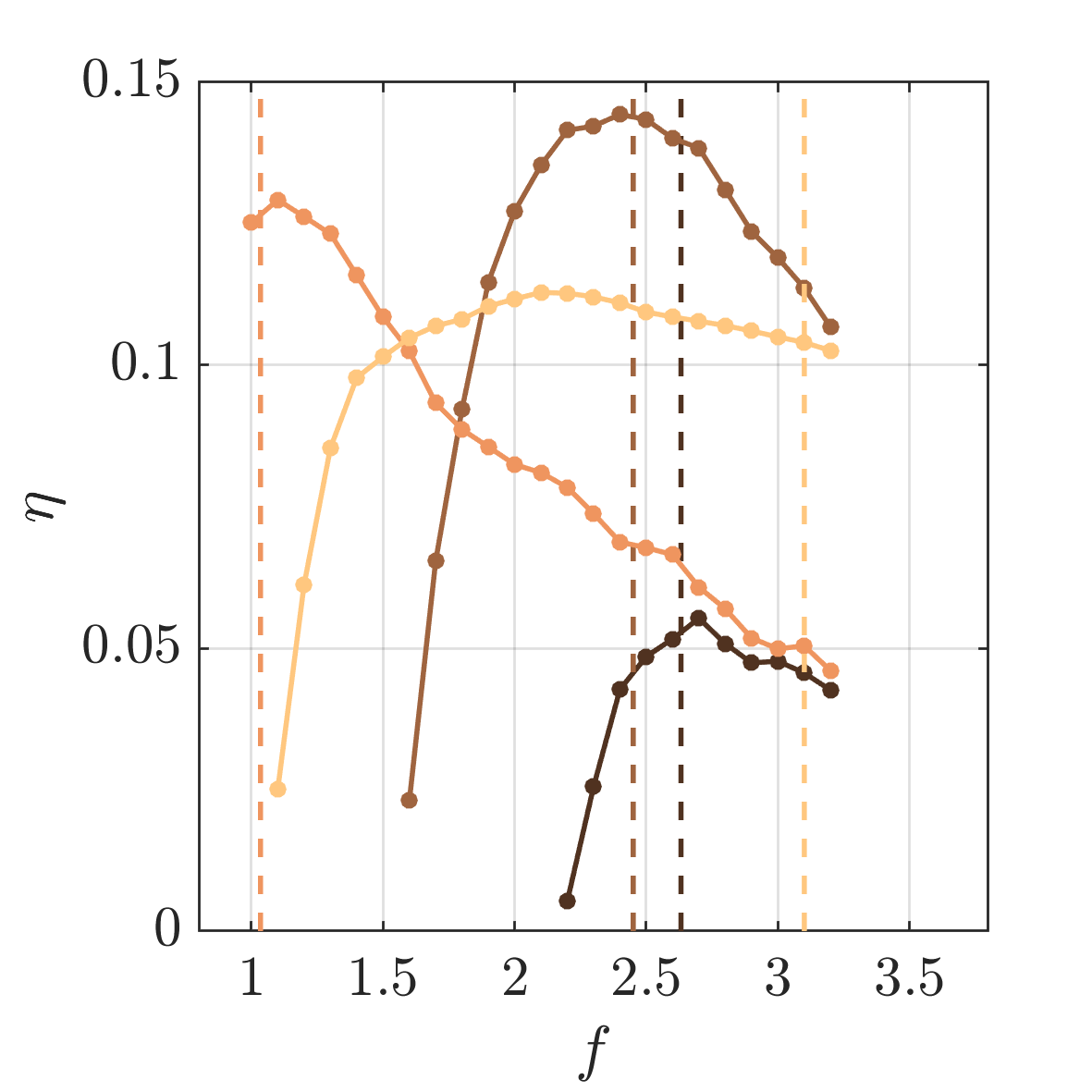}
	\end{subfigure}
    \caption{Maximum transverse displacement of the trailing edge (left), mean thrust (second from left), mean input power (second from right), and Froude efficiency (right) versus dimensionless frequency, $f$, obtained from the nonlinear simulations for $h_0=0.1$. Each plot contains four separate colors corresponding to four different stiffnesses: $S = 0.02$ (\protect\solidrule{black!60!orange!100!}), $S = 0.2$ (\protect\solidrule{black!30!orange!100!}), $S = 2$ (\protect\solidrule{white!20!orange!100!}), and $S = 20$ (\protect\solidrule{white!50!orange!100!}). The dashed vertical lines correspond to the natural frequency defined using the imaginary part of the global eigenvalues presented in section \ref{sec:GMresults}.}
	\label{fig:global_performance_nonlin}
\end{figure}

Figure~\ref{fig:global_performance_nonlin} provides various quantities related to performance. Some features of these performance plots appear reflective of resonant behavior. For example, for every stiffness except the stiffest case ($S = 20$), there are peaks in trailing-edge amplitude, mean thrust and efficiency at roughly the natural frequency of the global linear system (though note that in the case of mean thrust, the peak near the resonant frequency is only a local maximum for $S = 2$). 

However, there are also differences between the nonlinear and linear results that suggest the presence of non-resonant mechanisms. First, even when there are peaks in trailing-edge amplitude, mean thrust, and efficiency for a given stiffness, the peaks occur at slightly different frequencies for each quantity. For example, for $S = 2$ the maximum trailing-edge amplitude, mean thrust, and efficiency occur at $f = 1, 1.3,$ and~$1.1$, respectively. Second, the swimming performance for $S = 20$ is entirely different from what one would expect from resonance-based arguments; \emph{e.g.}, there are broad peaks in maximum trailing-edge amplitude and efficiency, with maxima occurring at $f = 2.4$ and $f = 2.1$, respectively (well below the resonant frequency of $f \approx 3.1$). Third, for $S = 2$, there is an increase in maximum trailing-edge amplitude and mean thrust for $f\ge 2.4$ despite the fact that there are no eigenvalues corresponding to this frequency range (or to a mode at higher frequency up to $f=4$) for this stiffness. 

\begin{figure}
\centering
	\begin{subfigure}[b]{0.24\textwidth}
		\hspace*{-3mm}
        		\includegraphics[scale = 0.42,trim={0cm 1.45cm 0cm 0cm},clip]{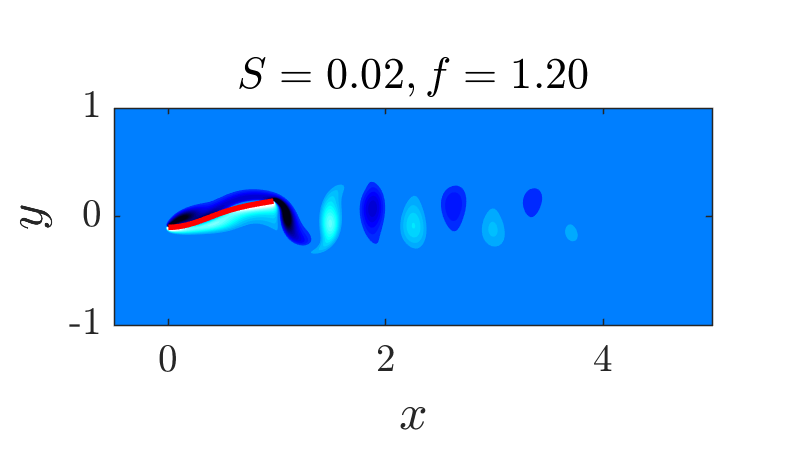}
	\end{subfigure}
	\begin{subfigure}[b]{0.24\textwidth}
		\hspace*{1.9mm}
        		\includegraphics[scale = 0.42,trim={1.4cm 1.45cm 0cm 0cm},clip]{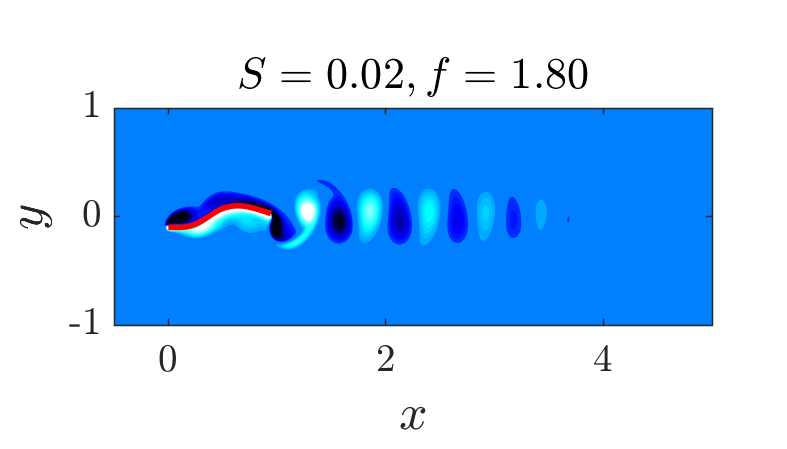}
	\end{subfigure}
    	\begin{subfigure}[b]{0.24\textwidth}
		\hspace*{0.95mm}
        		\includegraphics[scale = 0.42,trim={1.4cm 1.45cm 0cm 0cm},clip]{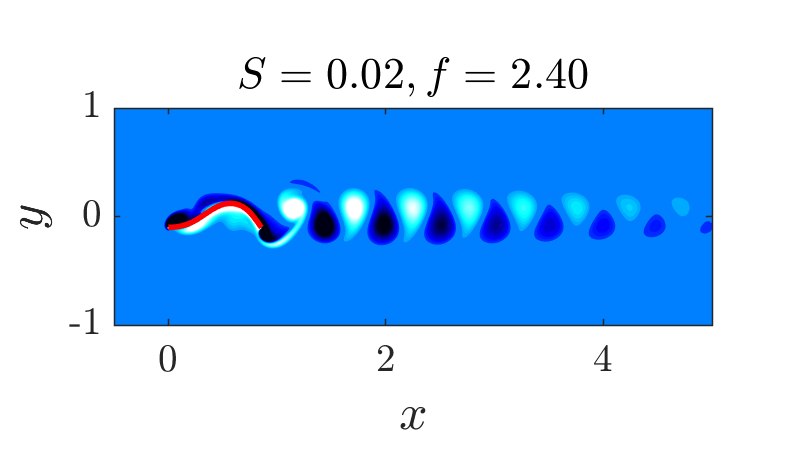}
	\end{subfigure}
	\begin{subfigure}[b]{0.24\textwidth}
		\hspace*{0mm}
        		\includegraphics[scale = 0.42,trim={1.4cm 1.45cm 0cm 0cm},clip]{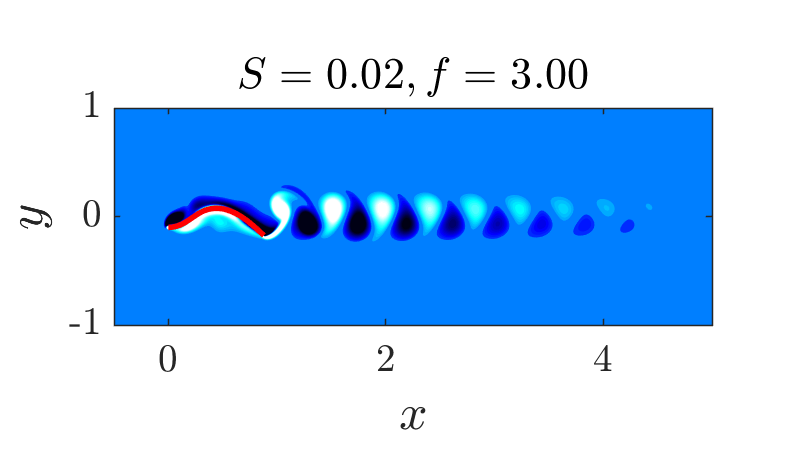}
	\end{subfigure}

\begin{subfigure}[b]{0.24\textwidth}
		\hspace*{-3mm}
        		\includegraphics[scale = 0.42,trim={0cm 1.45cm 0cm 0cm},clip]{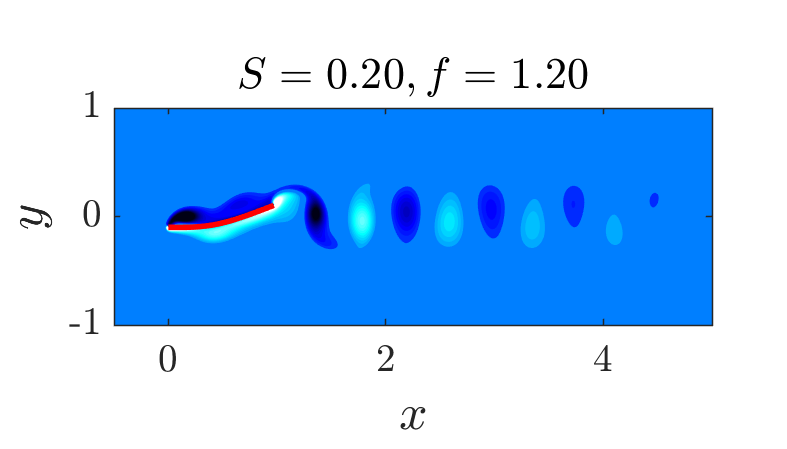}
	\end{subfigure}
	\begin{subfigure}[b]{0.24\textwidth}
		\hspace*{1.9mm}
        		\includegraphics[scale = 0.42,trim={1.4cm 1.45cm 0cm 0cm},clip]{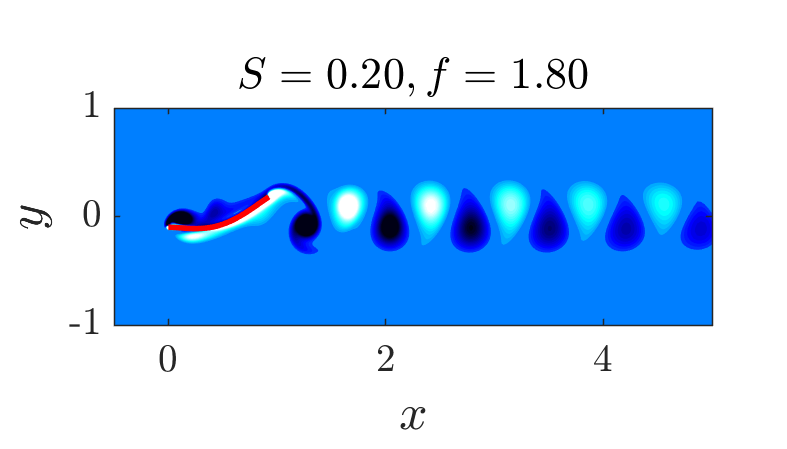}
	\end{subfigure}
    	\begin{subfigure}[b]{0.24\textwidth}
		\hspace*{0.95mm}
        		\includegraphics[scale = 0.42,trim={1.4cm 1.45cm 0cm 0cm},clip]{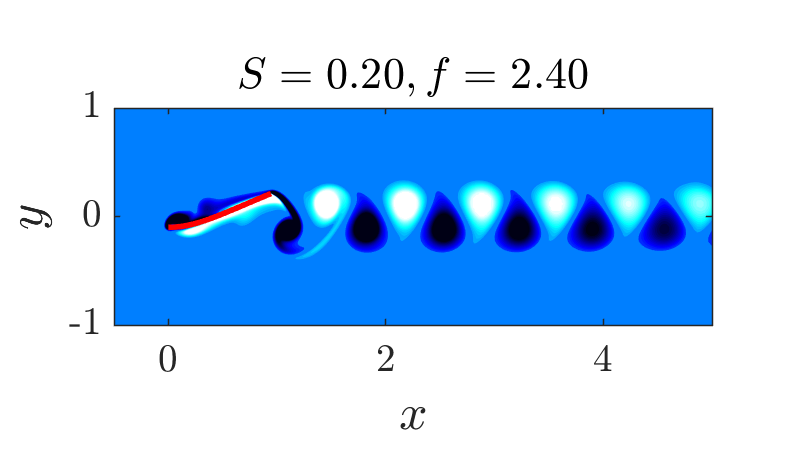}
	\end{subfigure}
	\begin{subfigure}[b]{0.24\textwidth}
		\hspace*{0mm}
        		\includegraphics[scale = 0.42,trim={1.4cm 1.45cm 0cm 0cm},clip]{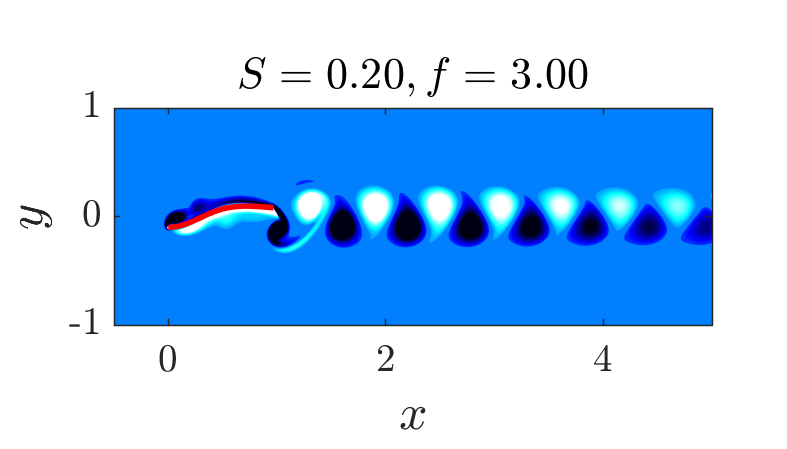}
	\end{subfigure}
	
	\begin{subfigure}[b]{0.24\textwidth}
		\hspace*{-3mm}
        		\includegraphics[scale = 0.42,trim={0cm 1.45cm 0cm 0cm},clip]{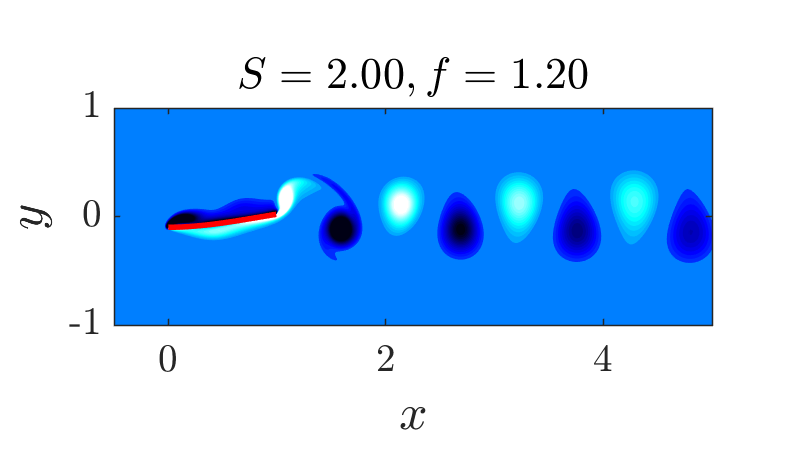}
	\end{subfigure}
	\begin{subfigure}[b]{0.24\textwidth}
		\hspace*{1.9mm}
        		\includegraphics[scale = 0.42,trim={1.4cm 1.45cm 0cm 0cm},clip]{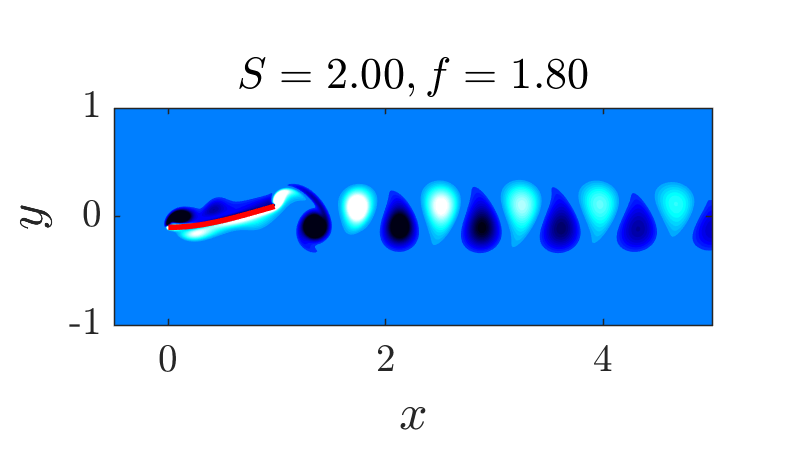}
	\end{subfigure}
    	\begin{subfigure}[b]{0.24\textwidth}
		\hspace*{0.95mm}
        		\includegraphics[scale = 0.42,trim={1.4cm 1.45cm 0cm 0cm},clip]{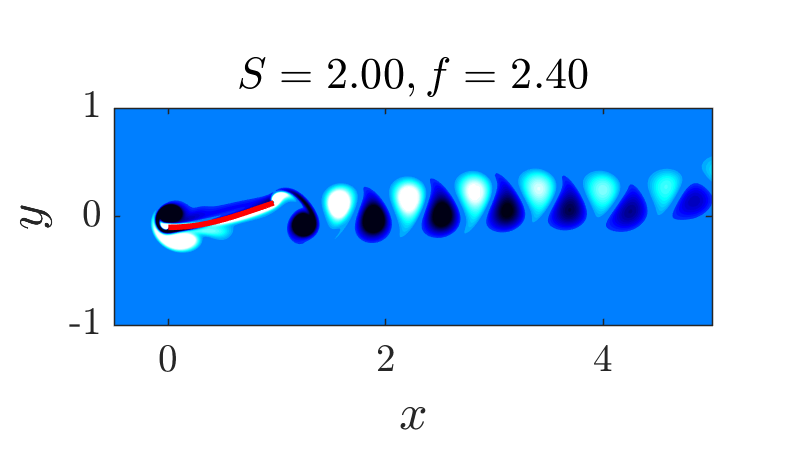}
	\end{subfigure}
	\begin{subfigure}[b]{0.24\textwidth}
		\hspace*{0mm}
        		\includegraphics[scale = 0.42,trim={1.4cm 1.45cm 0cm 0cm},clip]{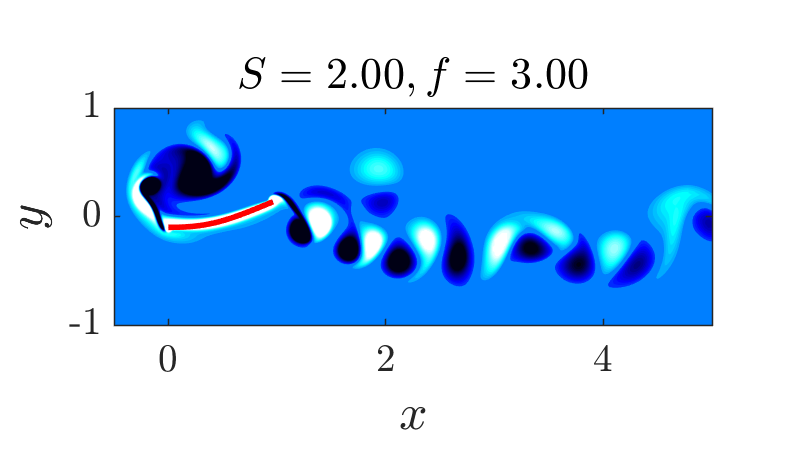}
	\end{subfigure}

	\begin{subfigure}[b]{0.24\textwidth}
		\hspace*{-3mm}
        		\includegraphics[scale = 0.42,trim={0cm 0cm 0cm 0cm},clip]{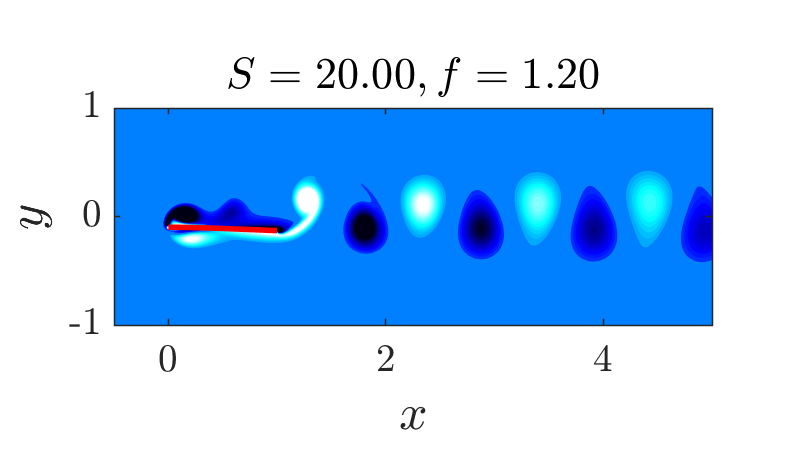}
	\end{subfigure}
	\begin{subfigure}[b]{0.24\textwidth}
		\hspace*{1.9mm}
        		\includegraphics[scale = 0.42,trim={1.4cm 0cm 0cm 0cm},clip]{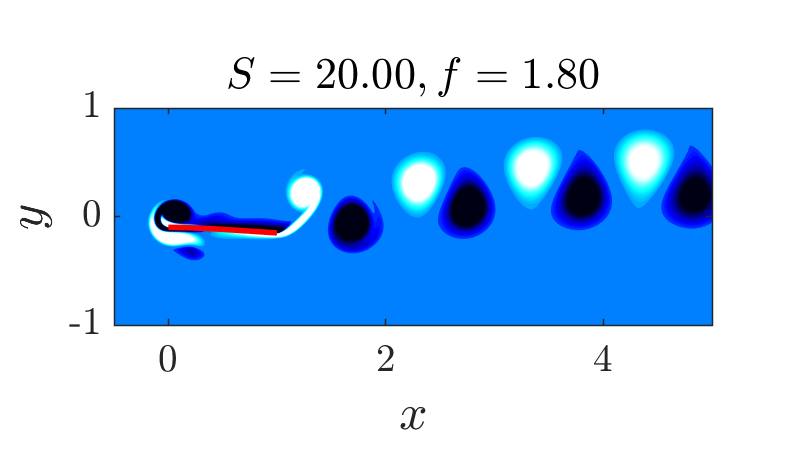}
	\end{subfigure}
    	\begin{subfigure}[b]{0.24\textwidth}
		\hspace*{0.95mm}
        		\includegraphics[scale = 0.42,trim={1.4cm 0cm 0cm 0cm},clip]{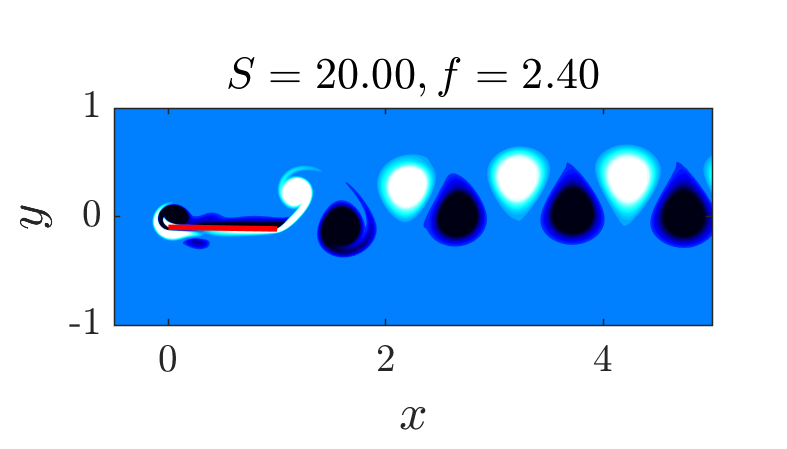}
	\end{subfigure}
	\begin{subfigure}[b]{0.24\textwidth}
		\hspace*{0mm}
        		\includegraphics[scale = 0.42,trim={1.4cm 0cm 0cm 0cm},clip]{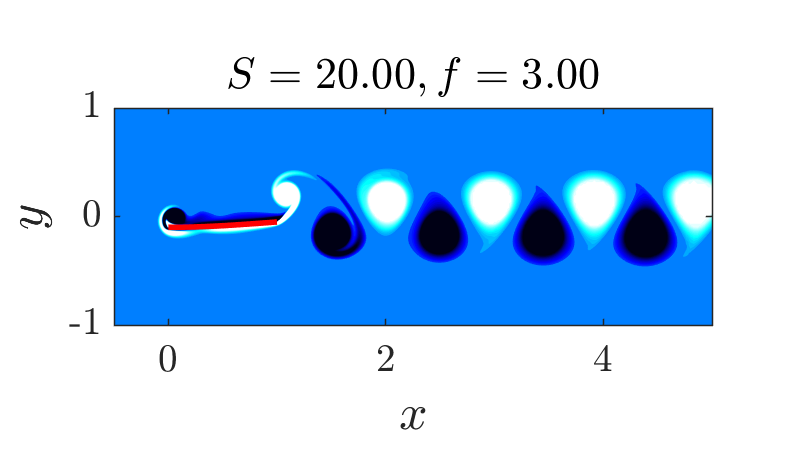}
	\end{subfigure}
	\caption{Snapshots during a flapping period, with each row containing a different stiffness and each column containing a different frequency. In each plot, the snapshot corresponds to the time instance where the plate's leading edge is at its negative peak. }
	\label{fig:snapshots_nonlin}
\end{figure}

Connections to and distinctions from resonant behavior are also suggested by the snapshots of large-amplitude swimming presented in figure~\ref{fig:snapshots_nonlin}. One connection to the resonant predictions is that near the resonant frequency, the plate shape in the nonlinear simulations resembles that of the global mode. However, the vortex dynamics in the nonlinear simulations are distinct from those observed in the global linear analysis: the finite-amplitude motion of the plate generates vortices at the leading and trailing edges, and these vortices subsequently interact in a manner not seen in the linear setting.

These observations motivate several questions about the relevance of resonance in large amplitude swimming. First, for stiffnesses where there are performance peaks near the natural frequency of the global linear system, how similar are the large-amplitude dynamics to the linear stability modes? That is, to what extent is resonance responsible for optimal performance at finite amplitudes? Second, what dynamics are associated with the boosts in trailing-edge amplitude, mean thrust, and mean input power at high frequency for $S = 2$? Finally, for $S = 20$ where there is a noticeable departure from resonance-based performance predictions, what dynamics are associated with disappearance in performance peaks? We probe these questions in the remainder of this section by considering nonlinear simulations of increasing heave amplitude, $h_0 = 0.001, 0.01,$ and~$0.1$.

\subsection{Probing the effect of increasing nonlinearity}

In this section, we consider stiffnesses $S = 0.02, 0.2, 2,$ and~$20$ separately. For each stiffness, we investigate the effect of increasing the heave amplitude from $h_0 = 0.001$, where the linear analysis of section \ref{sec:GMresults} is relevant, to $h_0 = 0.1$, which better represents fish swimming. Through this systematic variation in amplitude, we identify connections to and distinctions from resonance in swimming performance at finite heave amplitudes.

\begin{figure}
\centering
	\begin{subfigure}[b]{0.24\textwidth}
		\hspace*{-2mm}
        		\includegraphics[scale=0.24,trim={0cm 0cm 0cm 0cm},clip]{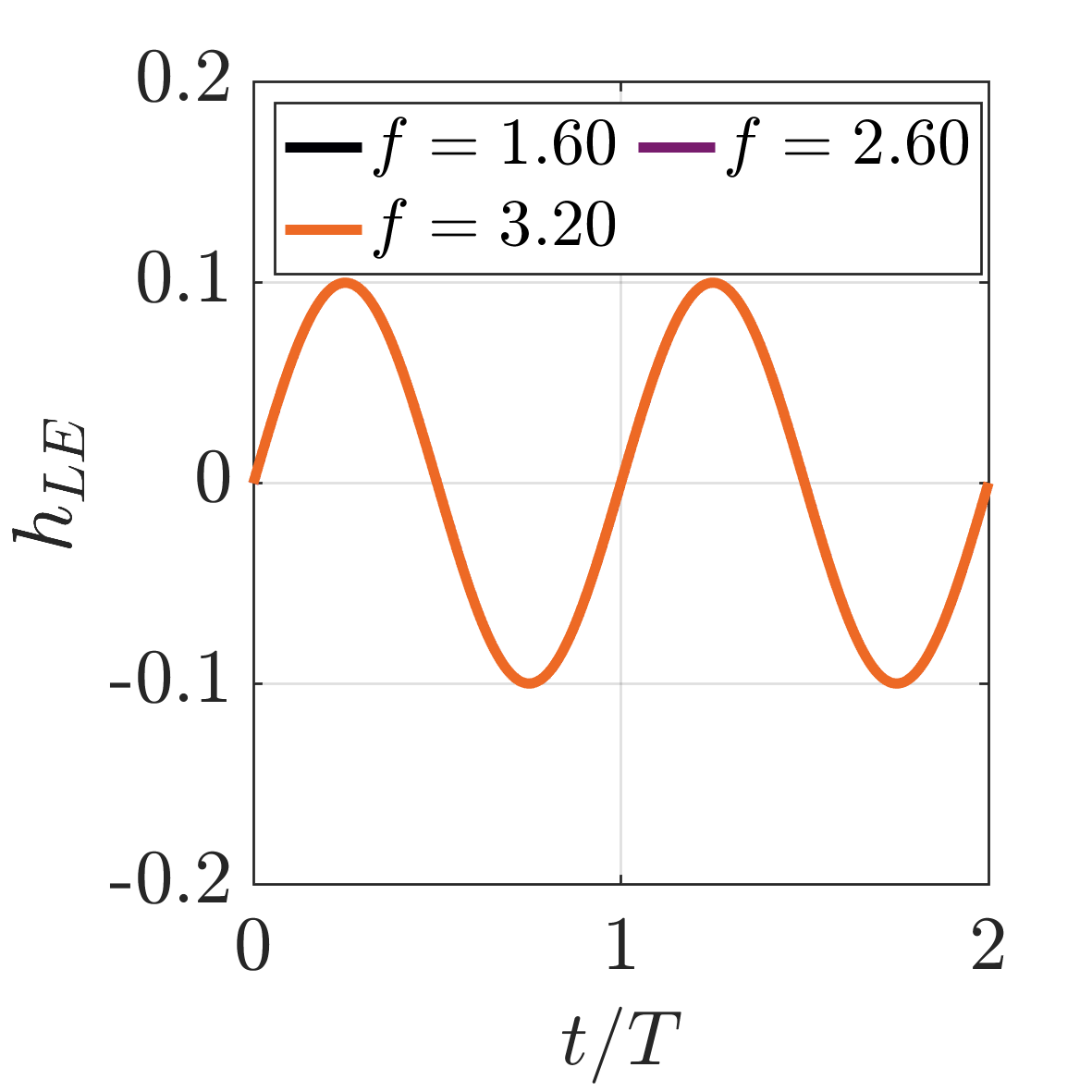}
	\end{subfigure}
	\begin{subfigure}[b]{0.24\textwidth}
		\hspace*{-1.7mm}
        		\includegraphics[scale=0.24,trim={0cm 0cm 0cm 0cm},clip]{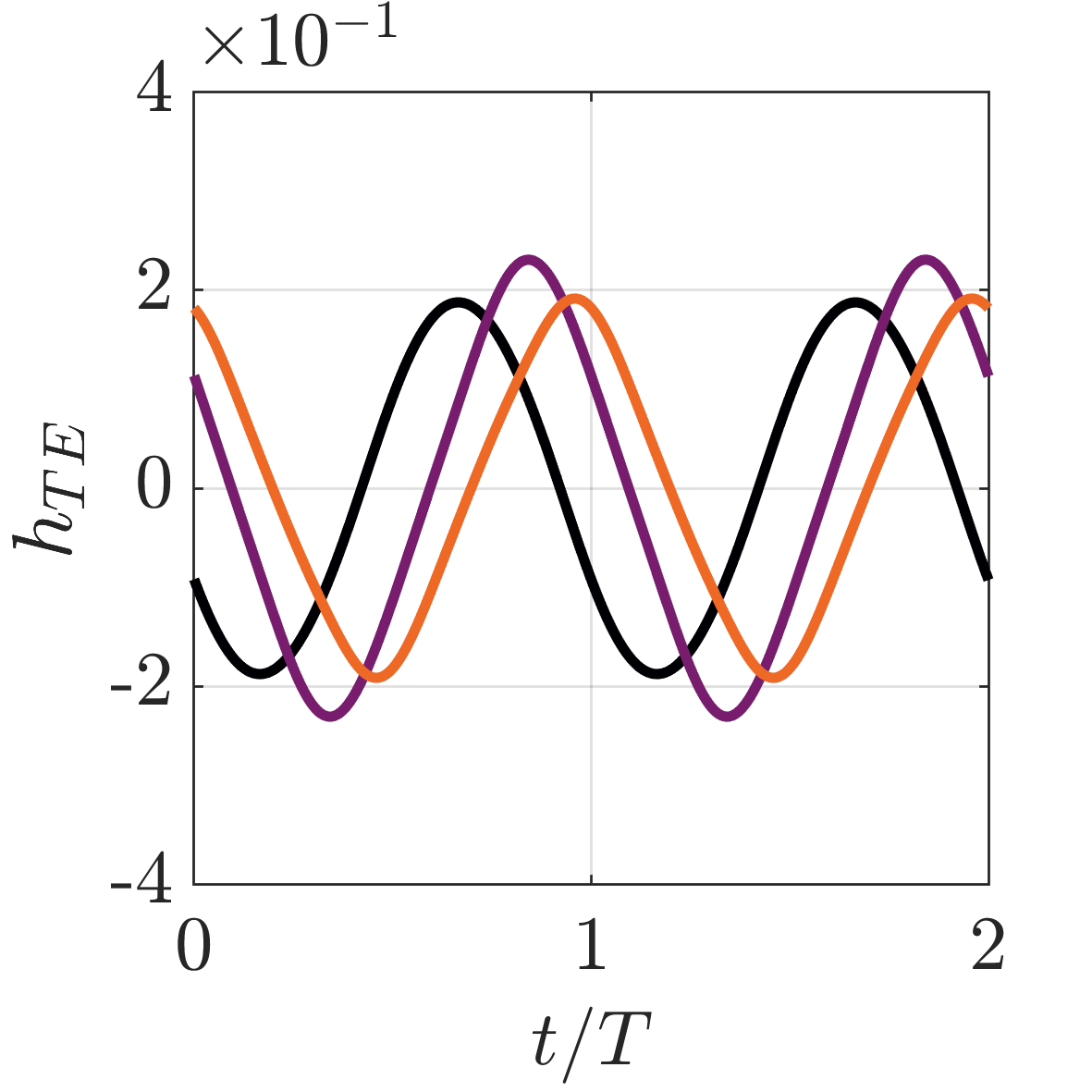}
	\end{subfigure}
	\begin{subfigure}[b]{0.24\textwidth}
		\hspace*{-1.2mm}
        		\includegraphics[scale=0.24,trim={0cm 0cm 0cm 0cm},clip]{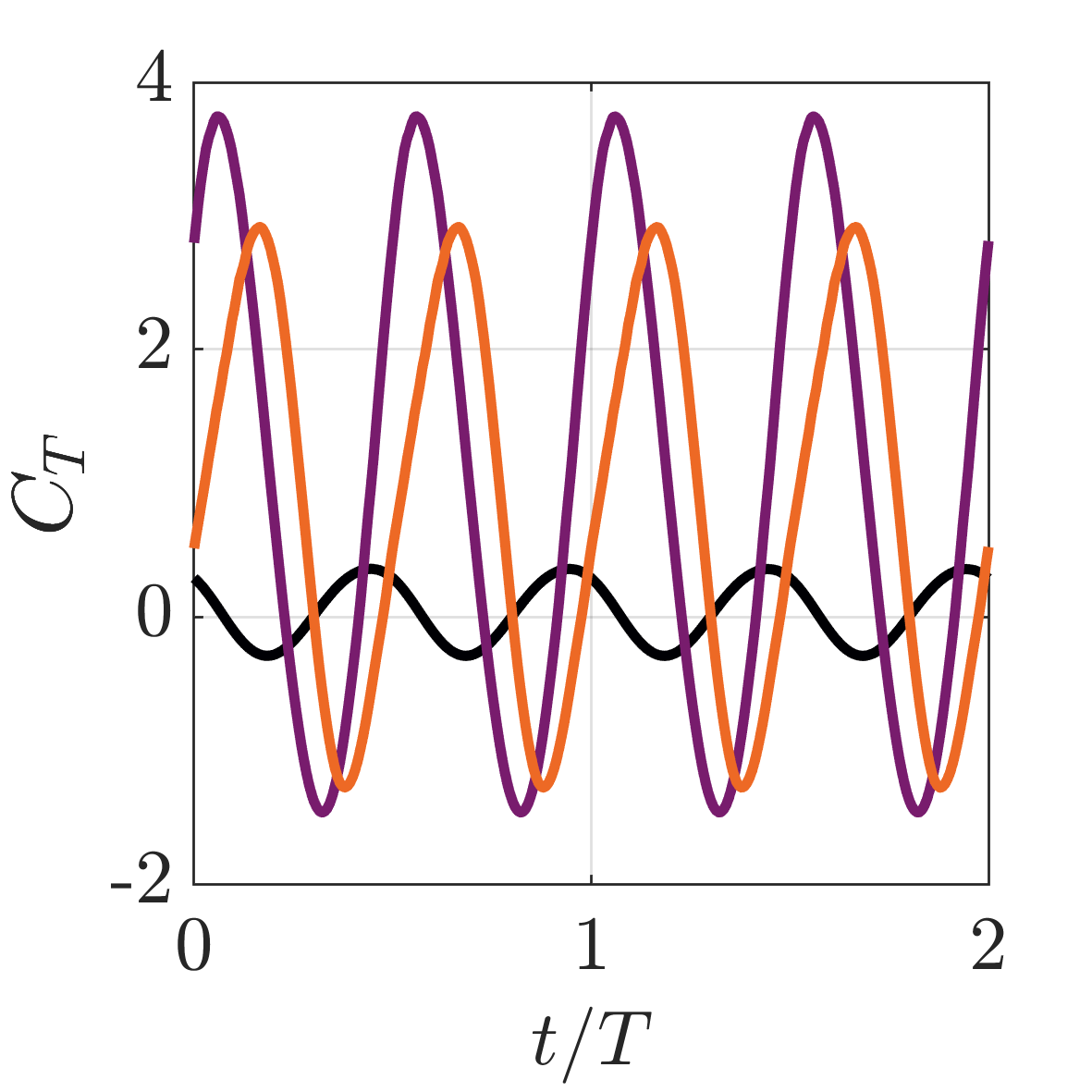}
	\end{subfigure}
    	\begin{subfigure}[b]{0.24\textwidth}
		\hspace*{-0.3mm}
        		\includegraphics[scale=0.24,trim={0cm 0cm 0cm 0cm},clip]{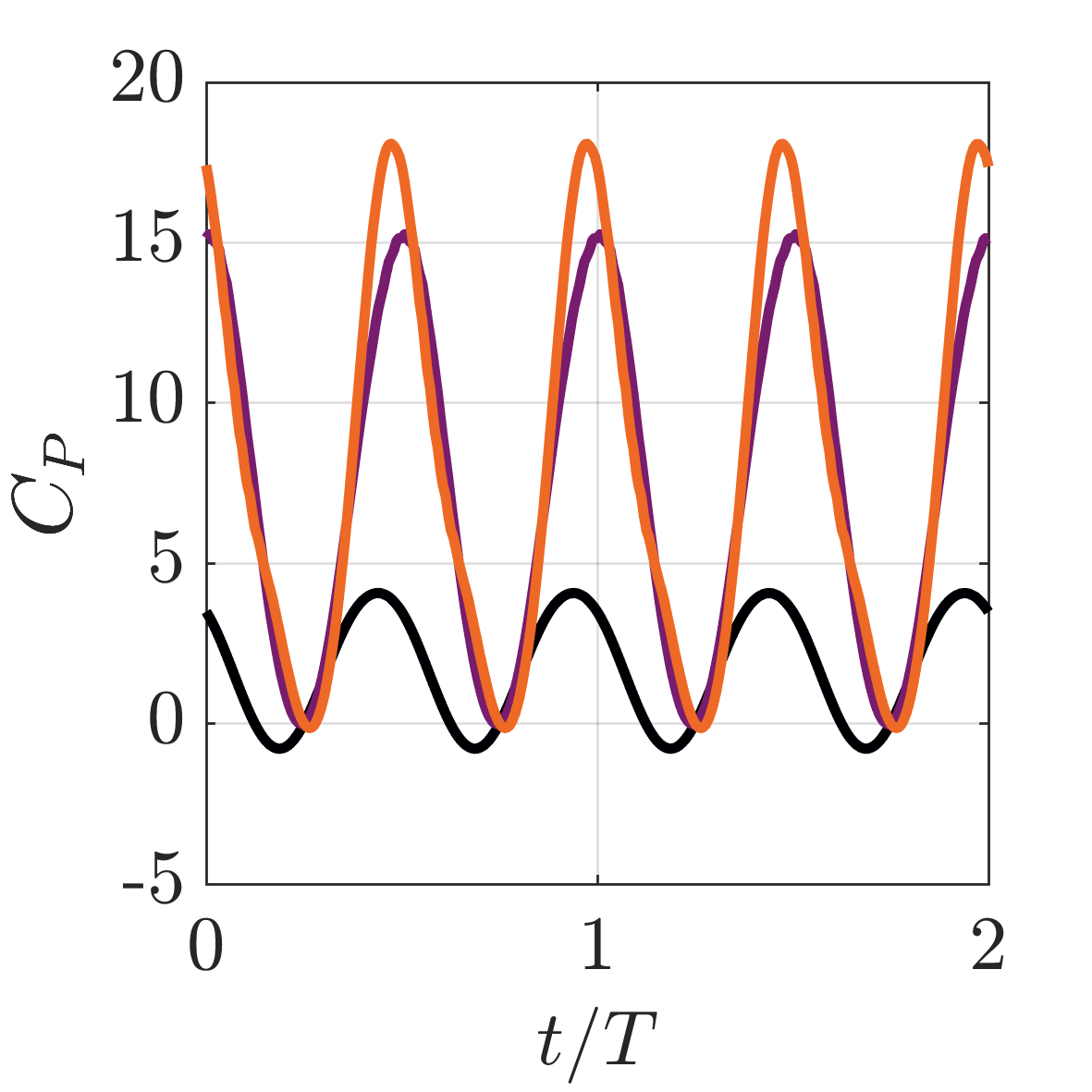}
	\end{subfigure}
    \caption{Time traces of various quantities for $S = 0.2, h_0 = 0.1$. Only the $f=3.2$ curve is visible in the plot of $h_{LE}$, as the prescribed leading-edge kinematics are identical. The limit-cycle behavior observed here is indicative of the plate-flow behavior except for $f\ge2.4$ at $S=2$.}
	\label{fig:dynamics_sample}
\end{figure}

To give context for the results below, we note that for the majority of parameters considered, the plate-fluid system exhibits limit-cycle dynamics (exceptions to this will be discussed further below). Figure~\ref{fig:dynamics_sample} depicts time traces of various quantities for $S=0.2$ and three different frequencies, which are qualitatively representative of the typical behavior of the system. To enable a meaningful comparison across frequencies, time is shifted so that $t=0$ coincides with the instance when the leading-edge displacement has maximal derivative (\emph{i.e.}, velocity), and scaled by $T$ so that the leading-edge motion has a period of unity. Note that the thrust and input power have twice the frequency of the trailing and leading-edge displacement. This fact is intuitive, since one would expect a peak in thrust or input power for both the top and bottom portions of a flapping cycle.

In the shifted and scaled representation of time, any of the quantities in figure~\ref{fig:dynamics_sample} can be written as $aw((t + \varphi)/T)$, where $a$ is the peak amplitude of the function, $w(t/T)$ is a periodic function (with a period of unity for $h_{TE}$ and $h_{LE}$ and of one half for $C_T$ and $C_P$) constructed so that $w(0)$ corresponds to the signal having a maximal derivative value, and $\varphi$ is the phase shift to account for the fact that the signal does not, in general, exhibit its maximal derivative at $t/T=0$. The expression is left in terms of a generic periodic function $w(t/T)$, since the signals are not generally sinusoids for every frequency and stiffness considered (due to a nonzero offset and the possible presence of additional harmonics). The convention of $w(0)$ coinciding with the time instant for which the derivative $dw/dt$ is maximal is used to enable an intuitive interpretation of $\varphi$: $\varphi=0$ corresponds to motion that is in phase with the prescribed leading-edge motion characterized by $h_{TE}(t)$. The prevalence of the periodic behavior in this FSI system forms the basis for much of the analysis below.

\subsubsection{Minimal stiffness: $S = 0.02$}

\begin{figure}
\centering
	\begin{subfigure}[b]{0.3\textwidth}
		\hspace*{-6mm}
        		\includegraphics[scale=0.315,trim={0cm 2.39cm 0cm 0cm},clip]{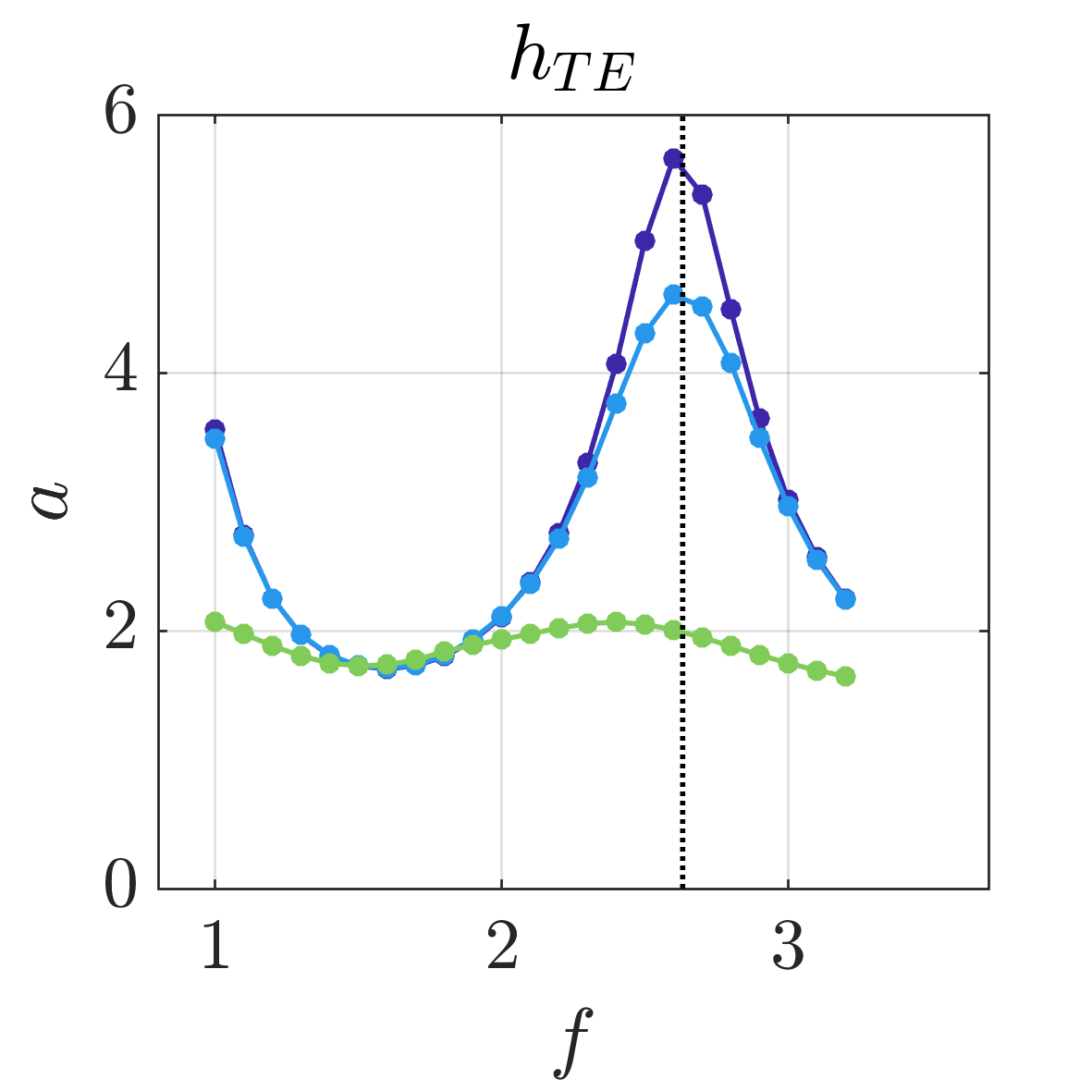}
	\end{subfigure}
	\begin{subfigure}[b]{0.3\textwidth}
		\hspace*{1mm}
        		\includegraphics[scale=0.315,trim={1cm 2.39cm 0cm 0cm},clip]{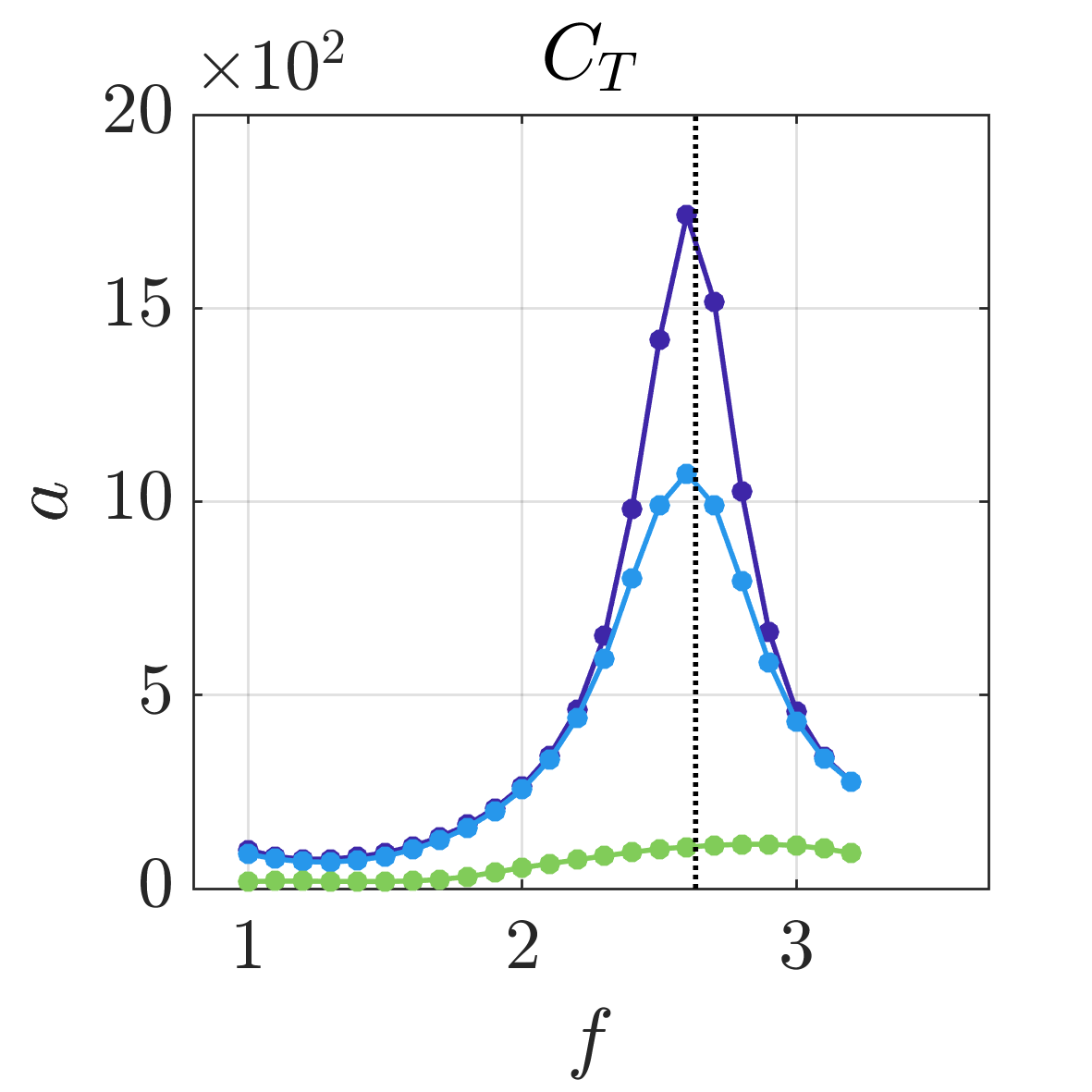}
	\end{subfigure}
	\begin{subfigure}[b]{0.3\textwidth}
		\hspace*{4.2mm}
        		\includegraphics[scale=0.315,trim={1cm 2.39cm 0cm 0cm},clip]{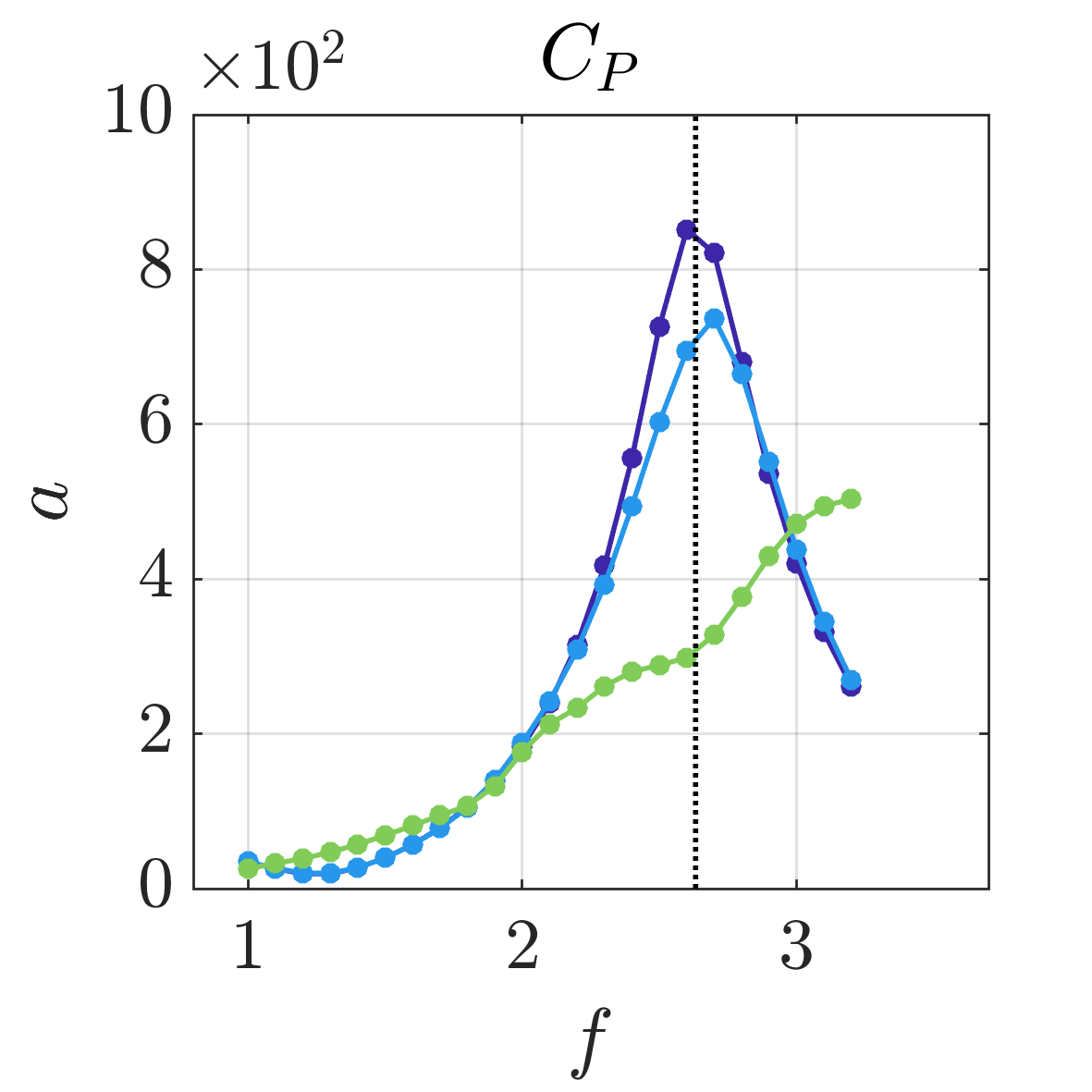}
	\end{subfigure}
	
	\begin{subfigure}[b]{0.3\textwidth}
		\hspace*{-8.5mm}
        		\includegraphics[height=4.73cm, width=5cm, trim={0cm 0cm 0cm 0cm},clip]{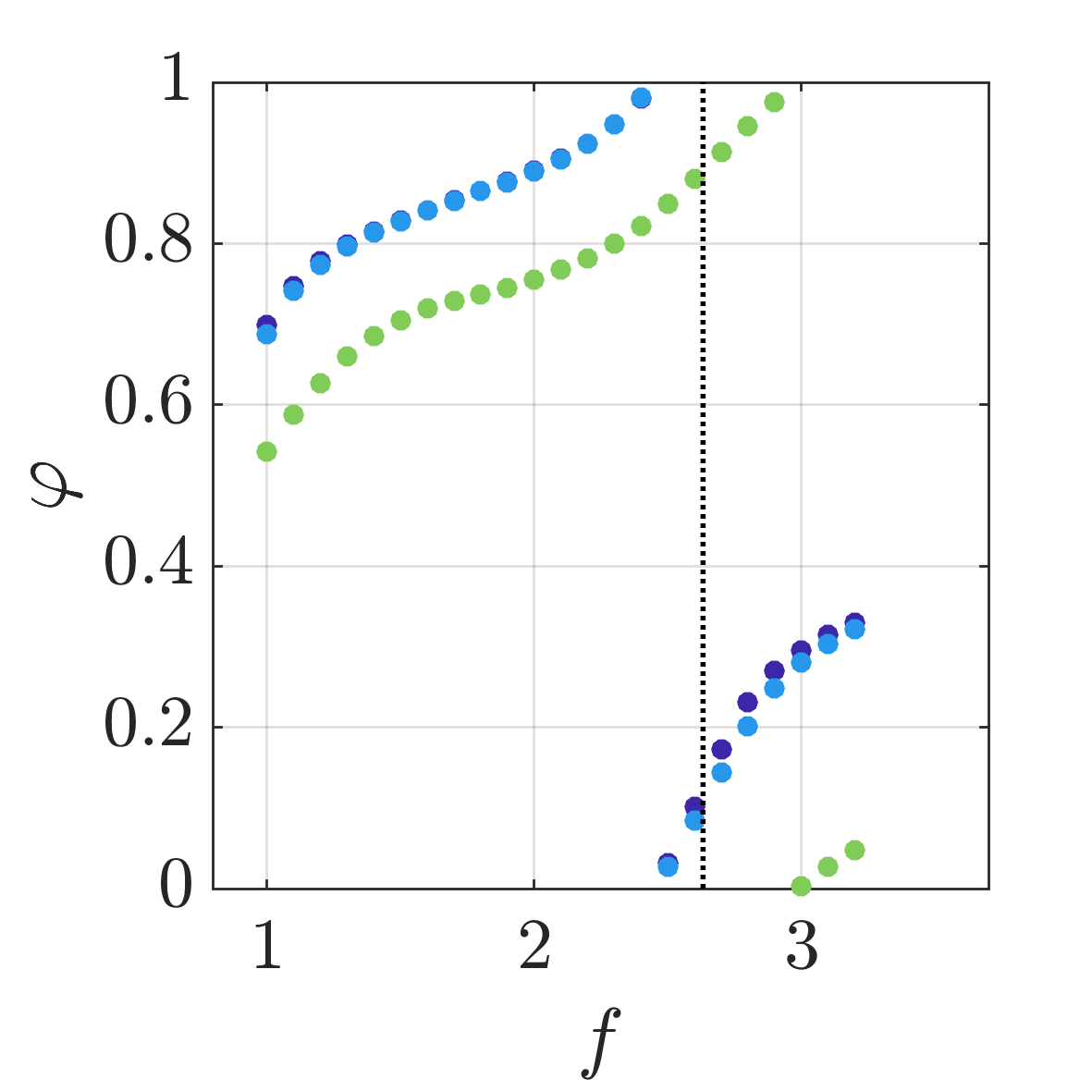}
	\end{subfigure}
	\begin{subfigure}[b]{0.3\textwidth}
		\hspace*{1.5mm}
        		\includegraphics[scale=0.315, trim={1.2cm 0cm 0cm 0cm},clip]{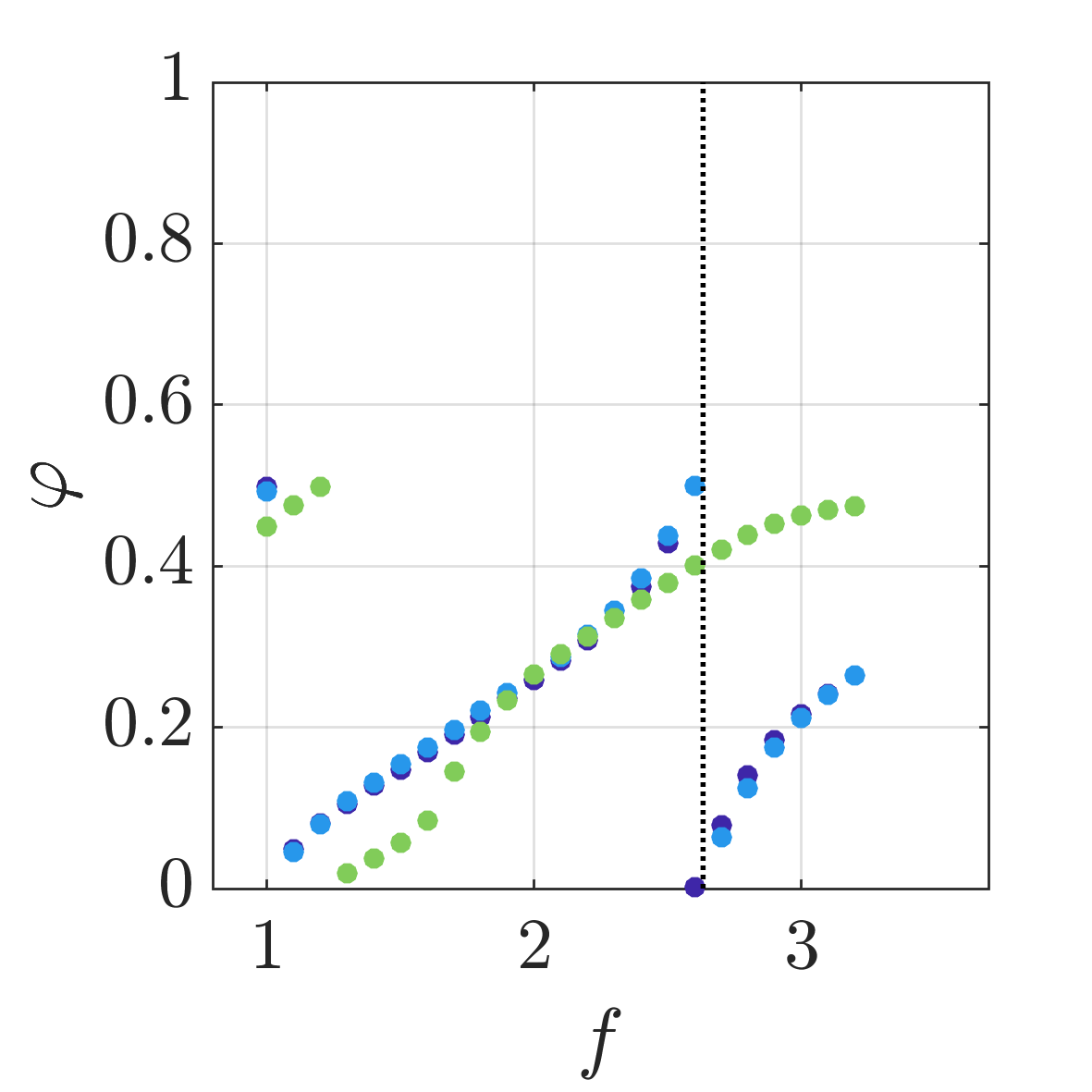}
	\end{subfigure}
	\begin{subfigure}[b]{0.3\textwidth}
		\hspace*{4.5mm}
        		\includegraphics[scale=0.315, trim={1.2cm 0cm 0cm 0cm},clip]{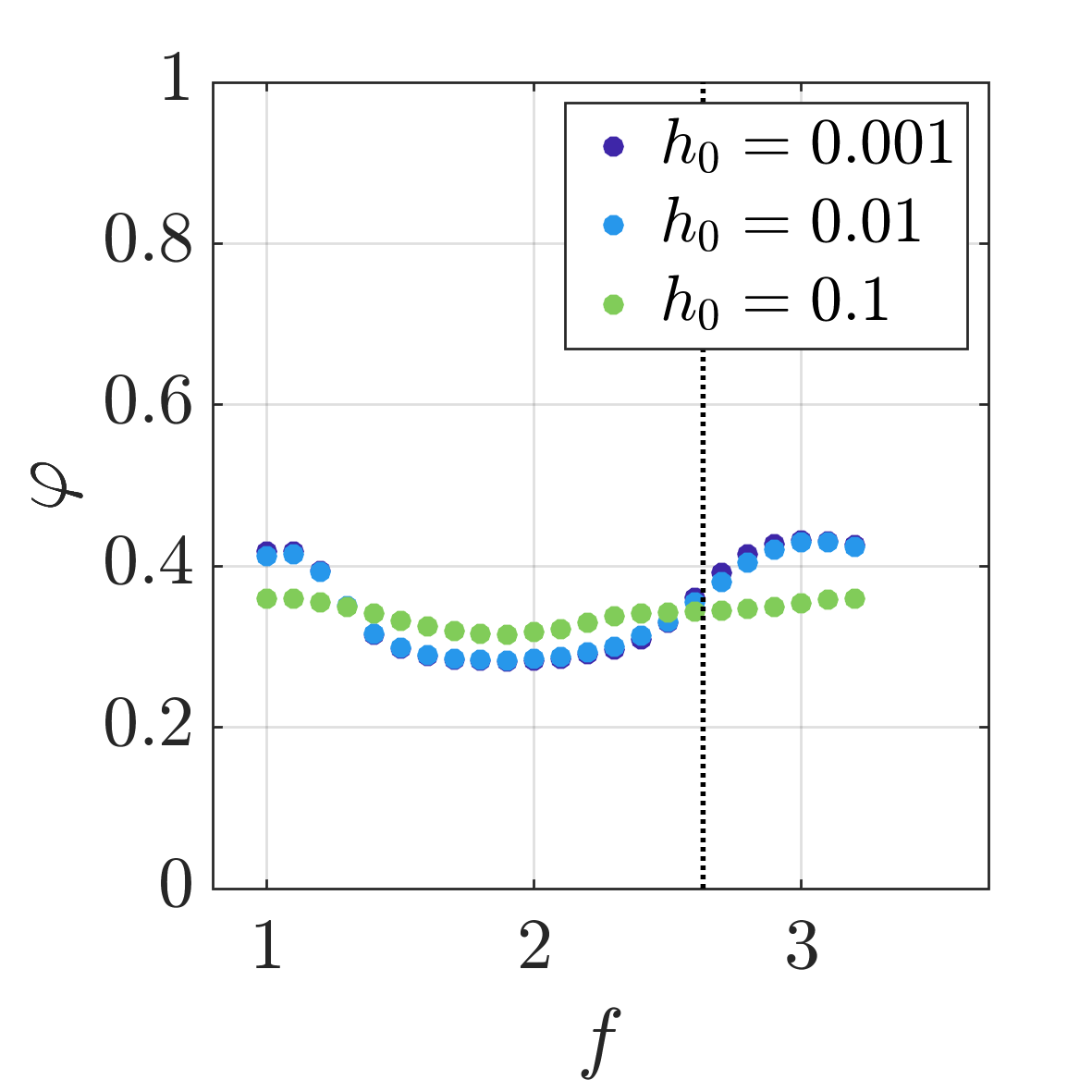}
	\end{subfigure}
    \caption{The amplitude (top row), $a$, and phase shift (bottom row), $\varphi$, of the trailing-edge displacement (left column), thrust (middle column), and input power (right column) versus frequency for $S=0.02$. Each plot contains three sets of markers corresponding to different heave amplitudes. To facilitate a comparison of the maximum amplitude across different values of $h_0$, $a$ is scaled as described in table \ref{tab:amp_scalings}. The vertical line denotes the natural frequency for this stiffness.}
	\label{fig:ampphaseSp02}
\end{figure}

Here we study the minimally stiff case ($S=0.02$) in more detail. Figure~\ref{fig:ampphaseSp02} shows the phase shift, $\varphi$, and amplitude, $a$, associated with the periodic signals $h_{TE}, C_T,$ and $C_P$ at a given frequency. Note that the phase shift exhibits a discontinuous jump at $\varphi=1$ for $h_{TE}$ and $\varphi=0.5$ for $C_T$ and $C_P$, which is the period (in terms of the scaled time $t/T$) of each of these quantities. 

\begin{table}
\centering
\begin{tabular}{ c c c c c }
 & & $h_{TE}$ & $C_T$ & $C_P$ \\ \hline
 $a$ & & $ \displaystyle{\frac{\max(h_{TE})}{h_0}}$ & $\displaystyle{\frac{\max\left(C_T - \overline{C_T}\right) }{h_0^2}}$ & $\displaystyle{\frac{\max\left(C_P - \overline{C_P}\right) }{h_0^2}}$ 
 \end{tabular}
 \caption{Definition of the oscillation amplitude, $a$, for the quantities $h_{TE}$, $C_T$, $C_P$. }
\label{tab:amp_scalings}
\end{table}


To enable a meaningful comparison across the various heave amplitudes, the amplitude, $a$, of a quantity is scaled as described in table \ref{tab:amp_scalings}. If the dynamics were entirely linear, the scaled amplitudes in table \ref{tab:amp_scalings} would not change with increasing heave amplitude $h_0$. This fact is readily apparent for the trailing-edge displacement, which scales linearly with $h_0$ under linear dynamics; the quadratic scaling of $C_T$ and $C_P$ under linear dynamics is explained in, \emph{e.g.}, \citet{Wu1961,Floryan2018}.  


Figure~\ref{fig:ampphaseSp02} demonstrates that resonance plays a significant role in swimming performance for $h_0 = 0.001$, where the swimming behavior is essentially linear. In particular, $f = 2.6$ (the frequency nearest to the natural frequency of the mode that maximizes the energy ratio $E_r$) is associated with the largest peaks in trailing-edge amplitude, thrust, and input power. It is intuitive from the linear stability analysis of section \ref{sec:GMresults} that actuating near this natural frequency leads to maximal energy put into the plate, resulting in the largest displacements and forces on the plate. We note that there is a local maximum in trailing-edge amplitude at $f = 1$, which is near the frequency range associated with the flow-driven modes. Yet, consistent with the fact that these modes correspond to low values of $E_r$, these large plate displacements do not coincide with peaks in performance.

It is also clear from figure~\ref{fig:ampphaseSp02} that the linear dynamics associated with $h_0 = 0.001$ largely persist at the larger heave amplitude of $h_0 = 0.01$. The largest peak in trailing-edge amplitude, thrust, and input power continues to occur at $f = 2.6$. Moreover, the dynamics of a given quantity at $h_0 = 0.01$ are of extremely similar phase to those at $h_0=0.001$. At the same time, nonlinear effects are also apparent: the peak value of $a$ for $C_T$ and $C_P$ is smaller than that associated with $h_0=0.001$.

For $h_0 = 0.1$, small peaks in the amplitude $a$ persist for $h_{TE}$ and $C_T$ near---but no longer at---the natural frequency. The slight change in the frequency corresponding to maximal trailing-edge amplitude and thrust does not preclude the possibility of resonant-type behavior; that is, a modified natural frequency could occur due to finite-amplitude effects of added mass and viscosity. At the same time, the smaller, broader peaks in trailing-edge amplitude and thrust at $h_0=0.1$ compared with $h_0=0.001,0.01$, as well as the qualitatively different phase behavior at high frequencies for $h_0=0.1$, suggest non-resonant mechanisms.

Whereas the trailing-edge displacement and thrust appear to exhibit a mixture of resonant and non-resonant behavior for $h_0=0.1$, the input power is markedly different in both $a$ and $\varphi$ at this higher heave amplitude. The phase is roughly constant for $h_0=0.1$, and the maximum amplitude of the input power does not exhibit a peak. 

\begin{figure}
\centering
	\begin{subfigure}[b]{0.3\textwidth}
		\hspace*{0mm}
        		\includegraphics[scale = 0.28,trim={0cm 0cm 0cm 0cm},clip]{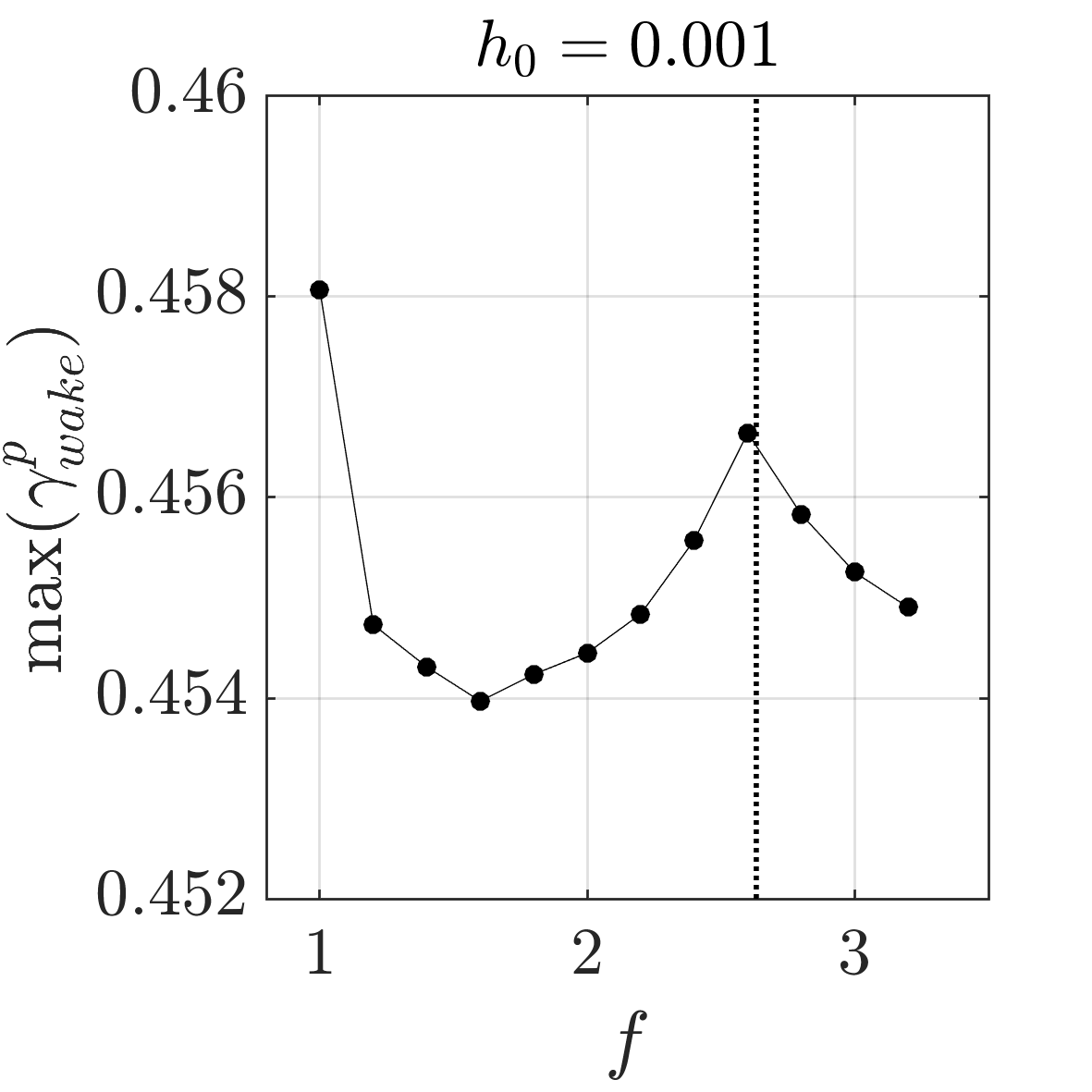}
	\end{subfigure}
	\begin{subfigure}[b]{0.3\textwidth}
		\hspace*{0mm}
        		\includegraphics[scale = 0.28,trim={1.17cm 0cm 0cm 0cm},clip]{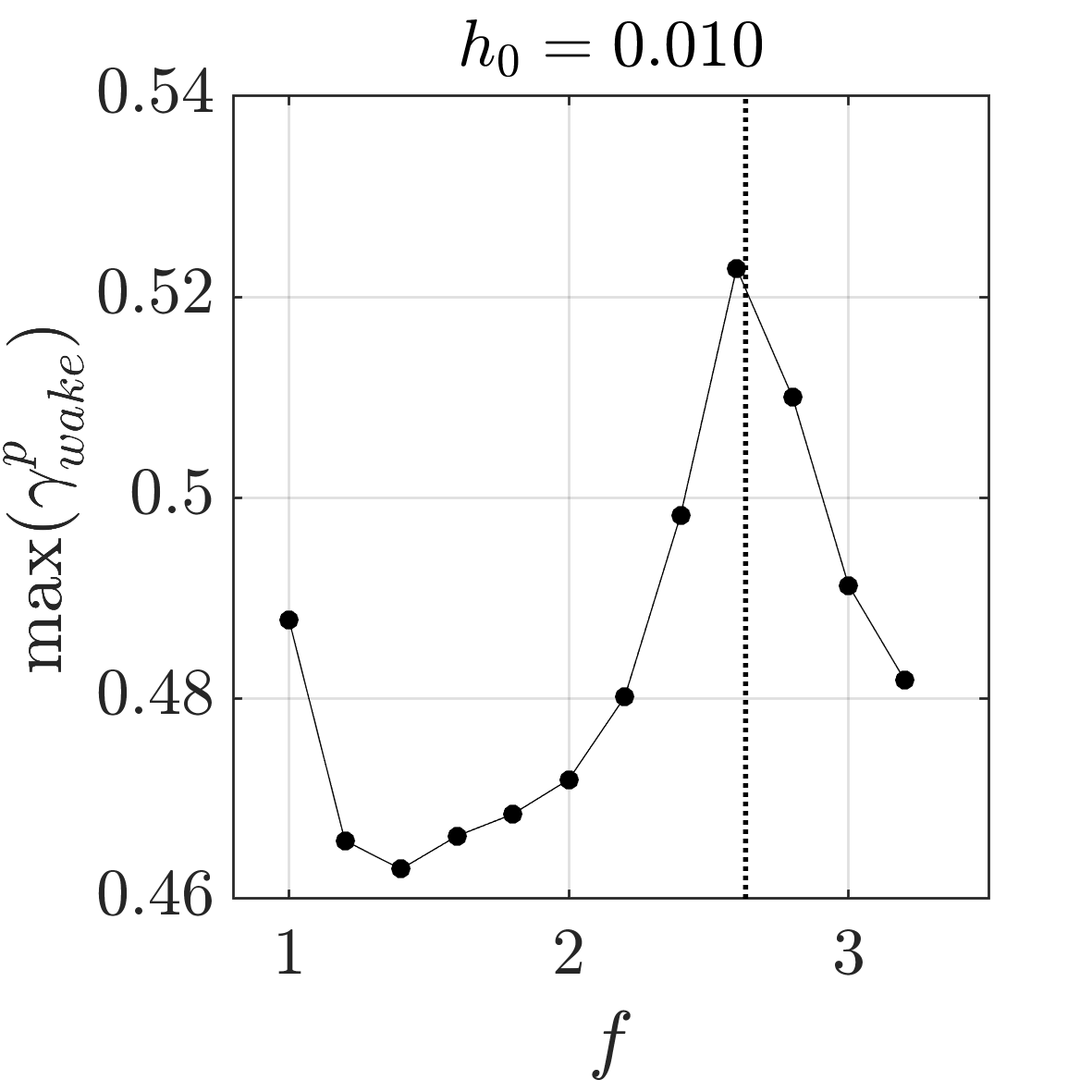}
	\end{subfigure}
	\begin{subfigure}[b]{0.3\textwidth}
		\hspace*{0mm}
        		\includegraphics[scale = 0.28,trim={1.17cm 0cm 0cm 0cm},clip]{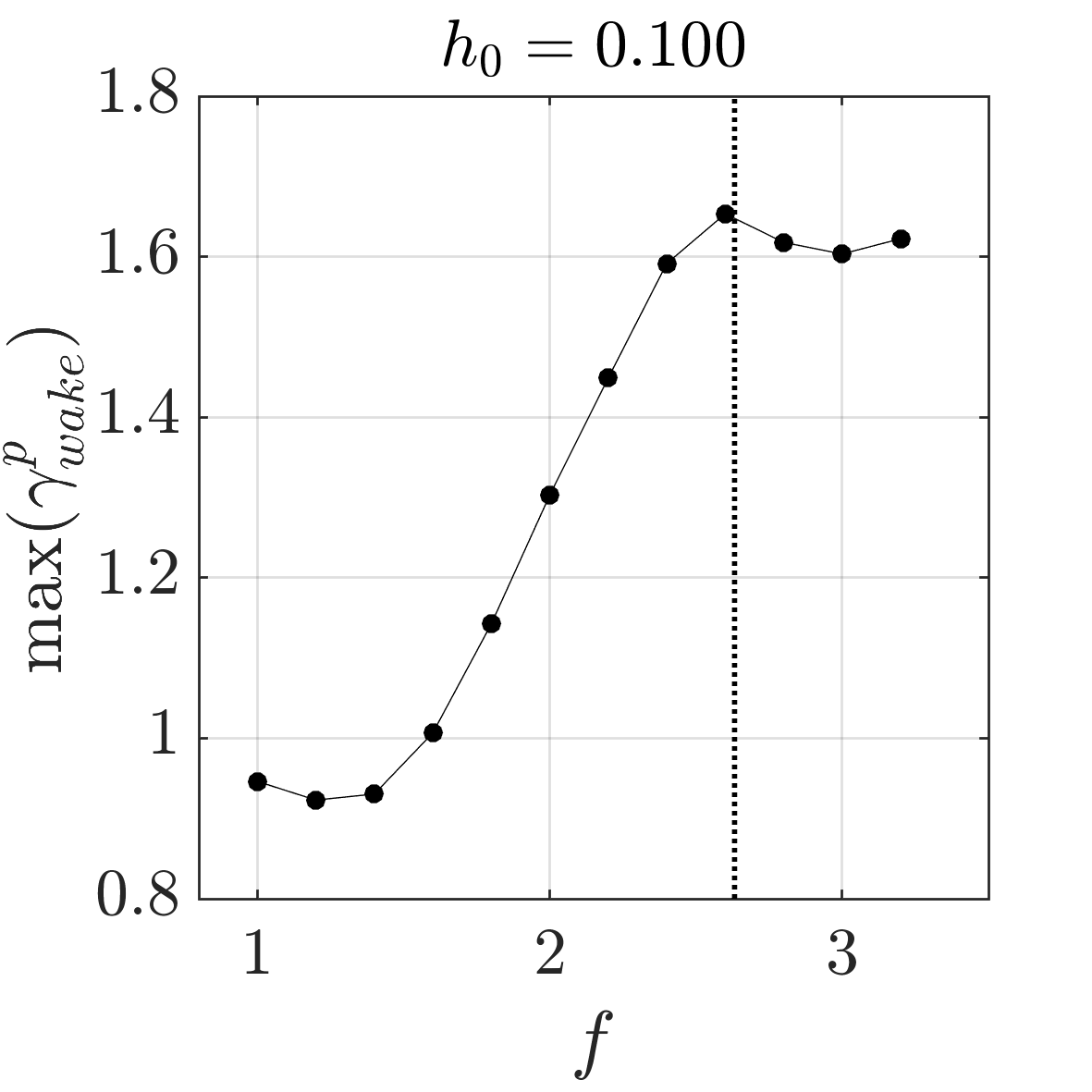}
	\end{subfigure}
	 \caption{Maximum positive circulation in the plate's wake versus frequency for $S=0.02$ and various heave amplitudes (see main text for details on how the maximum circulation is defined). The vertical dashed line depicts the resonant frequency at this stiffness.}
	\label{fig:gamma_max_S0.02}
\end{figure}

To continue to explore the role of resonance, we show in figure~\ref{fig:gamma_max_S0.02} the maximum circulation in the wake of the plate. The positive wake circulation, $\gamma^p_{wake}$, is defined as the integrated positive vorticity within a rectangle defined by $1 \le x \le 5/3$ and $-1/2\le y \le 1/2$ (where the leading edge of the plate is positioned at $(x,y)=(0,0)$ when $h_{LE}=0$). The maximum circulation, $\max(\gamma^p_{wake})$, is defined as the maximum value of $\gamma^p_{wake}$ over 5 flapping periods. The analogous plots associated with the negative circulation are nearly identical to the ones provided here for the positive circulation. The figure demonstrates that for all amplitudes, the maximal wake circulation occurs near the resonant frequency associated with the mode of largest $E_r$ value. The fact that maximal circulation occurs near the resonant frequency for all heave amplitudes suggests that, at this stiffness, resonance plays a role in the circulation formed in the plate's wake. The mean thrust is maximal at this same frequency (\emph{cf.}, figure~\ref{fig:global_performance_nonlin}), indicating that this connection between resonance and circulation extends to mean swimming performance.  

To summarize the results for this stiffness, the role of resonance is less straightforward than what one might conclude from figure~\ref{fig:global_performance_nonlin}, even for $S = 0.02$, which exhibits peaks near the natural frequency for several mean performance quantities. The thrust dynamics are different at large heave amplitudes from their small-amplitude counterparts, particularly at high frequencies. Moreover, the input power transitions to qualitatively different dynamics at large amplitudes. Together, these facts indicate that non-resonant, nonlinear effects play a nontrivial role in performance, though the persistence of a peak in mean thrust  and maximal wake circulation near the resonant frequency suggests that features of resonance may continue to be important in finite-amplitude swimming at this stiffness. Regarding this resonant phenomena, the general trend is that the peaks broaden and shorten as heave amplitude is increased. This could be reflective of a number of phenomena not accounted for by the global linear analysis and small-amplitude simulations, including added-mass effects from finite-amplitude heaving or formation and interaction of vortical structures along the plate's length and in the plate's wake. Interestingly, introducing damping into a spring-mass system has a similar effect of broadening and weakening the frequency response of the system, and it is possible that the aforementioned effects could be providing similar damping mechanisms. Probing this possibility in more depth is left to future study.

\subsubsection{Moderate stiffness: $S = 0.2$}

\begin{figure}
\centering
	\begin{subfigure}[b]{0.3\textwidth}
		\hspace*{-6mm}
        		\includegraphics[scale=0.315,trim={0cm 2.39cm 0cm 0cm},clip]{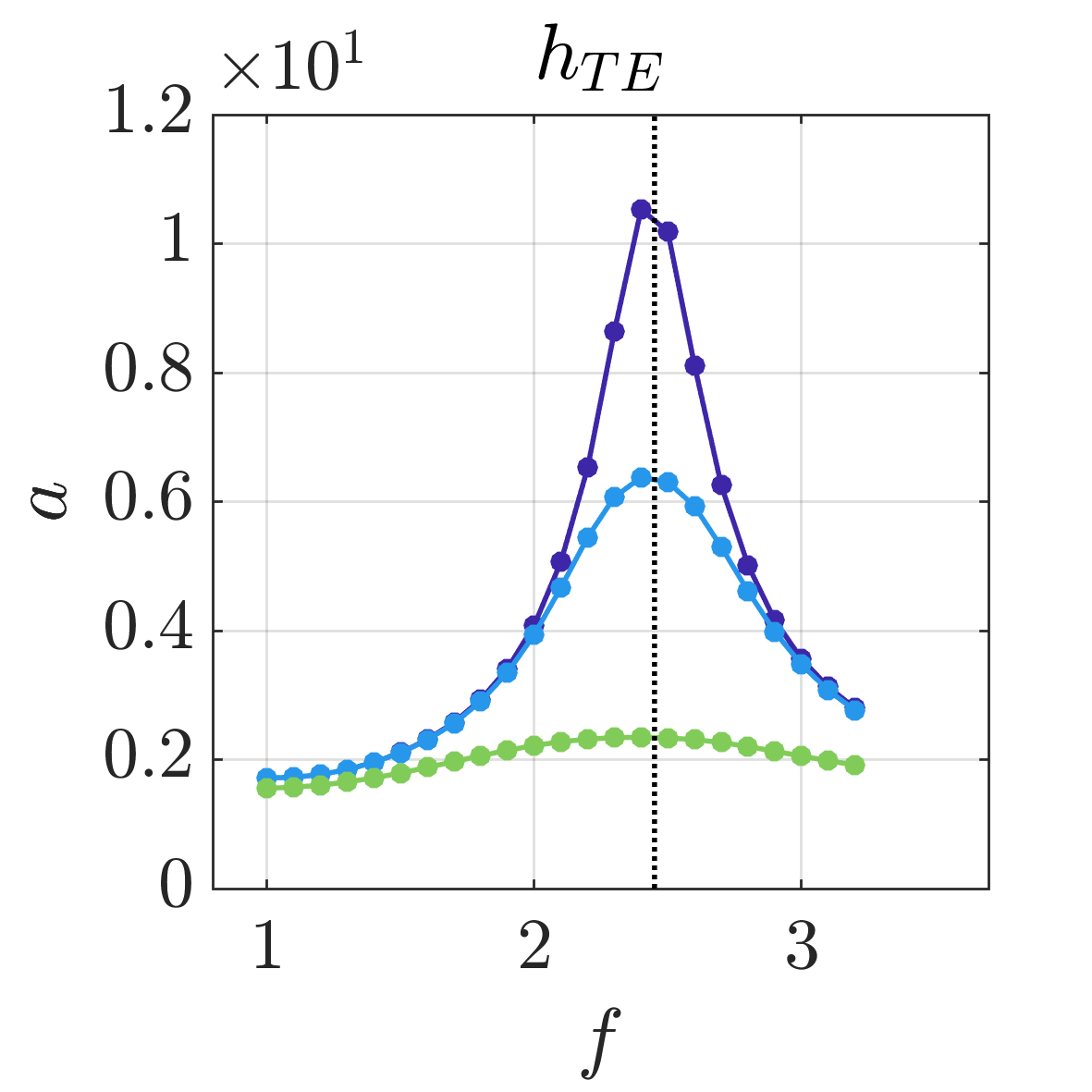}
	\end{subfigure}
	\begin{subfigure}[b]{0.3\textwidth}
		\hspace*{1mm}
        		\includegraphics[scale=0.315,trim={1cm 2.39cm 0cm 0cm},clip]{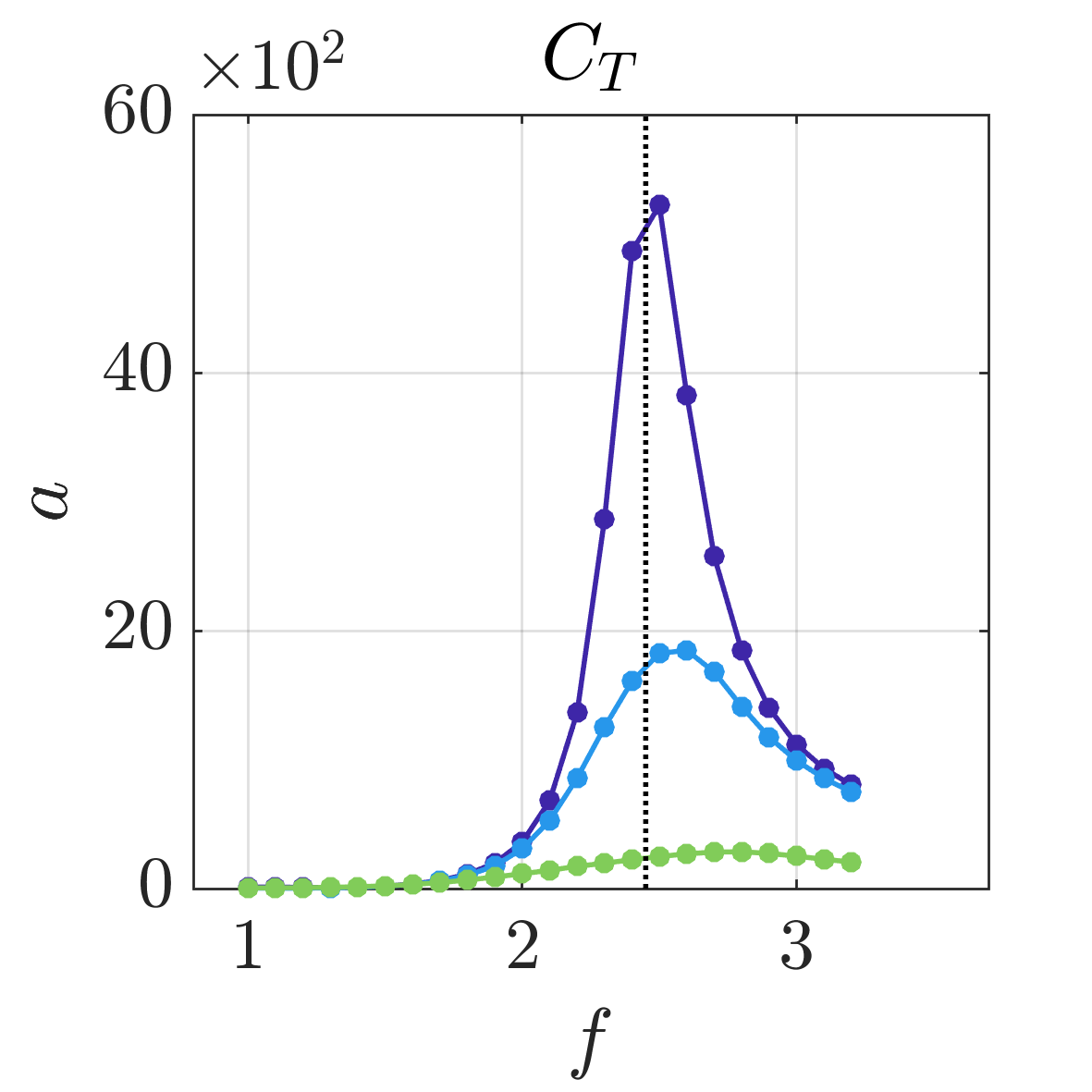}
	\end{subfigure}
	\begin{subfigure}[b]{0.3\textwidth}
		\hspace*{4.2mm}
        		\includegraphics[scale=0.315,trim={1cm 2.39cm 0cm 0cm},clip]{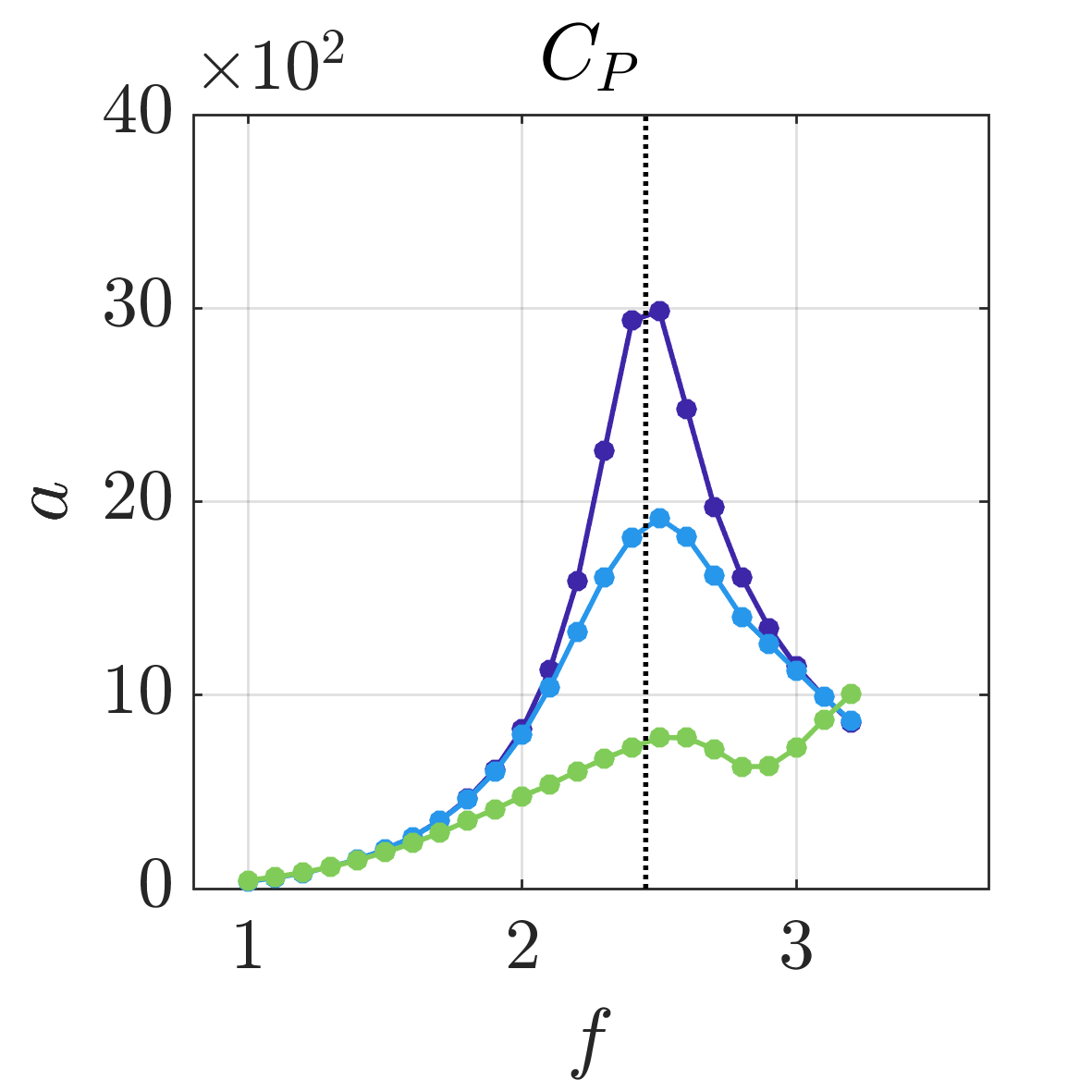}
	\end{subfigure}
	
	\begin{subfigure}[b]{0.3\textwidth}
		\hspace*{-6mm}
        		\includegraphics[scale=0.315, trim={0cm 0cm 0cm 0cm},clip]{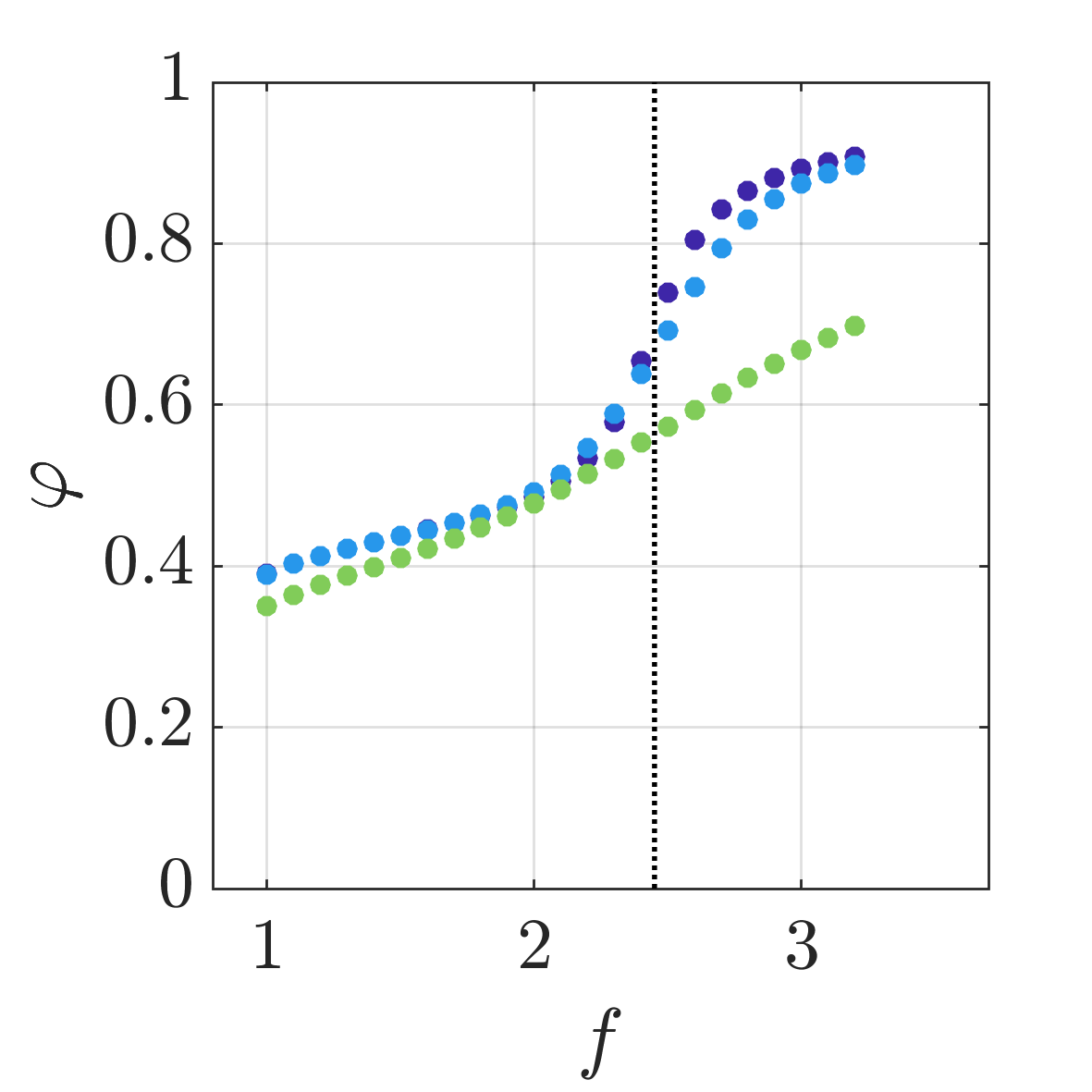}
	\end{subfigure}
	\begin{subfigure}[b]{0.3\textwidth}
		\hspace*{1.5mm}
        		\includegraphics[scale=0.315, trim={1.2cm 0cm 0cm 0cm},clip]{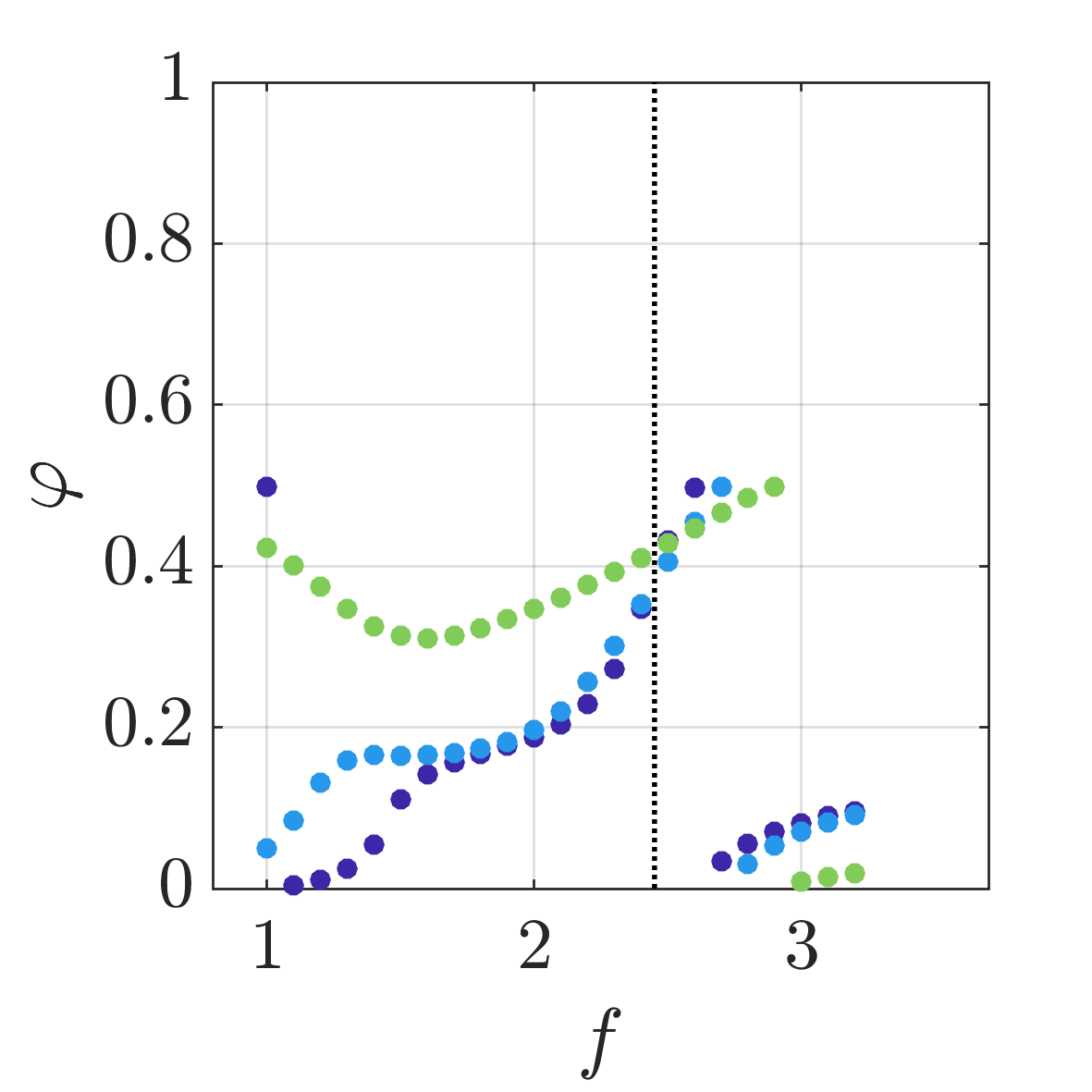}
	\end{subfigure}
	\begin{subfigure}[b]{0.3\textwidth}
		\hspace*{4.5mm}
        		\includegraphics[scale=0.315, trim={1.2cm 0cm 0cm 0cm},clip]{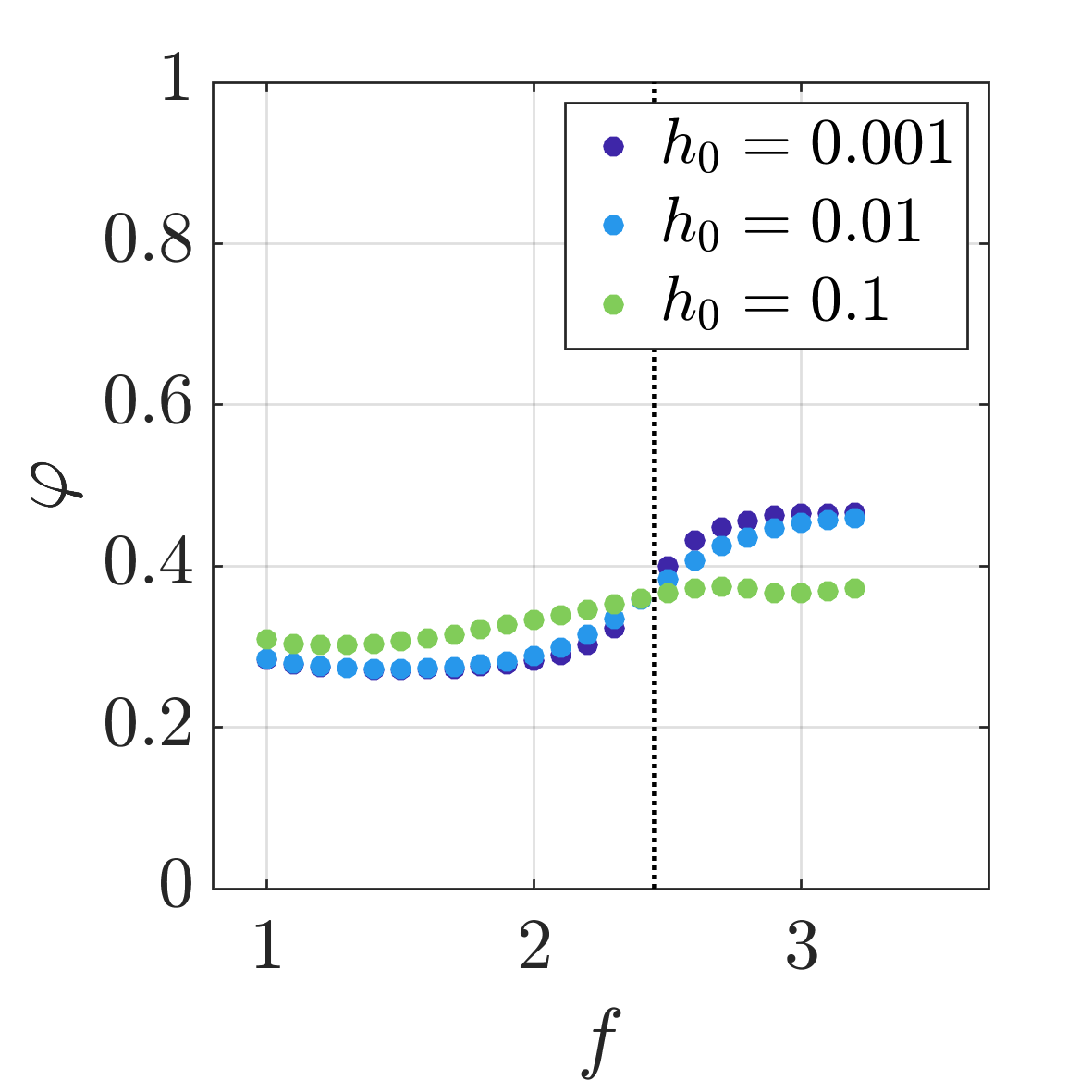}
	\end{subfigure}
    \caption{Analog of figure~\ref{fig:ampphaseSp02} for stiffness $S = 0.2$.}
	\label{fig:ampphaseSp2}
\end{figure}

We now consider increasing the stiffness to $S = 0.2$. Figure~\ref{fig:ampphaseSp2} shows the phase and maximum amplitude associated with $h_{TE}, C_T,$ and $C_P$ at a given frequency. As was observed for $S=0.02$, the figure suggests the presence of both resonant and non-resonant mechanisms. 

For small heave amplitudes ($h_0\le 0.01$), the dynamics appear to be primarily linear: the behavior of $a$ and $\phi$ is similar for all three quantities (though the smaller peaks in amplitude for $h_0 = 0.01$ than for $h_0 = 0.001$ demonstrate the appearance of nonlinear mechanisms). Moreover, resonance appears to play the primary role for these values of $h_0$, as there are peaks in all three quantities near the resonant frequency associated with the mode of largest $E_r$ value.

For $h_0=0.1$, either global or local peaks appear in all quantities near the resonant frequency. Moreover, the phase associated with the thrust is similar to that of the lower heave amplitudes near the resonant frequency. These facts suggest a nontrivial role of resonance in dictating performance of finite-amplitude swimming at this stiffness.  At the same time, the peaks are broader and smaller at this larger heave amplitude, and the phase behavior of all quantities reflect significant differences from the smaller-heave-amplitude cases (where resonance dominates). Moreover, while there is a local peak in $a$ for the input power, the overall shape of the curve is different from those associated with $h_0=0.001, 0.01$.

\begin{figure}
\centering
	\begin{subfigure}[b]{0.3\textwidth}
		\hspace*{0mm}
        		\includegraphics[scale = 0.28,trim={0cm 0cm 0cm 0cm},clip]{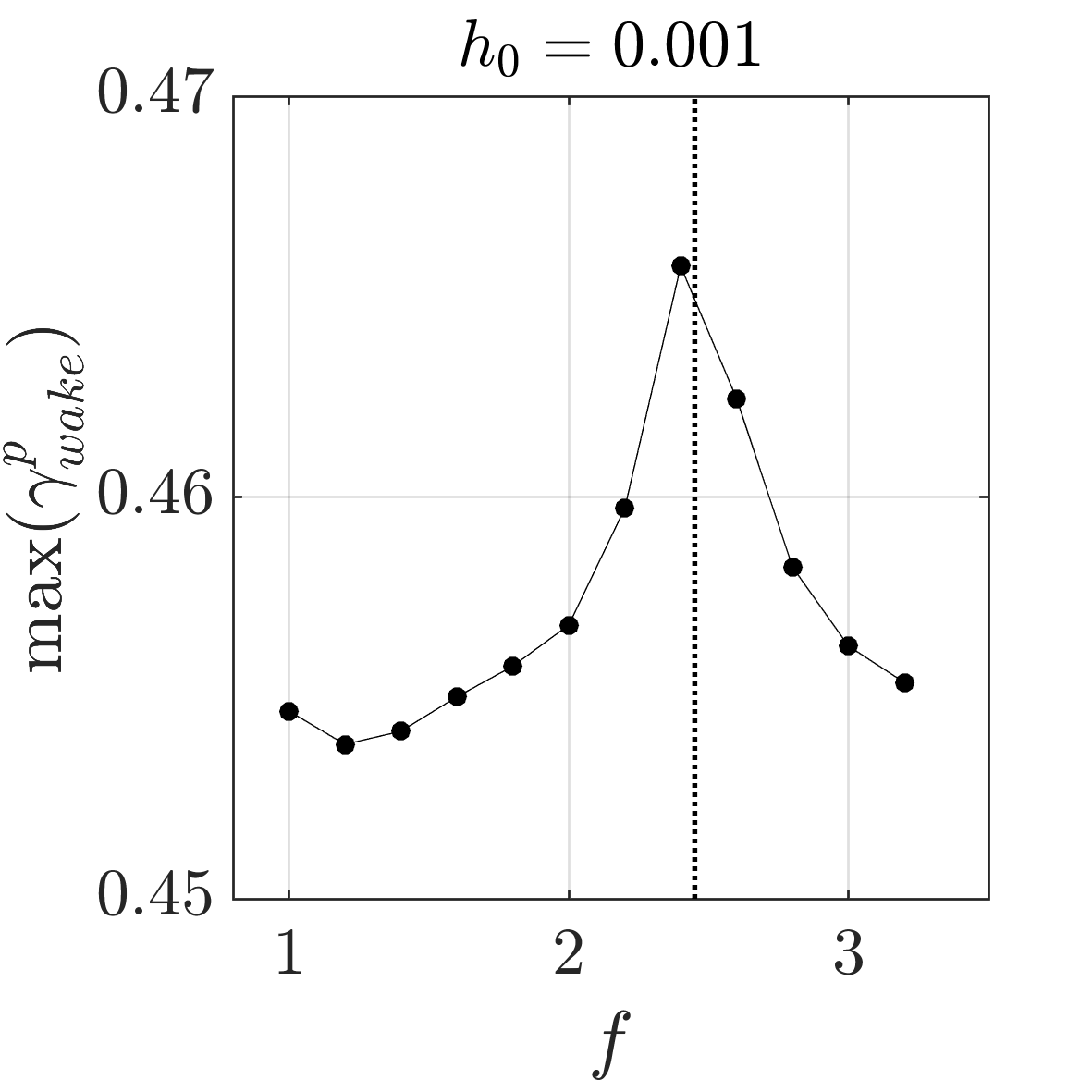}
	\end{subfigure}
	\begin{subfigure}[b]{0.3\textwidth}
		\hspace*{0mm}
        		\includegraphics[scale = 0.28,trim={1.17cm 0cm 0cm 0cm},clip]{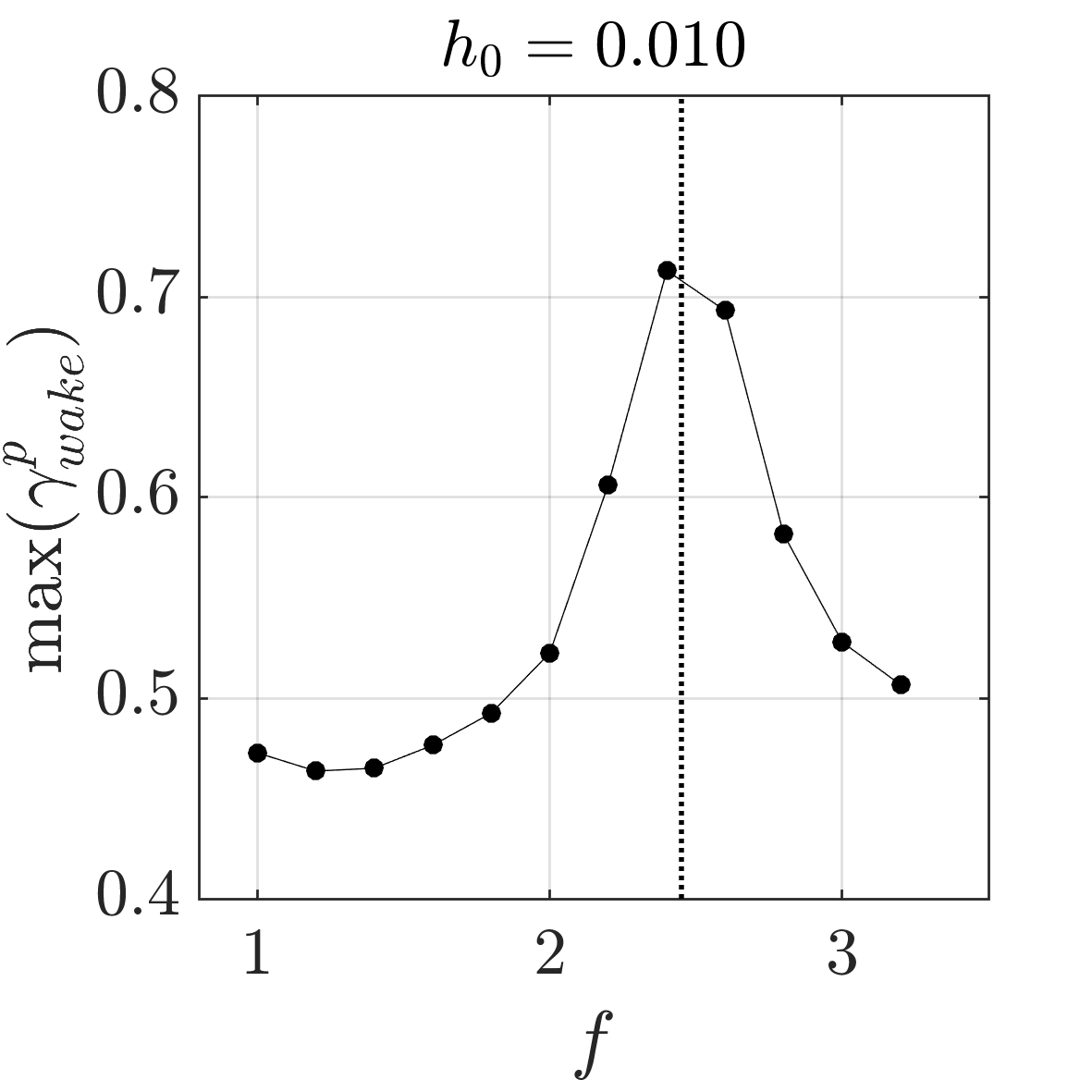}
	\end{subfigure}
    	\begin{subfigure}[b]{0.3\textwidth}
		\hspace*{0mm}
        		\includegraphics[scale = 0.28,trim={1.17cm 0cm 0cm 0cm},clip]{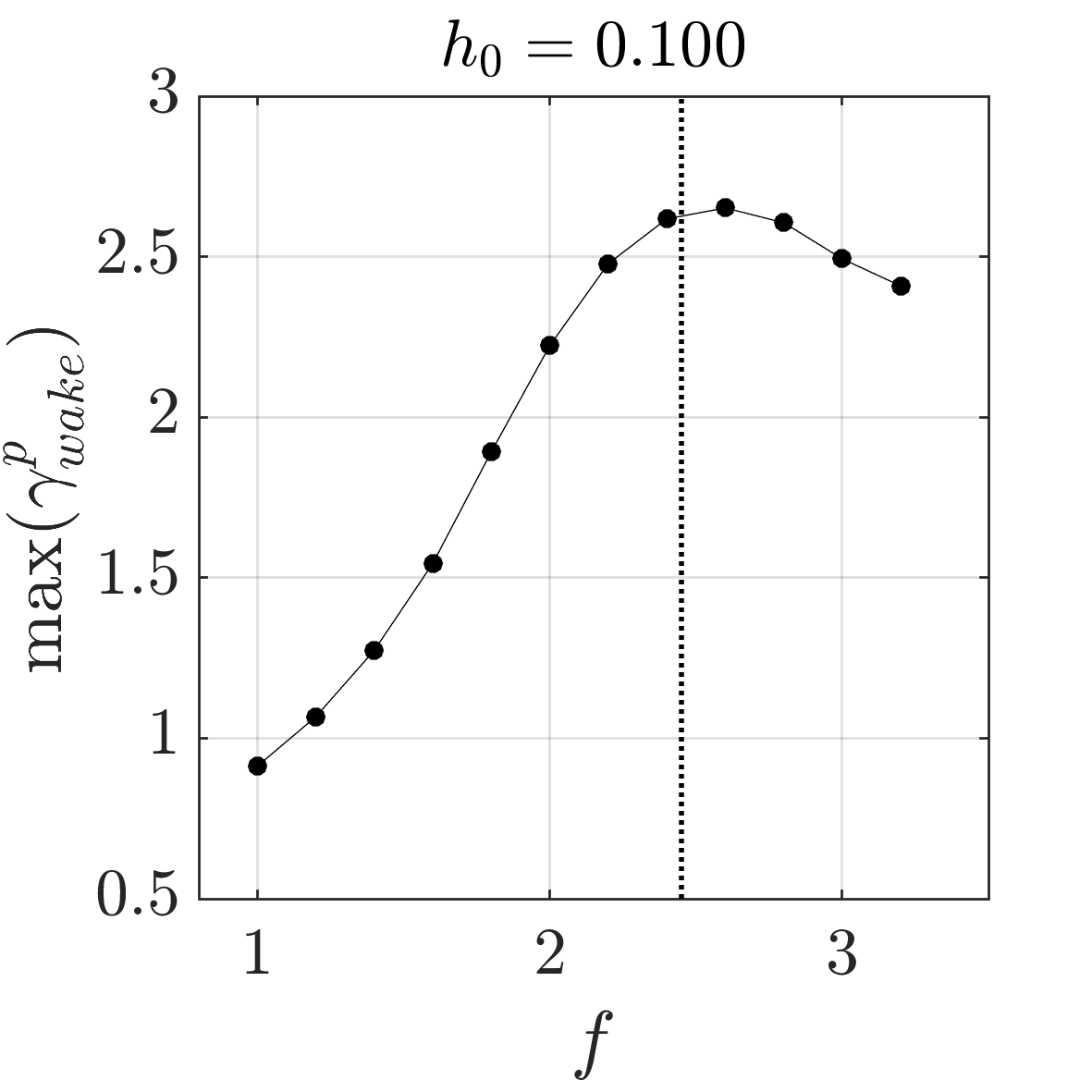}
	\end{subfigure}
  \caption{Analog of figure~\ref{fig:gamma_max_S0.02} for stiffness $S = 0.2$.}
	\label{fig:gamma_max_S0.2}
\end{figure}

To establish connections between vortex formation and performance, we provide in figure~\ref{fig:gamma_max_S0.2} the maximum wake circulation (see the text surrounding figure~\ref{fig:gamma_max_S0.02} for details on how this is defined) as a function of frequency for the various heave amplitudes. As with $S = 0.02$, the maximum wake circulation occurs near the resonant frequency for all heave amplitudes. The maximum mean thrust also occurs near this frequency, which again suggests that resonance could be connected to maximal mean thrust production at this stiffness.

\subsubsection{Moderate stiffness: $S = 2$}

\begin{figure}
\centering
	\begin{subfigure}[b]{0.3\textwidth}
		\hspace*{-6mm}
        		\includegraphics[scale=0.315,trim={0cm 2.39cm 0cm 0cm},clip]{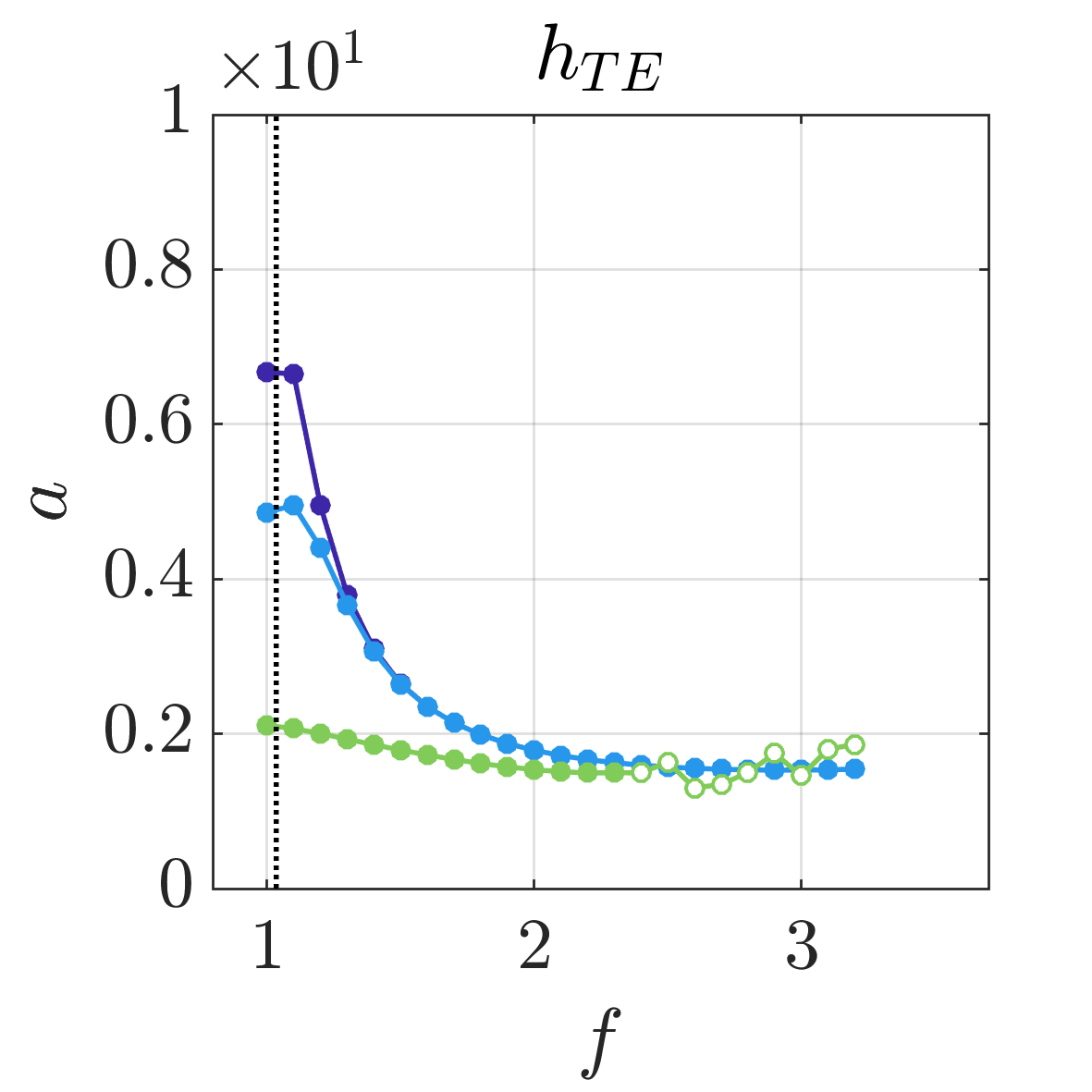}
	\end{subfigure}
	\begin{subfigure}[b]{0.3\textwidth}
		\hspace*{1mm}
        		\includegraphics[scale=0.315,trim={1cm 2.39cm 0cm 0cm},clip]{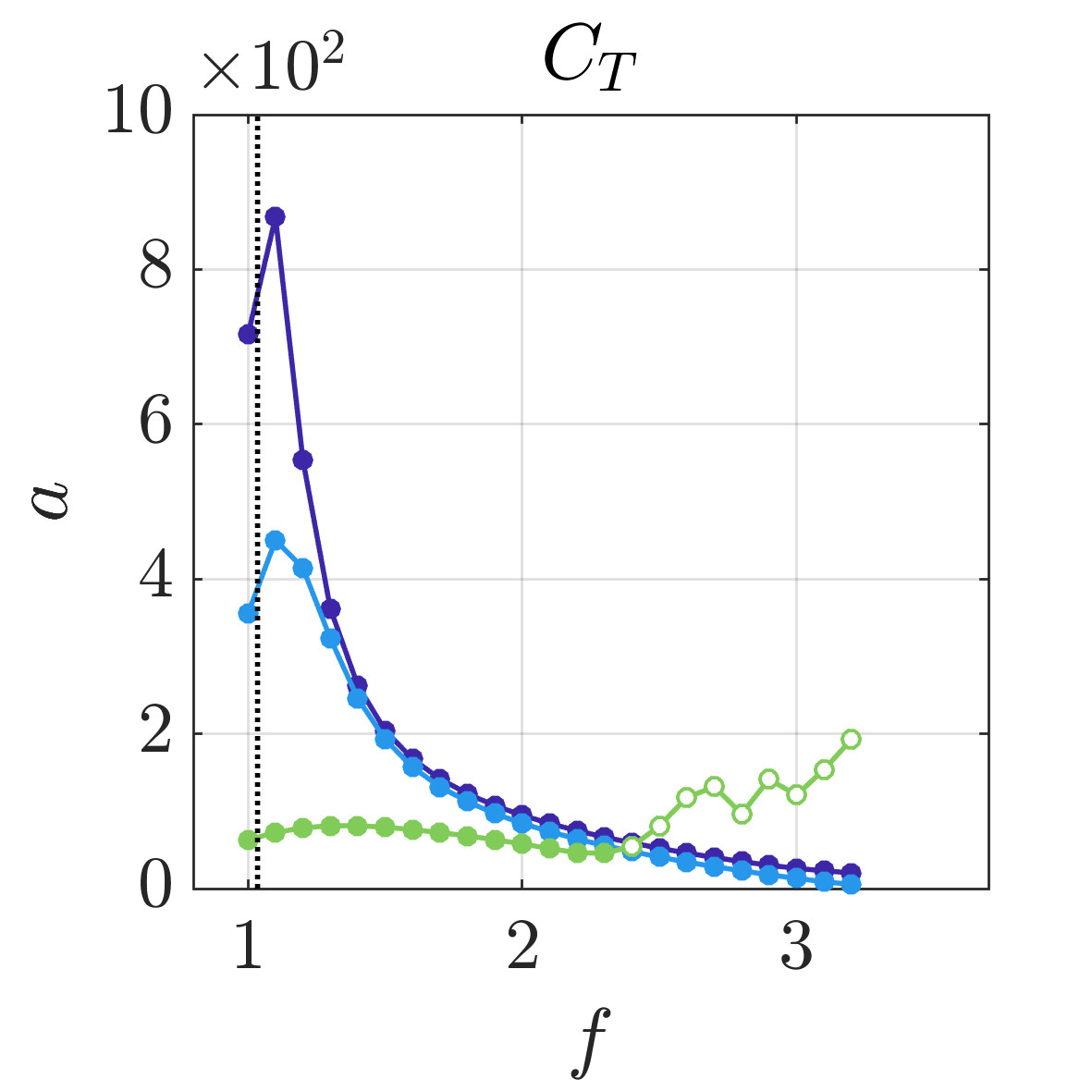}
	\end{subfigure}
	\begin{subfigure}[b]{0.3\textwidth}
		\hspace*{4.2mm}
        		\includegraphics[scale=0.315,trim={1cm 2.39cm 0cm 0cm},clip]{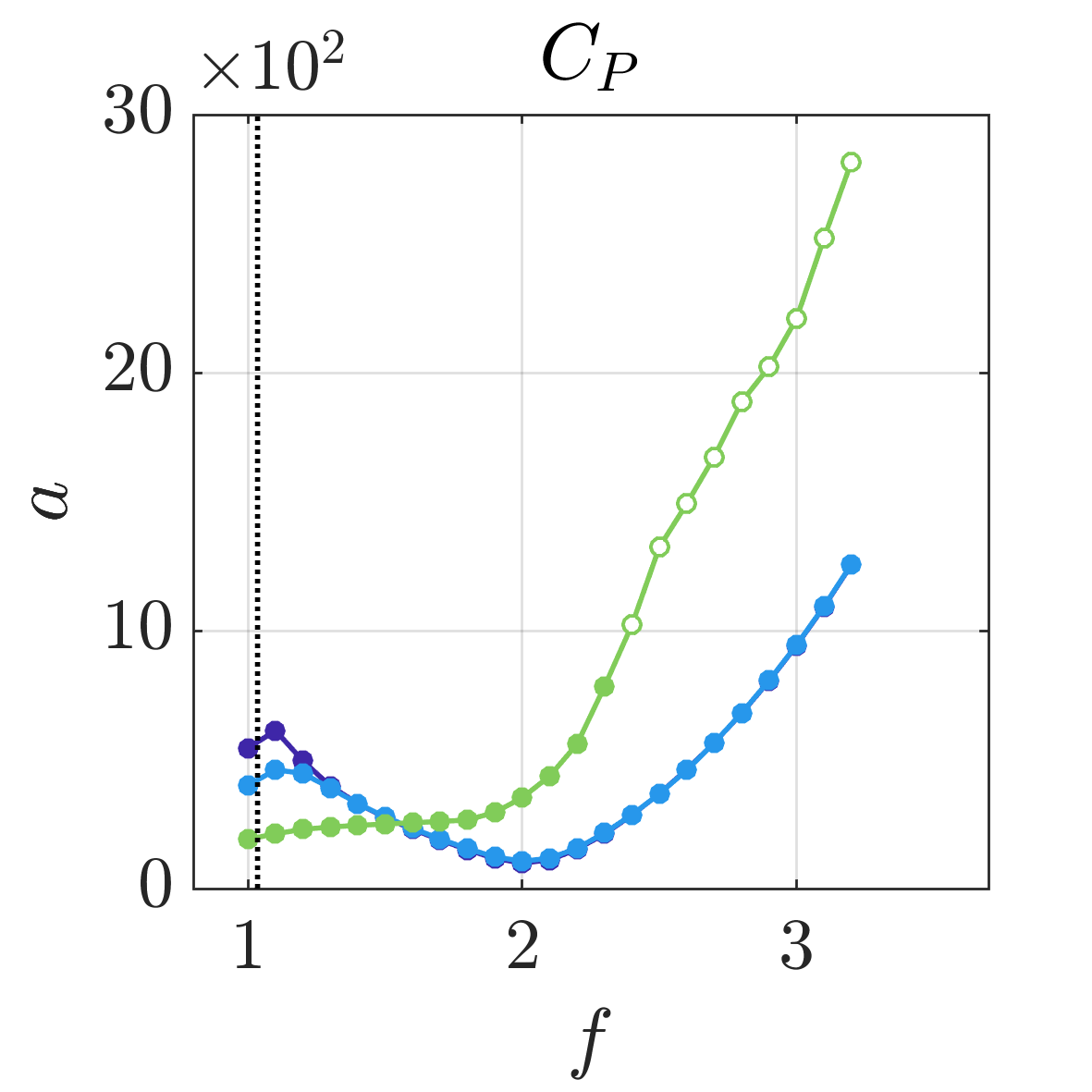}
	\end{subfigure}
	
	\begin{subfigure}[b]{0.3\textwidth}
		\hspace*{-6mm}
        		\includegraphics[scale=0.315, trim={0cm 0cm 0cm 0cm},clip]{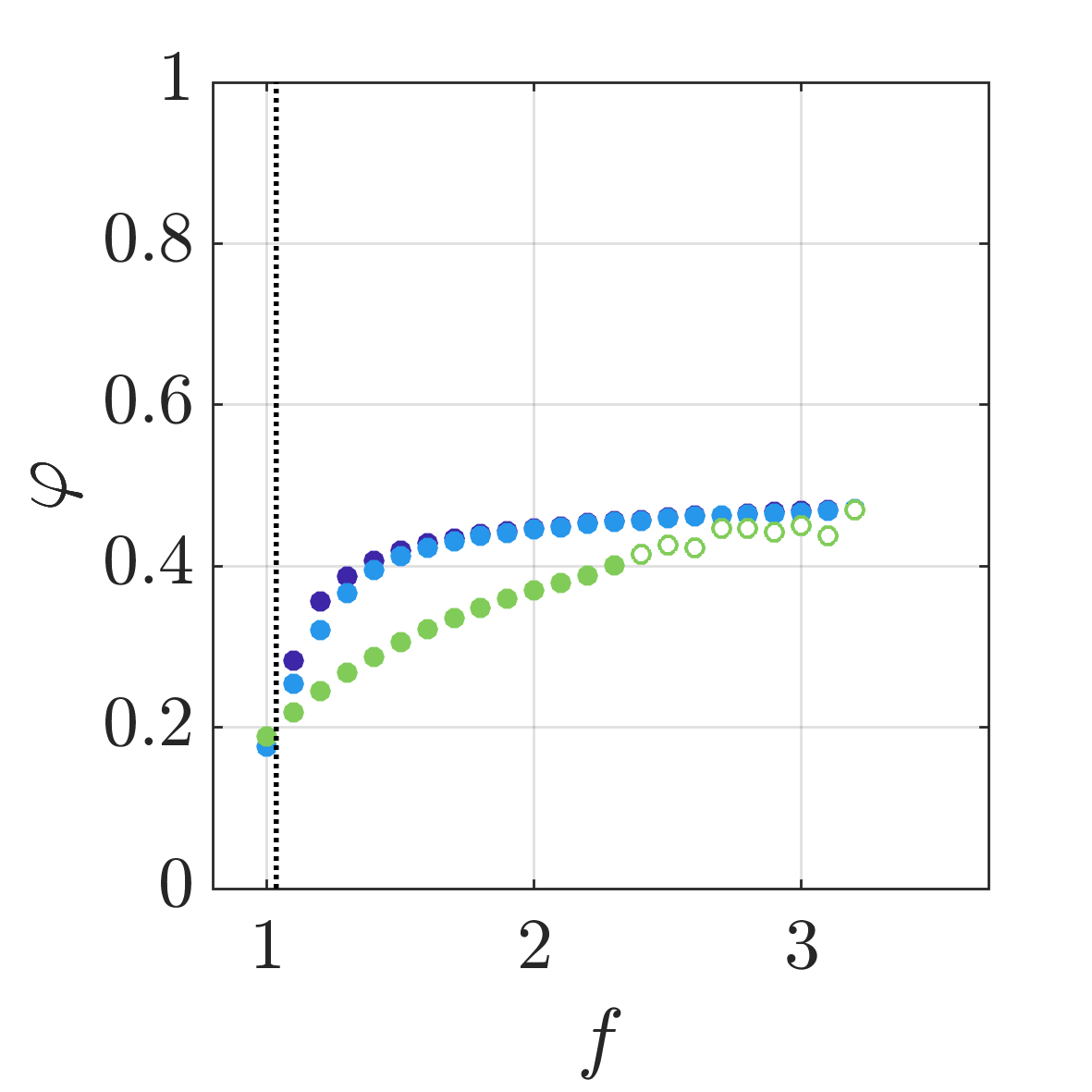}
	\end{subfigure}
	\begin{subfigure}[b]{0.3\textwidth}
		\hspace*{1.5mm}
        		\includegraphics[scale=0.315, trim={1.2cm 0cm 0cm 0cm},clip]{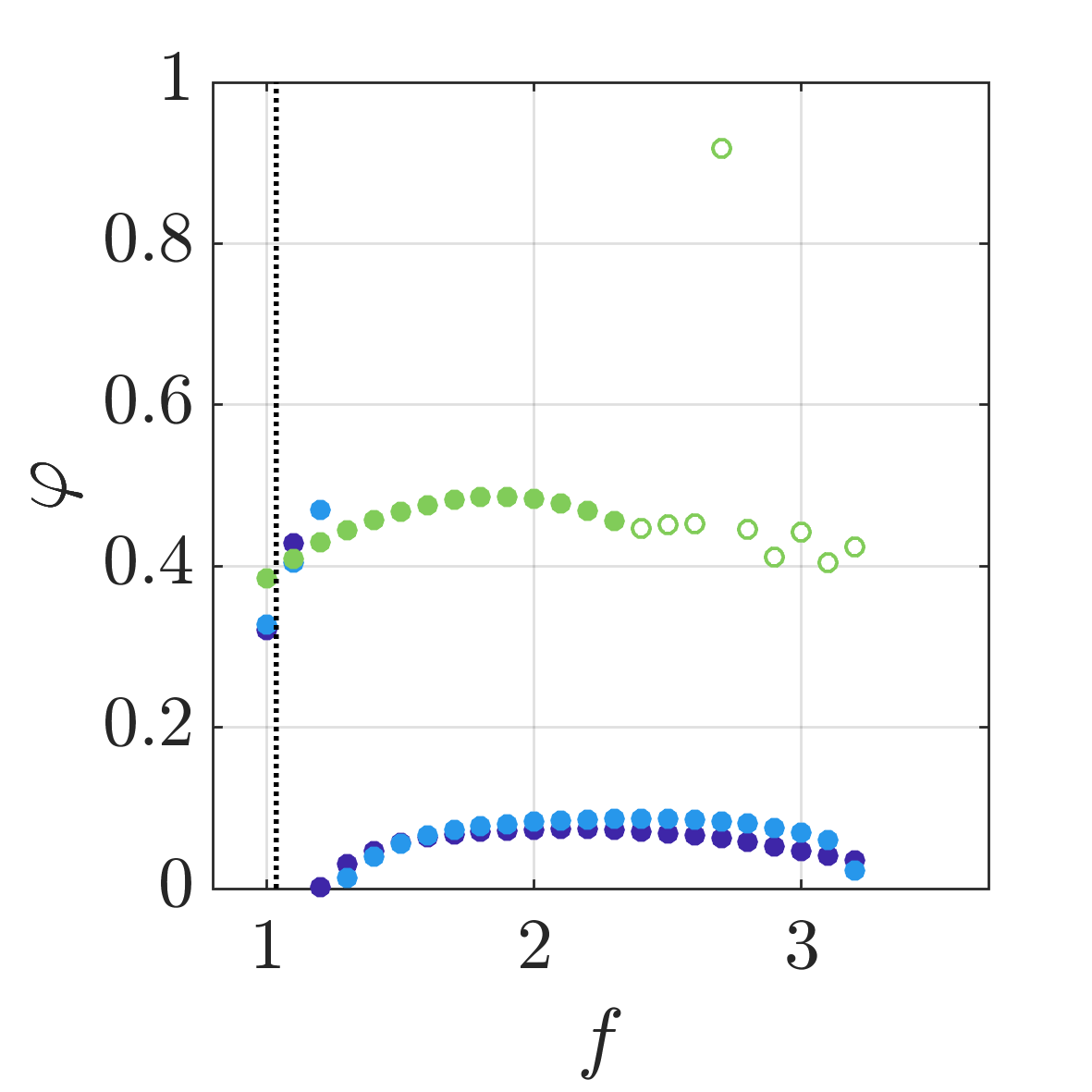}
	\end{subfigure}
	\begin{subfigure}[b]{0.3\textwidth}
		\hspace*{4.5mm}
        		\includegraphics[scale=0.315, trim={1.2cm 0cm 0cm 0cm},clip]{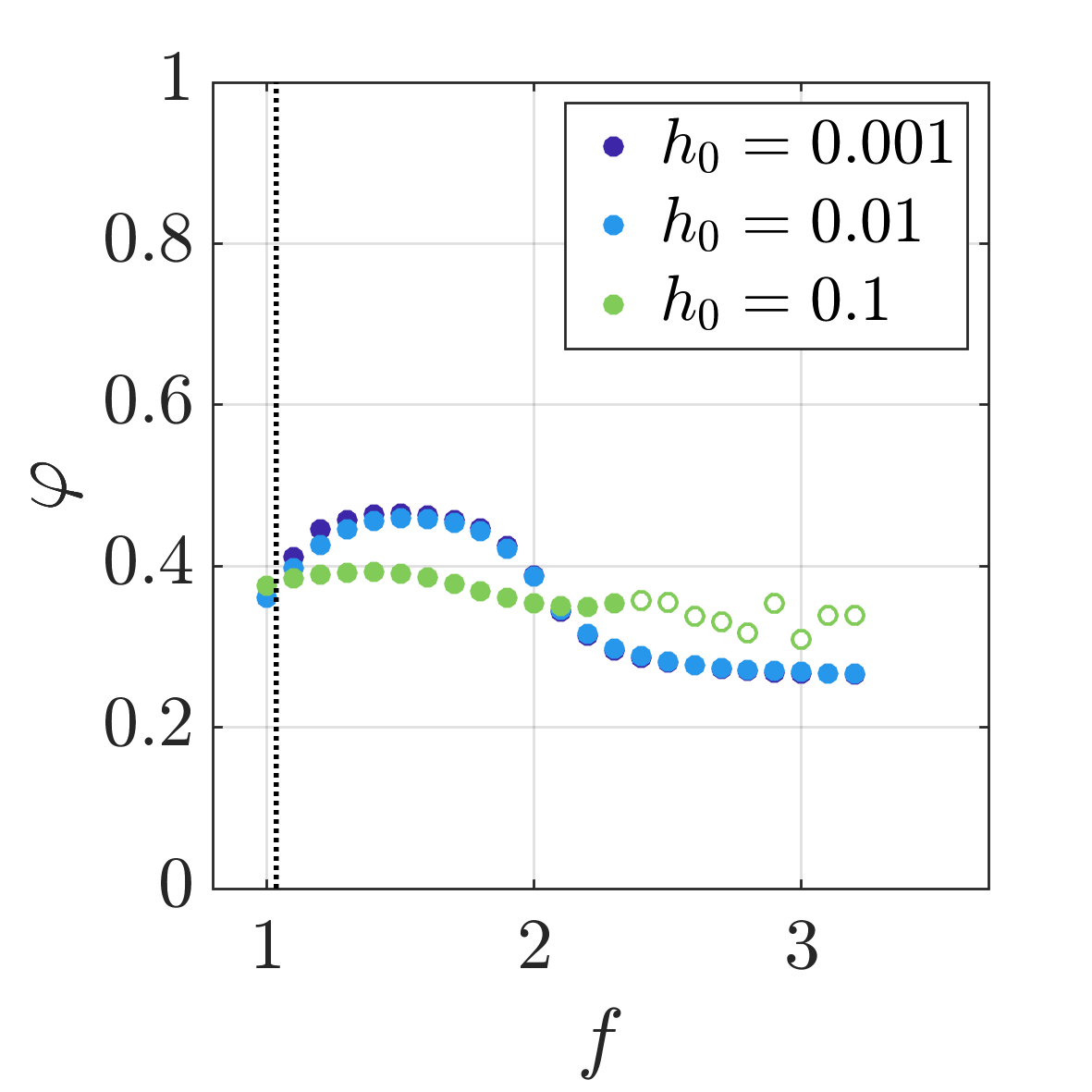}
	\end{subfigure}
    \caption{Analog of figure~\ref{fig:ampphaseSp02} for stiffness $S = 2$. The open markers for $h_0 = 0.1$, $f\ge 2.4$ correspond to aperiodic dynamics for which this amplitude-phase analysis is not valid.}
	\label{fig:ampphaseS2}
\end{figure}

Figure~\ref{fig:ampphaseS2} provides the phase and maximum amplitude associated with $h_{TE}, C_T,$ and $C_P$ at a given frequency for a stiffness $S = 2$. The figure indicates that the dynamics are essentially linear for the small heave amplitudes $h_0=0.001, 0.01$ (both $a$ and $\varphi$ exhibit similar behavior  at these smaller values of $h_0$), and that resonance continues to play a primary role for these values of $h_0$ (resonant peaks are present in all quantities). However, the behavior of $a$ for $C_P$ suggests that there are non-resonant mechanisms even in this linear regime: the maximum value of $a$ occurs at $f=3.2$---well away from the natural frequency for this stiffness\footnote{The global maximum may occur at a higher frequency, as the amplitude is still increasing at $f = 3.2$, though in either case this is away from the natural frequency.}.

For the largest heave amplitude, $h_0 = 0.1$, there are both similarities and differences to the cases with smaller heave amplitudes. The main similarity is that the thrust and trailing-edge amplitude exhibit either a local or global peak near the resonant frequency corresponding to the mode with largest $E_r$ value. This again points to resonant features present in the dynamics exhibited by this stiffness, particularly for low frequencies.

There are also several differences between the large-heave-amplitude dynamics and those of smaller amplitude. First, the phase associated with $h_0=0.1$ is different for all quantities to that associated with smaller heave amplitudes. An even more conspicuous difference is the significant variations in $a$ for $h_{TE}$ and $C_T$ when $f\ge2.4$. This variability is due to the fact that the signals are not periodic at this stiffness for these frequencies, rendering this amplitude-phase representation inappropriate.

\begin{figure}
\centering
	\begin{subfigure}[b]{0.3\textwidth}
		\hspace*{-1.7mm}
        		\includegraphics[scale=0.29,trim={0cm 0cm 0cm 0cm},clip]{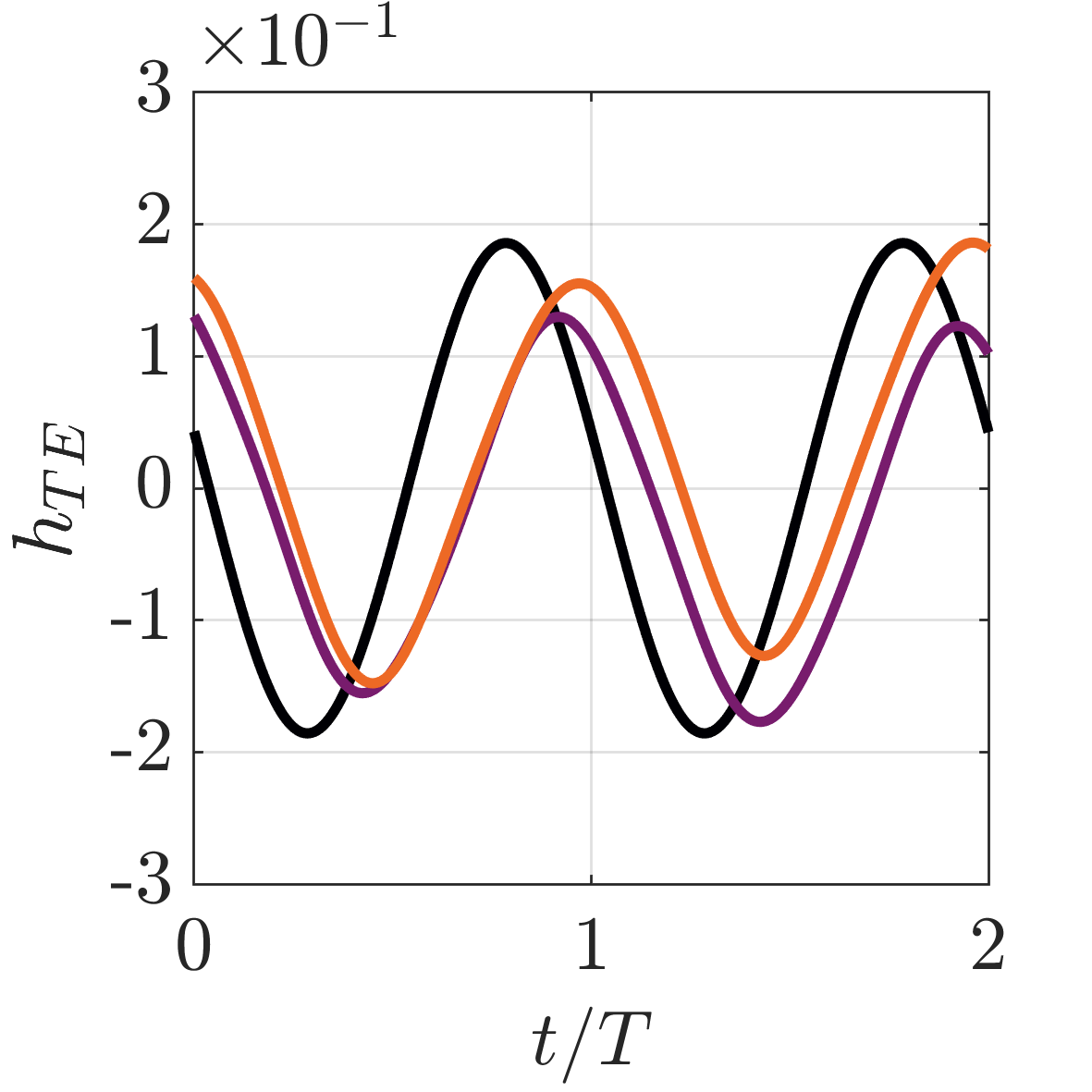}
	\end{subfigure}
	\begin{subfigure}[b]{0.3\textwidth}
		\hspace*{-1.2mm}
        		\includegraphics[scale=0.29,trim={0cm 0cm 0cm 0cm},clip]{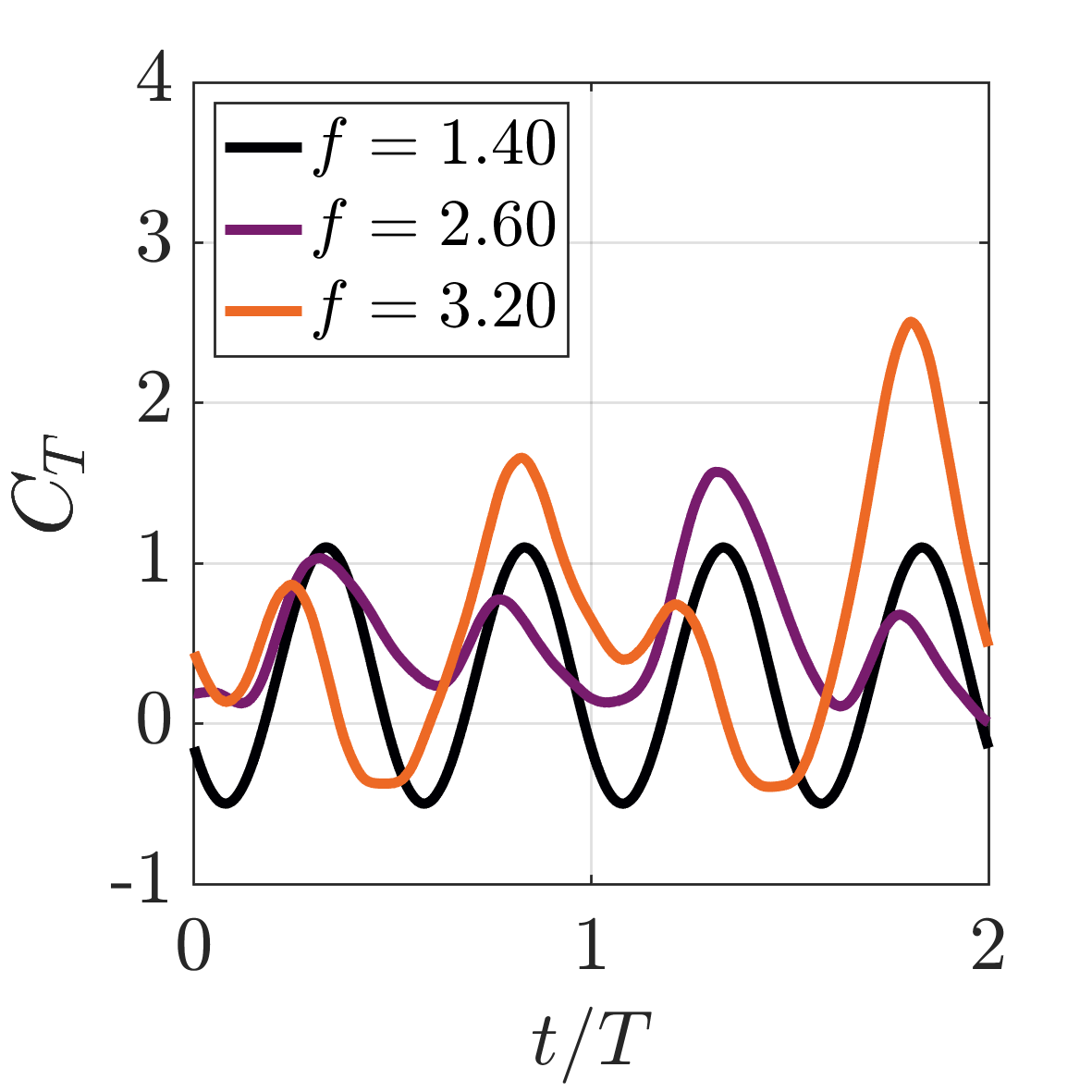}
	\end{subfigure}
    	\begin{subfigure}[b]{0.3\textwidth}
		\hspace*{-0.3mm}
        		\includegraphics[scale=0.29,trim={0cm 0cm 0cm 0cm},clip]{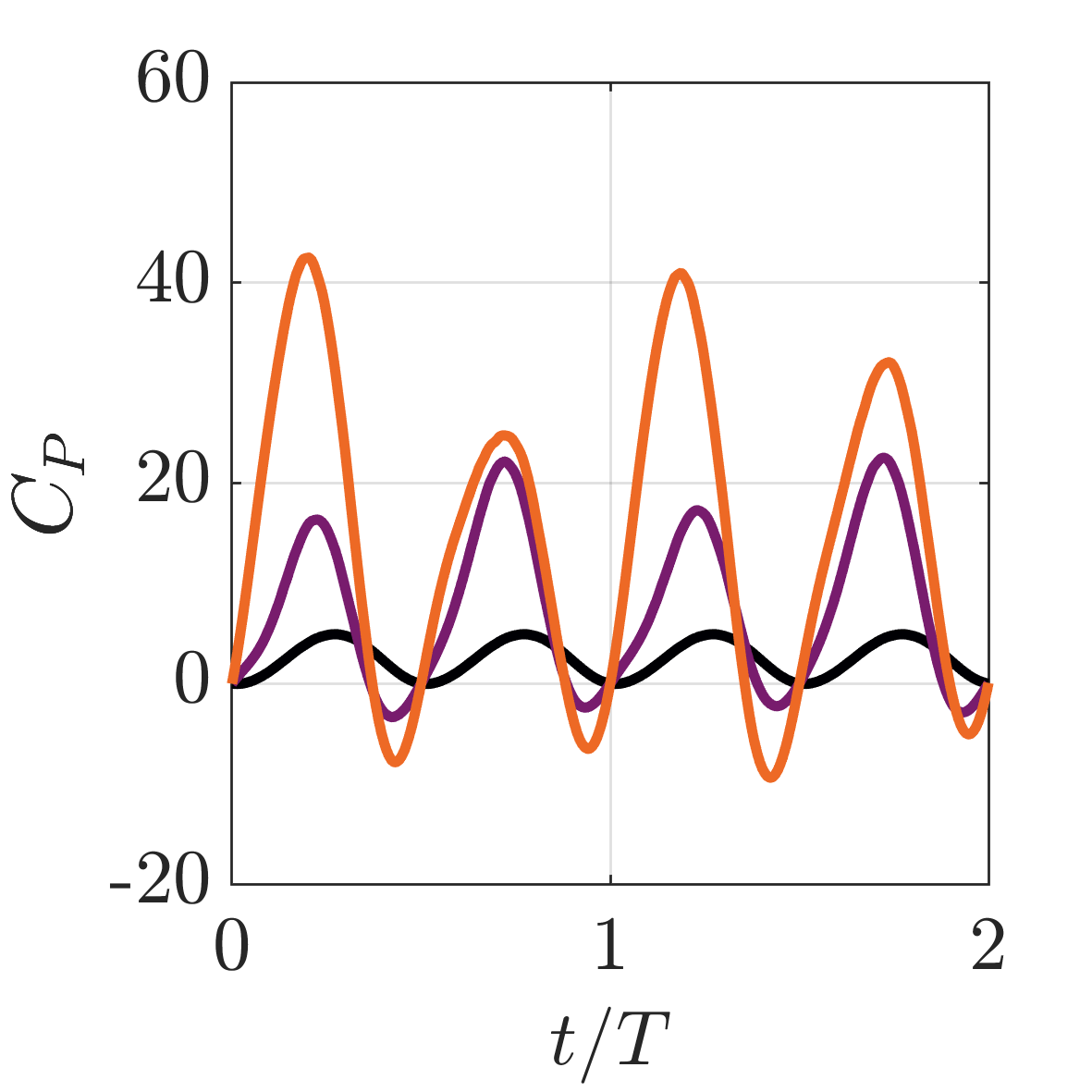}
	\end{subfigure}
    \caption{Time traces of various quantities for $S = 2, h_0 = 0.1$. Limit-cycle behavior is observed for $f=1.4$, but not for $f=2.6,3.2$. This illustrates the presence of non-linear, non-resonant mechanisms for $f>2.2$.}
	\label{fig:timetrace_aperiodic}
\end{figure}

Figure~\ref{fig:timetrace_aperiodic} provides time traces of $h_{TE}$, $C_T$, and $C_P$ for $f=1.4$ (for which the quantities exhibit periodic behavior), and for $f=2.6$ and $3.2$ (for which the quantities are aperiodic). These aperiodic dynamics are consistent with the pronounced leading-edge vortical structures (and associated separation at the leading edge) and asymmetric wake structures observed in figure~\ref{fig:snapshots_nonlin} for $S=2$ at $f = 2.4$ and $3$. These nonlinear, non-resonant mechanisms cause a significant departure from the essentially linear dynamics displayed in figure~\ref{fig:ampphaseS2} for small heave amplitudes. 

\begin{figure}
\centering
	\begin{subfigure}[b]{0.3\textwidth}
		\hspace*{0mm}
        		\includegraphics[scale = 0.28,trim={0cm 0cm 0cm 0cm},clip]{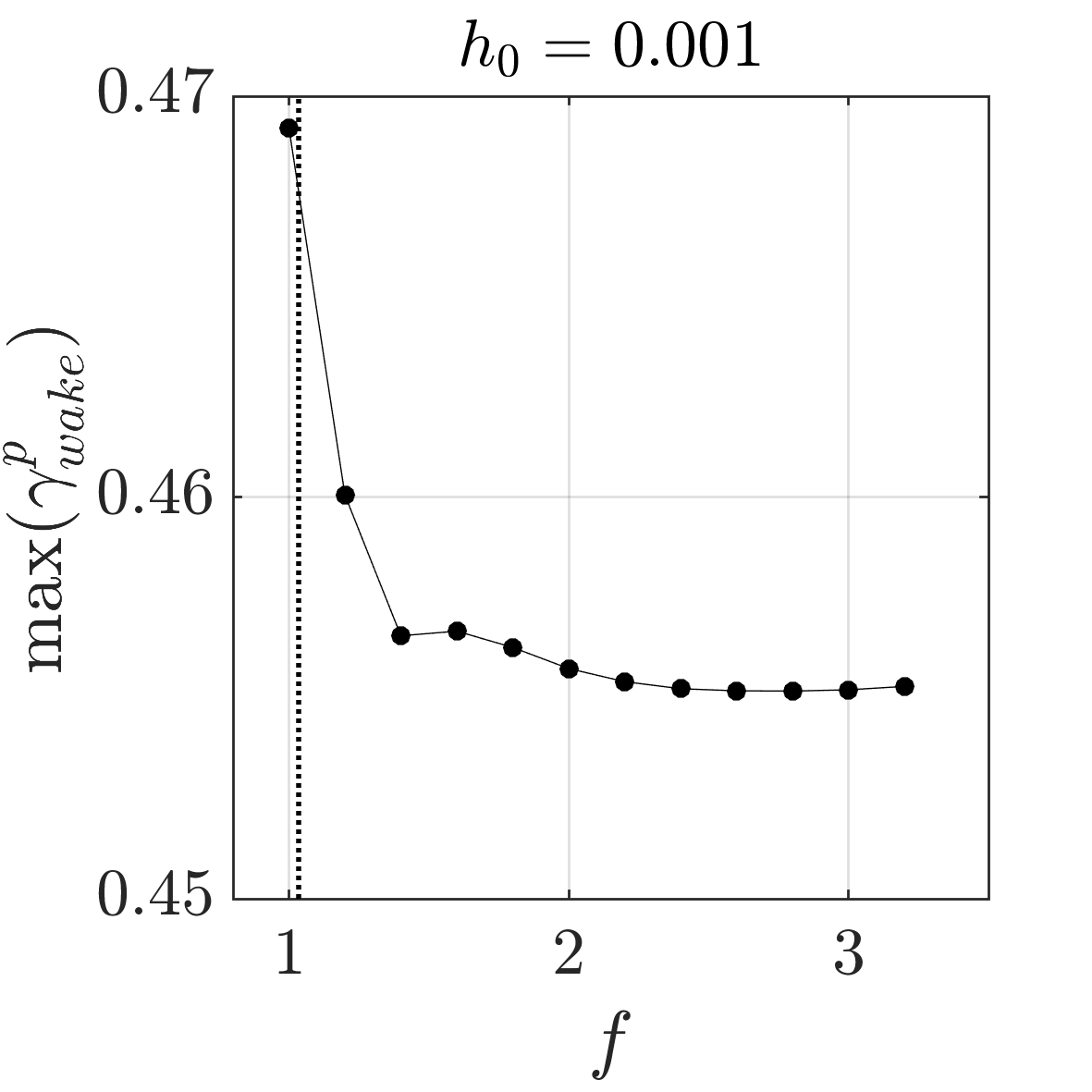}
	\end{subfigure}
	\begin{subfigure}[b]{0.3\textwidth}
		\hspace*{0mm}
        		\includegraphics[scale = 0.28,trim={1.17cm 0cm 0cm 0cm},clip]{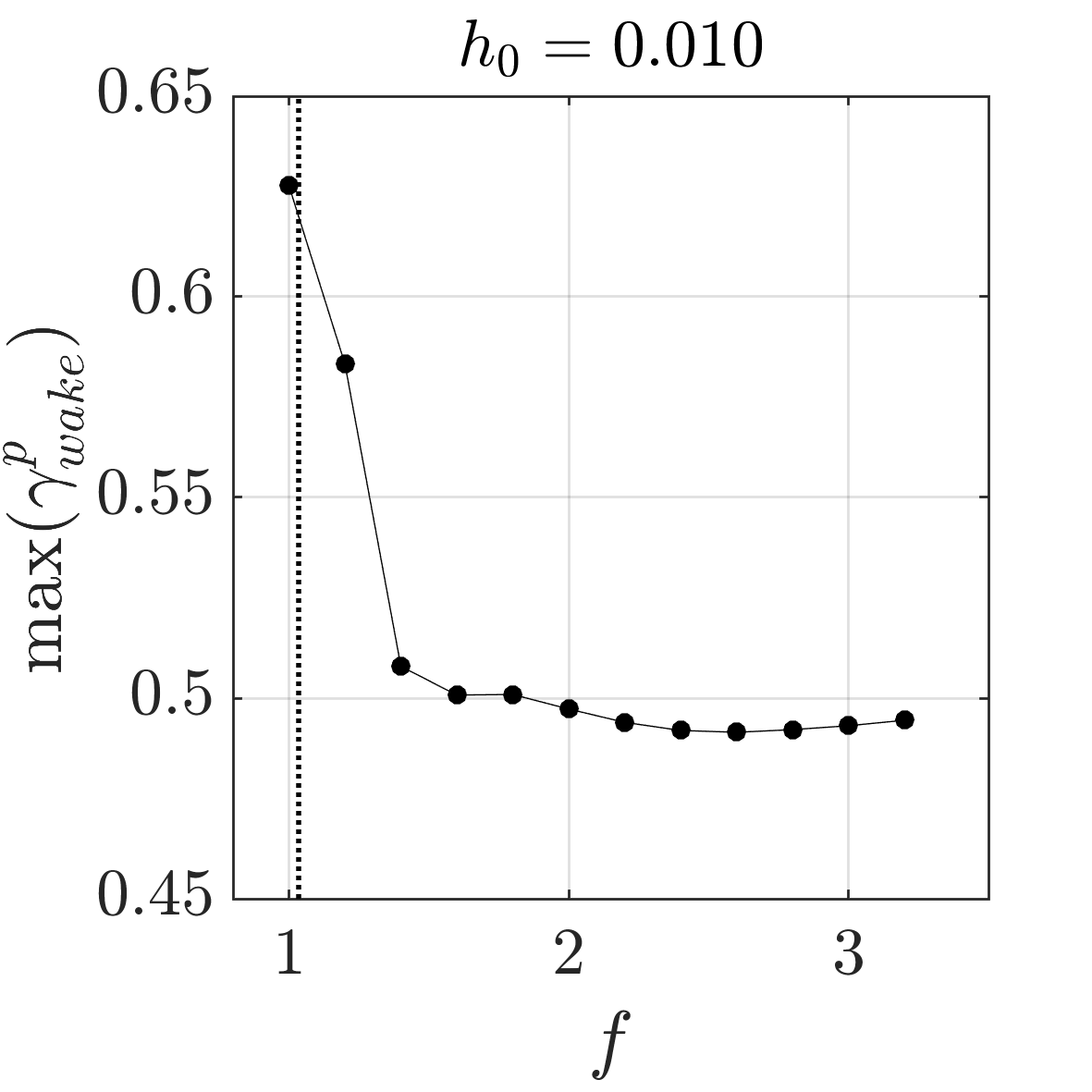}
	\end{subfigure}
    	\begin{subfigure}[b]{0.3\textwidth}
		\hspace*{0mm}
        		\includegraphics[scale = 0.28,trim={1.17cm 0cm 0cm 0cm},clip]{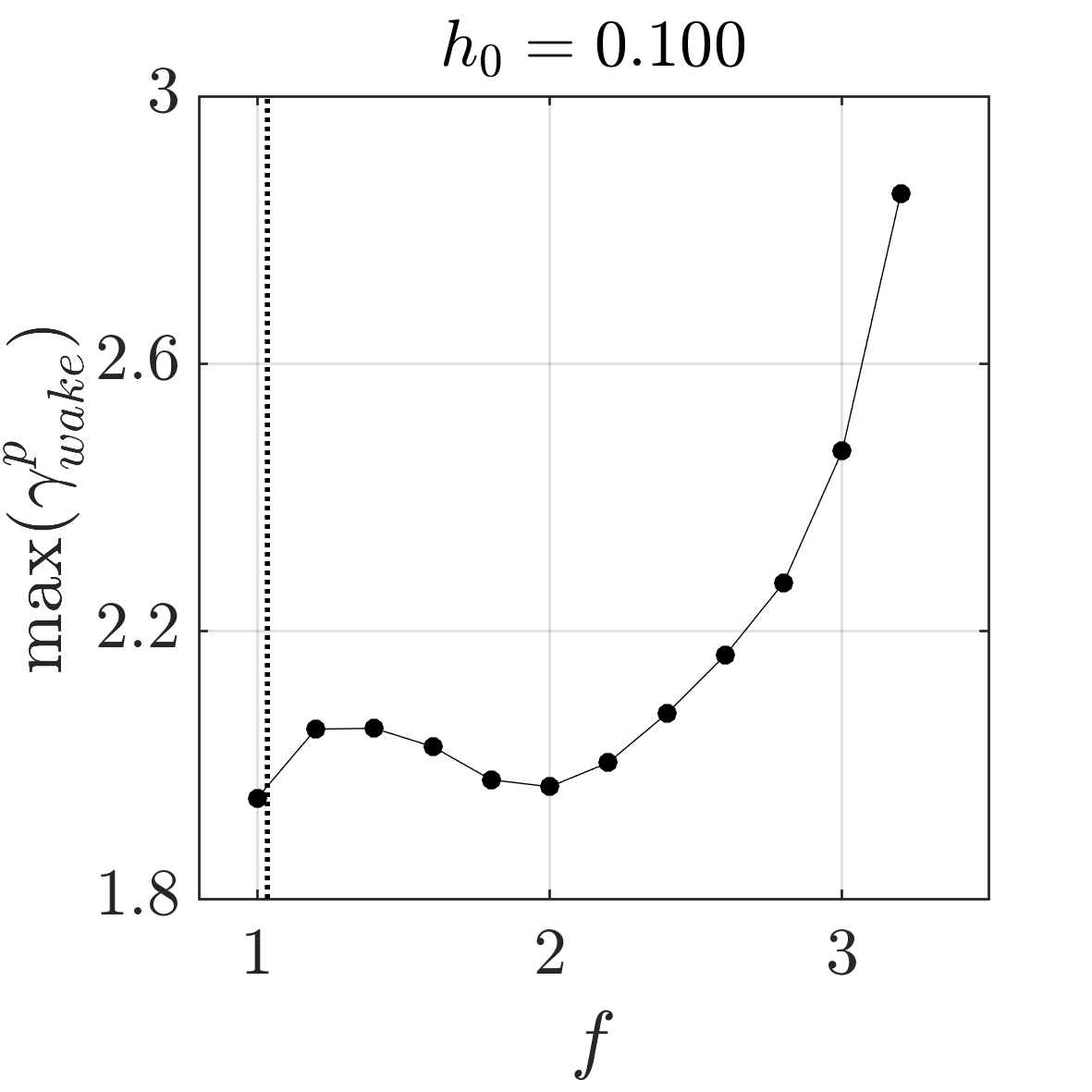}
	\end{subfigure}
	\caption{Analog of figure~\ref{fig:gamma_max_S0.02} for stiffness $S = 2$.}
	\label{fig:gamma_max_S2}
\end{figure}

To connect performance to wake circulation, we present in figure~\ref{fig:gamma_max_S2} the maximum circulation in the wake of the plate (see the text surrounding figure~\ref{fig:gamma_max_S0.02} for details on how this is defined) as a function of frequency for the various heave amplitudes. For $h_0 = 0.001, 0.01$ the maximum circulation is observed near the resonant frequency associated with the mode for which $E_r$ is maximal. For the larger heave amplitude of $h_0 = 0.1$, there is a local peak near this resonant frequency before the maximum circulation increases significantly at higher frequencies.

Together, the results contained in figures \ref{fig:ampphaseS2}--\ref{fig:gamma_max_S2} indicate that for large-amplitude swimming at $S = 2$, there is a distinction between low frequencies ($f<2.4$), for which leading-edge separation is sufficiently small, and high frequencies ($f\ge2.4$), for which leading-edge separation and vortex dynamics are prominent. At low frequencies, the results are analogous to those found for the lower stiffnesses. Many features of linear dynamics persist at large heave amplitudes, and thrust peaks are associated with actuating at the resonant frequency to obtain maximal wake circulation. At high frequencies, the results are substantially different from those associated with the more flexible plates. There are prominent leading-edge vortex dynamics that are associated with significantly larger thrust peaks and wake circulation than what is created by resonant mechanisms at lower frequencies.

\subsubsection{Maximal stiffness: $S = 20$}

\begin{figure}
\centering
	\begin{subfigure}[b]{0.3\textwidth}
		\hspace*{-6mm}
        		\includegraphics[scale=0.315,trim={0cm 2.39cm 0cm 0cm},clip]{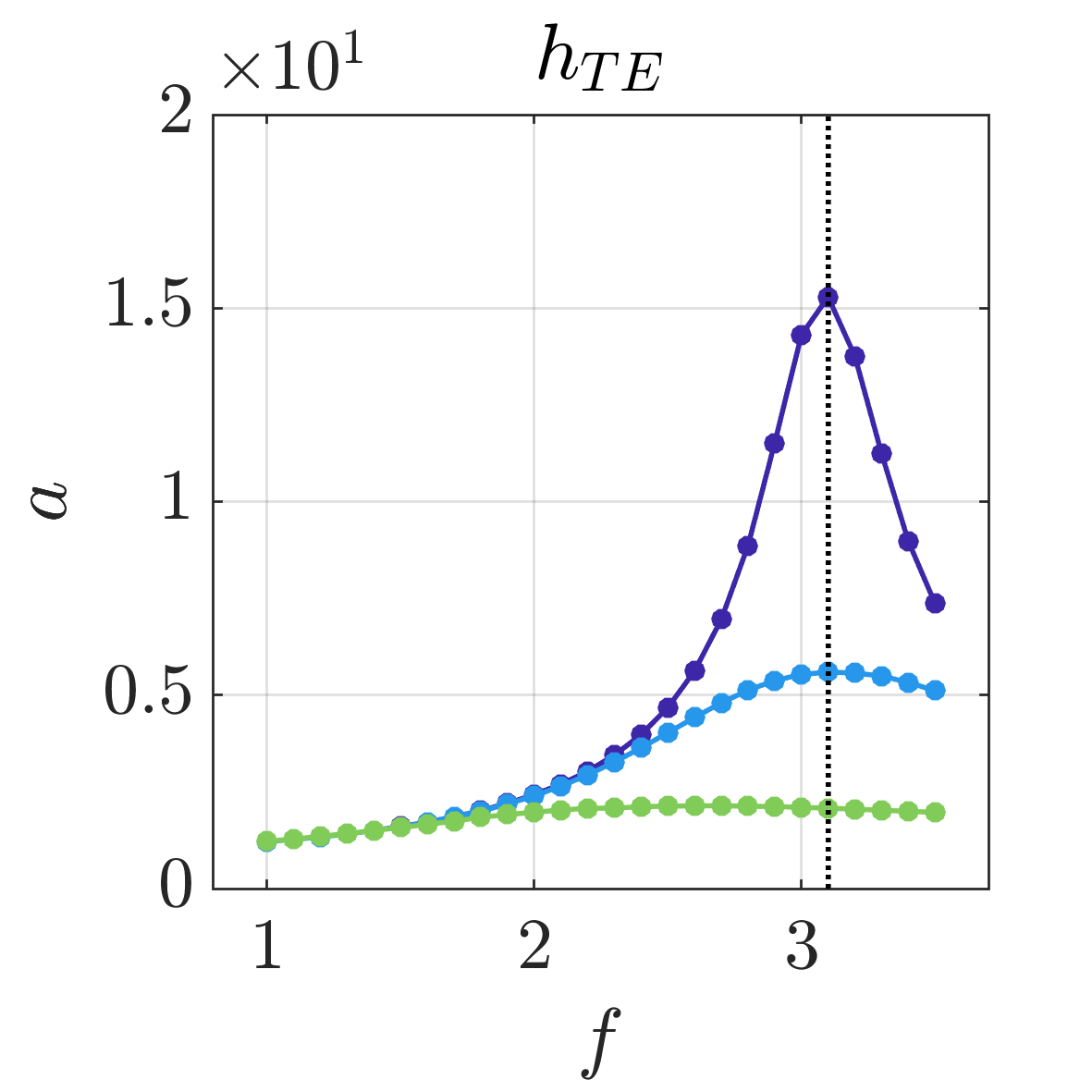}
	\end{subfigure}
	\begin{subfigure}[b]{0.3\textwidth}
		\hspace*{1mm}
        		\includegraphics[scale=0.315,trim={1cm 2.39cm 0cm 0cm},clip]{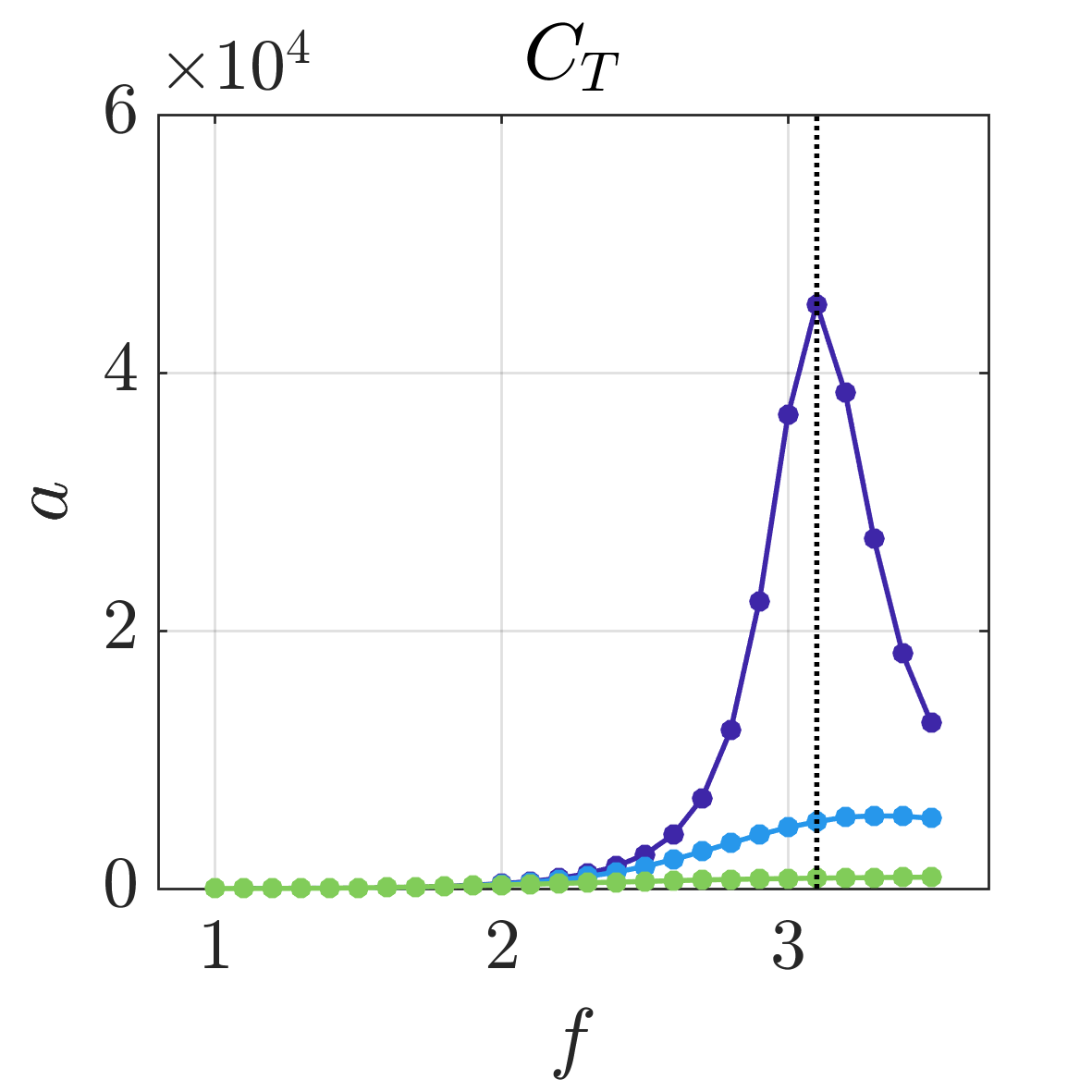}
	\end{subfigure}
	\begin{subfigure}[b]{0.3\textwidth}
		\hspace*{4.2mm}
        		\includegraphics[scale=0.315,trim={1cm 2.39cm 0cm 0cm},clip]{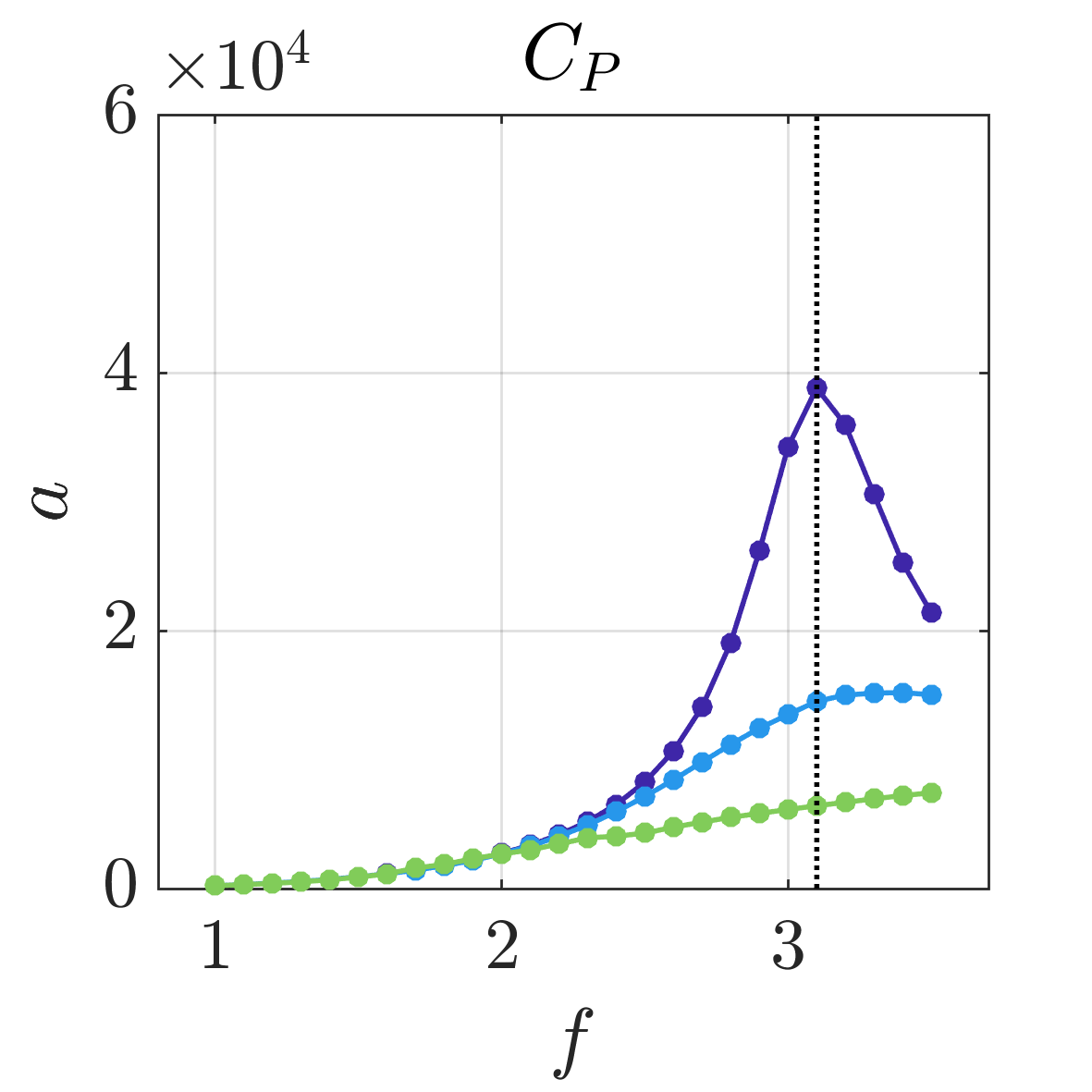}
	\end{subfigure}
	
	\begin{subfigure}[b]{0.3\textwidth}
		\hspace*{-6mm}
        		\includegraphics[scale=0.315, trim={0cm 0cm 0cm 0cm},clip]{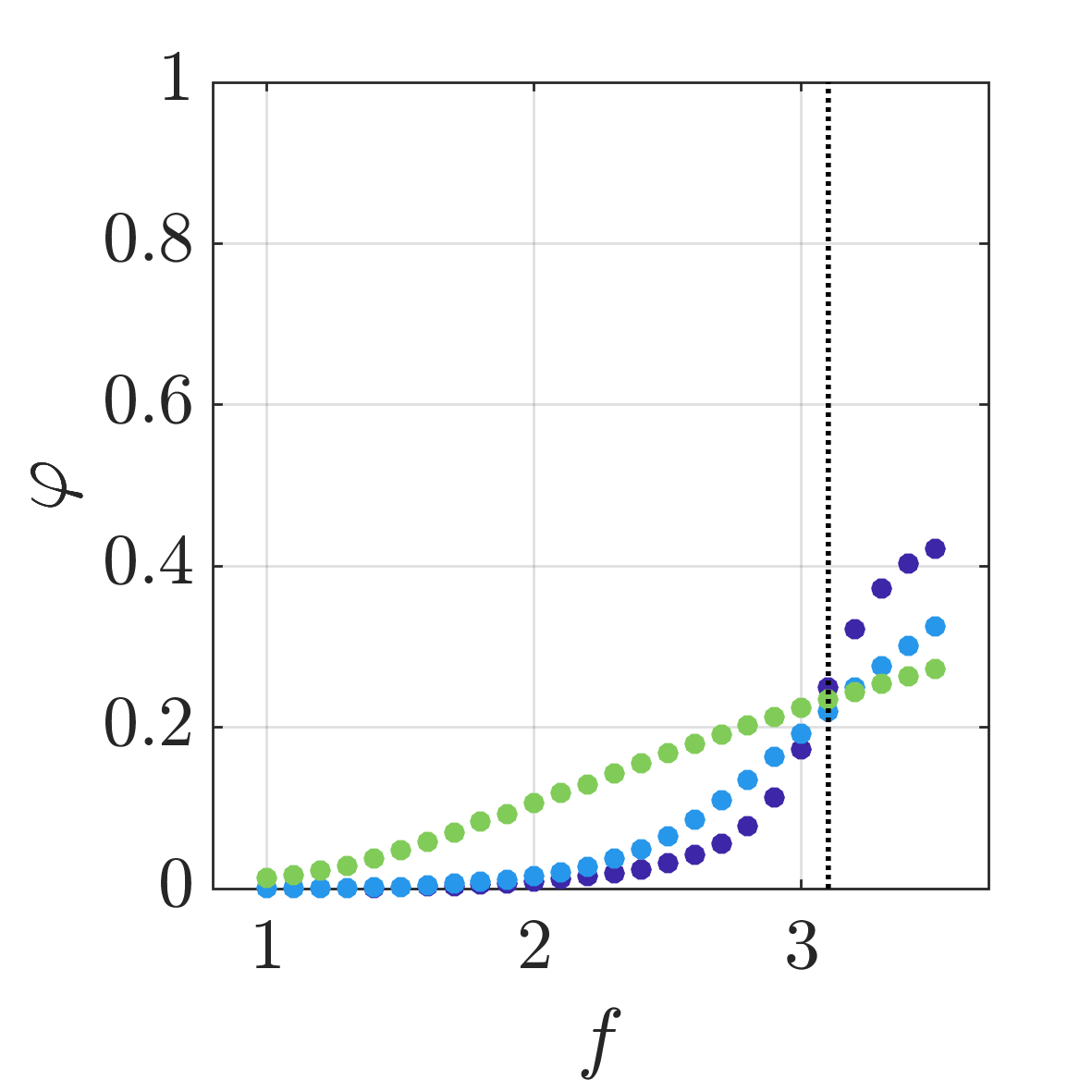}
	\end{subfigure}
	\begin{subfigure}[b]{0.3\textwidth}
		\hspace*{-1mm}
        		\includegraphics[height=4.73cm, width=4.62cm, trim={1.2cm 0cm 0cm 0cm},clip]{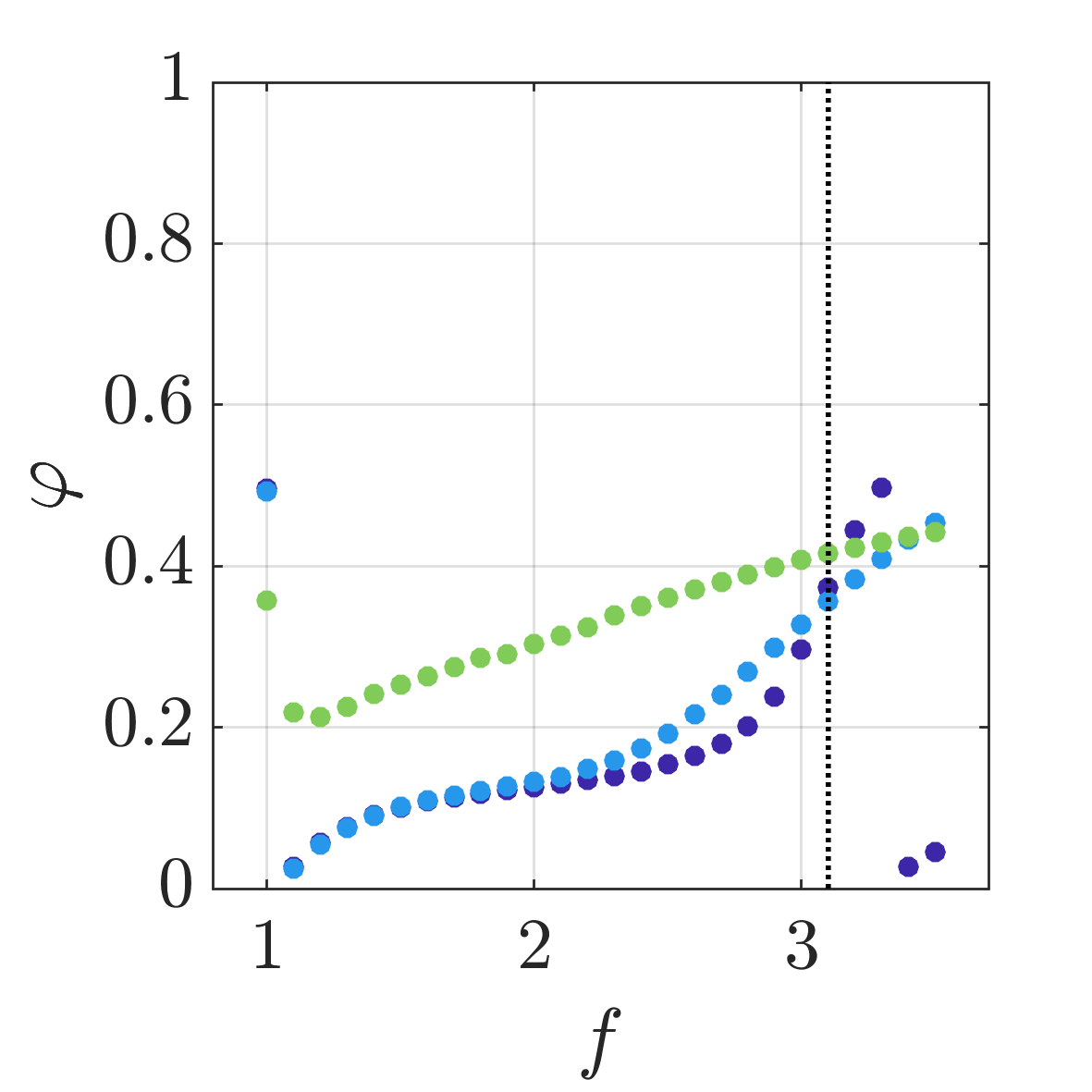}
	\end{subfigure}
	\begin{subfigure}[b]{0.3\textwidth}
		\hspace*{2.5mm}
        		\includegraphics[height=4.73cm, width=4.62cm, trim={1.2cm 0cm 0cm 0cm},clip]{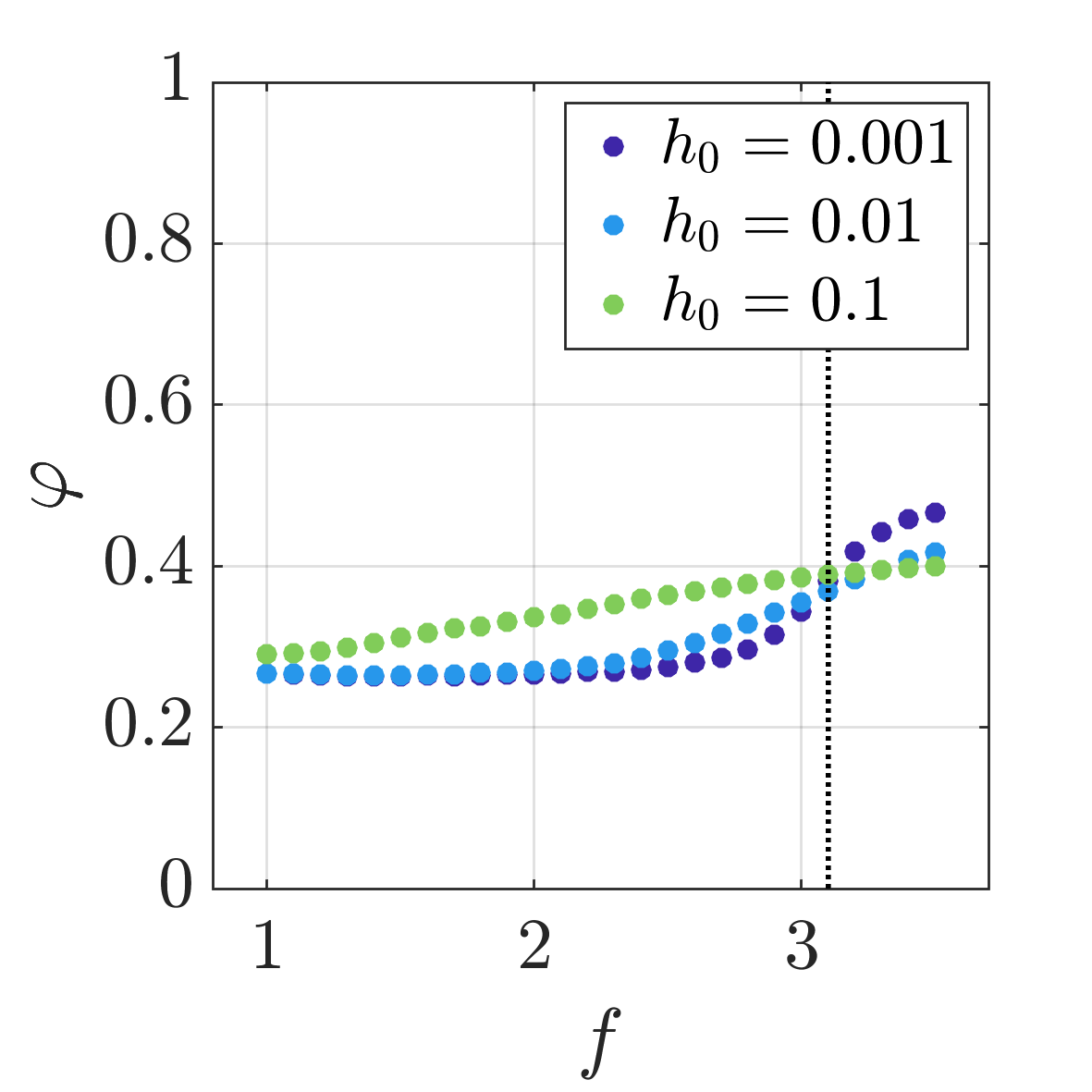}
	\end{subfigure}
    \caption{Analog of figure~\ref{fig:ampphaseSp02} for stiffness $S = 20$.}
	\label{fig:ampphaseS20}
\end{figure}

We now turn to the stiffest case considered, $S = 20$. Figure~\ref{fig:ampphaseS20} provides the phase and maximum amplitude associated with $h_{TE}, C_T,$ and $C_P$ at a given frequency. Note that for all results shown in this subsection (corresponding to $S=20$), the frequency range is $f\in[1,3.5]$ rather than $f\in[1,3.2]$. This is done to provide enough data points beyond the natural frequency of $f\approx3.1$ to enable meaningful conclusions about the role of resonance to be drawn.  

As was observed for the more flexible cases, resonance appears to be the dominant indicator of performance at the smallest heave amplitude ($h_0 = 0.001$), with peaks in $a$ occurring for all quantities near the natural frequency associated with the linear stability mode of largest $E_r$ value. These resonant peaks largely persist at the larger value of $h_0=0.01$, and the similar phases across $h_0=0.001$ and $h_0=0.01$ suggest that the dynamics remain primarily linear up to this larger amplitude.

The trends at the larger heave amplitude are similar to those observed for the other stiffness values. The peaks broaden and weaken significantly compared with the smaller heave amplitudes, but persist near the resonant frequency. In addition, the phase behavior is distinct from that corresponding to the lower heave amplitudes. 

We again probe the connection between thrust and wake circulation by showing the maximum wake circulation (see the text surrounding figure~\ref{fig:gamma_max_S0.02} for details on how this is defined) as a function of frequency for the various heave amplitudes. For $h_0 = 0.001, 0.01$, and $0.1$, the maximum circulation is observed near the resonant frequency associated with the mode for which $E_r$ is maximal (though it is difficult to arrive at concrete conclusions about the role of resonance in wake circulation at this stiffness because the resonant frequency is near the maximal frequency considered in this study). Irrespective of the connection to resonance, the frequency associated with maximal wake circulation is also associated with maximal mean thrust (\emph{cf.}, figure~\ref{fig:global_performance_nonlin}), as was observed for more flexible plates.

\begin{figure}
\centering
	\begin{subfigure}[b]{0.3\textwidth}
		\hspace*{0mm}
        		\includegraphics[scale = 0.28,trim={0cm 0cm 0cm 0cm},clip]{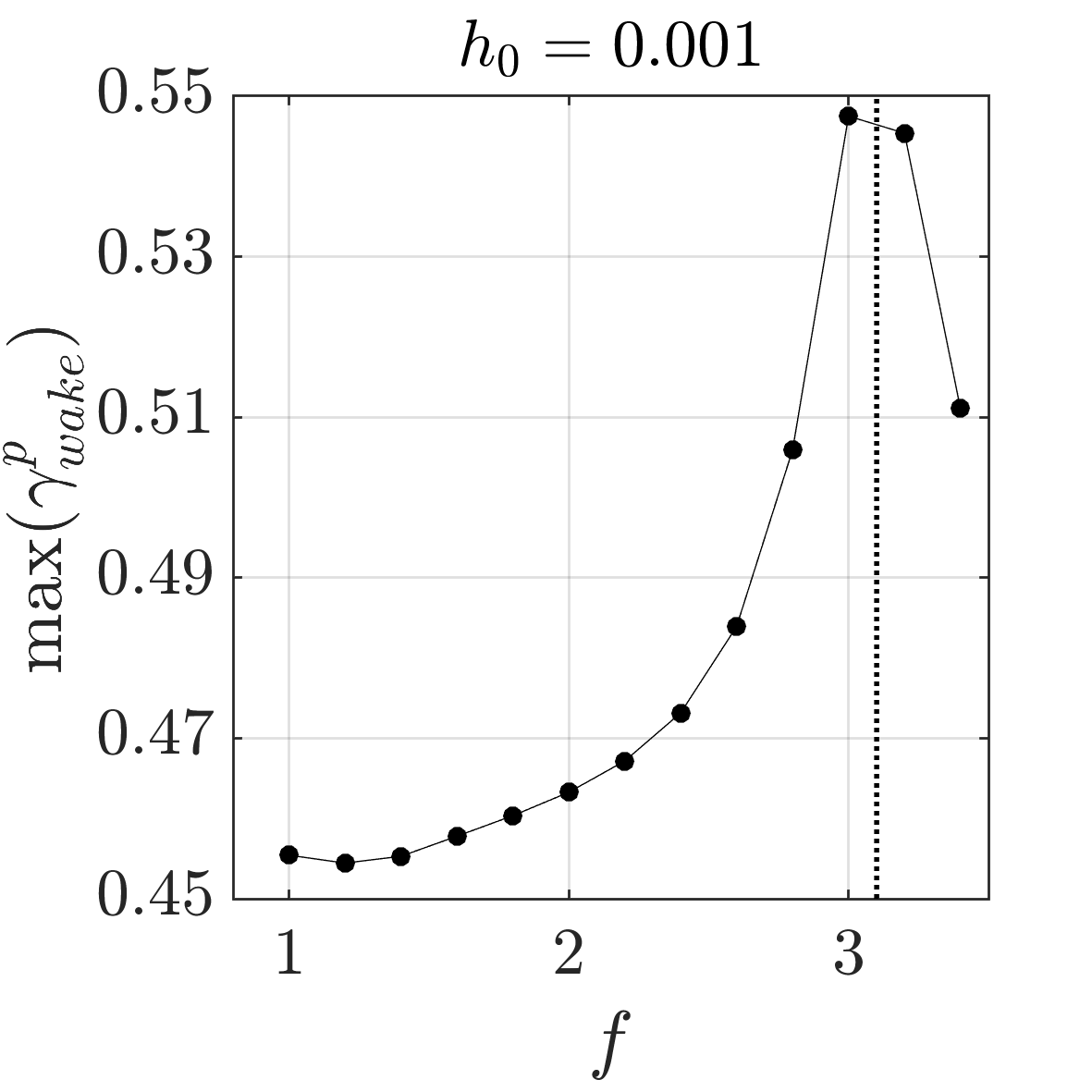}
	\end{subfigure}
	\begin{subfigure}[b]{0.3\textwidth}
		\hspace*{0mm}
        		\includegraphics[scale = 0.28,trim={1.17cm 0cm 0cm 0cm},clip]{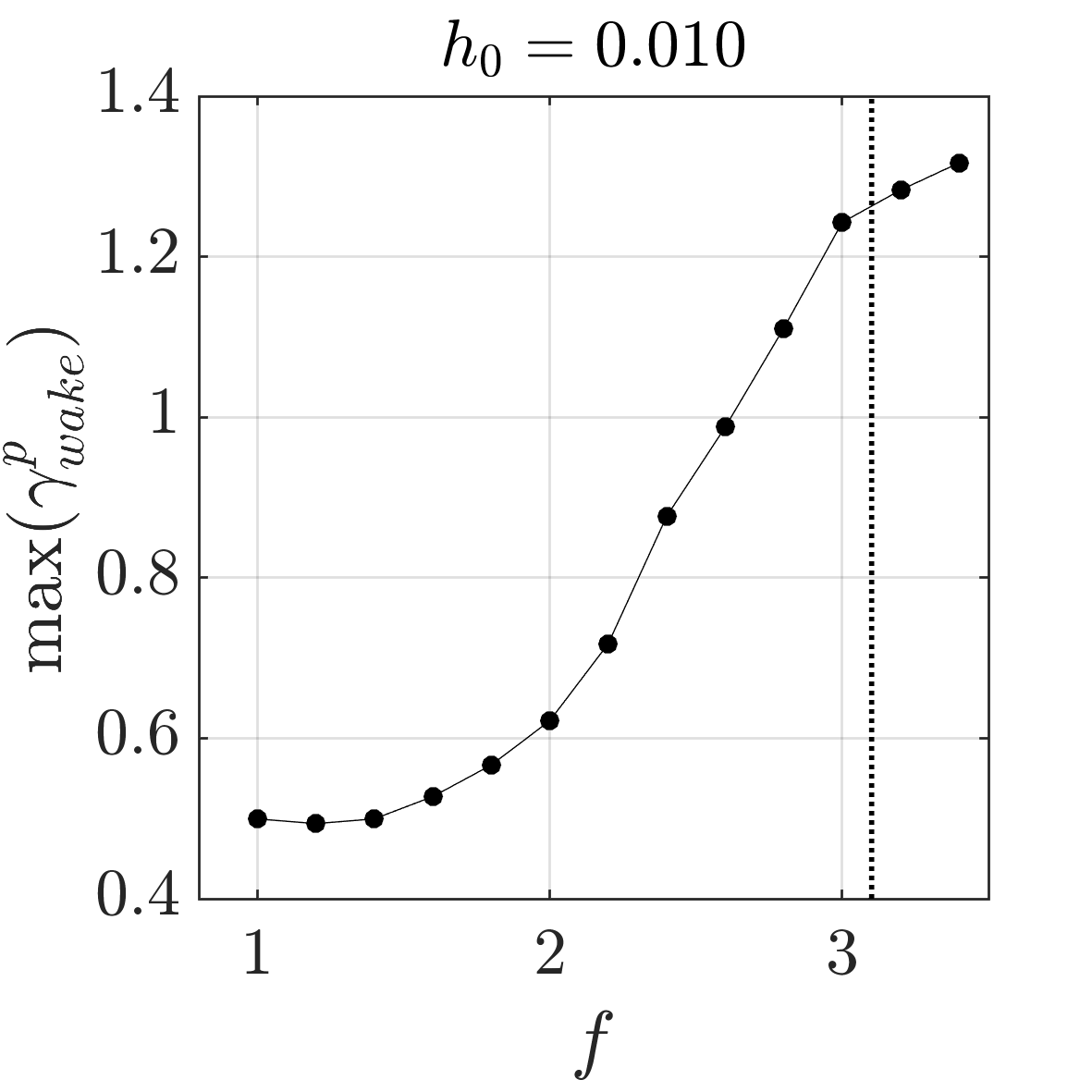}
	\end{subfigure}
    	\begin{subfigure}[b]{0.3\textwidth}
		\hspace*{0mm}
        		\includegraphics[scale = 0.28,trim={1.17cm 0cm 0cm 0cm},clip]{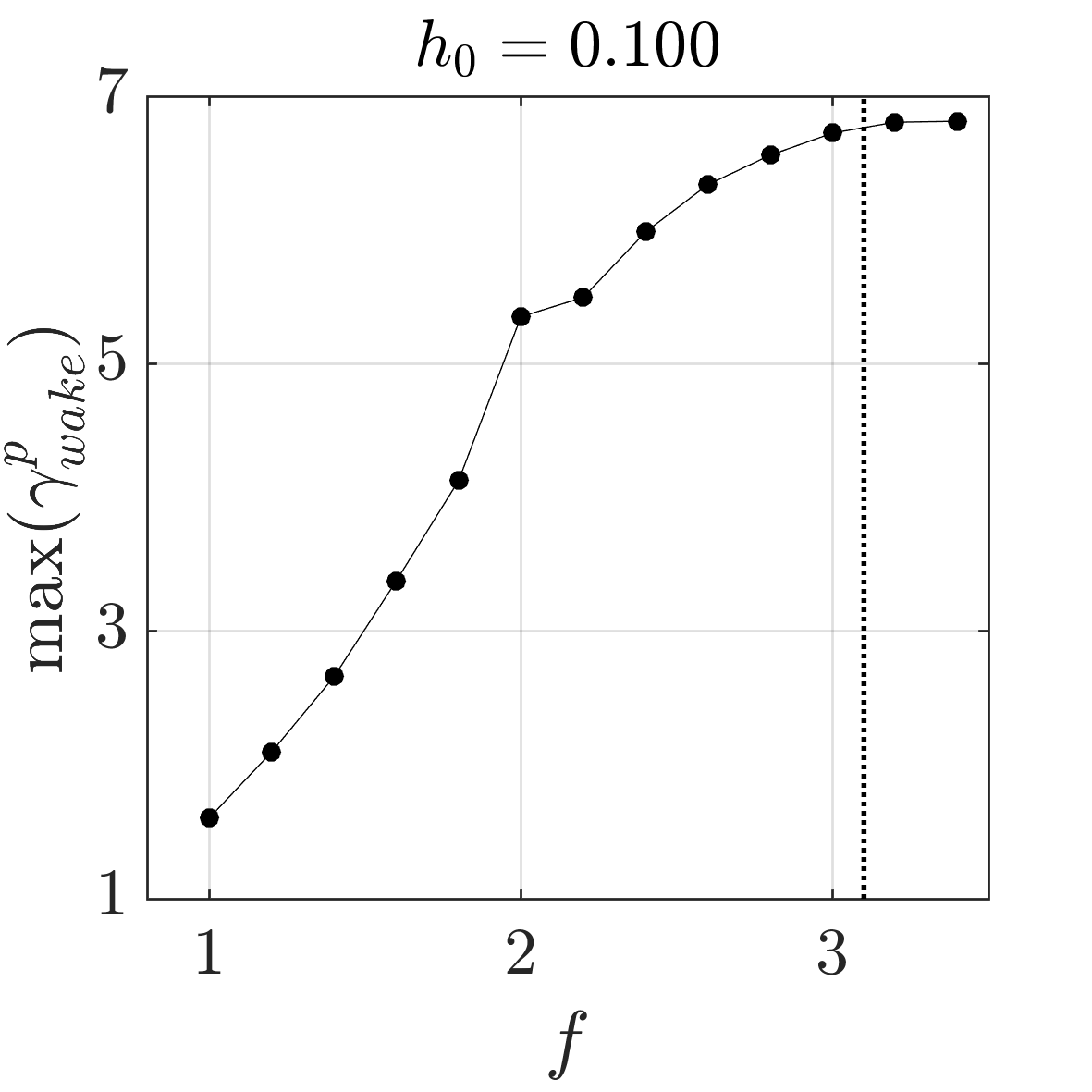}
	\end{subfigure}
	 \caption{Analog of figure~\ref{fig:gamma_max_S0.02} for stiffness $S = 20$.}
	\label{fig:gamma_max_S20}
\end{figure}

Finally, we note that for $h_0=0.1$ at this larger stiffness, the plate-fluid system contains significant leading-edge vorticity that indicates separation; \emph{cf.}, figure~\ref{fig:snapshots_nonlin}. This leading-edge vorticity is most pronounced for $S=20$ when $1.8\le f \le2.4$. From figure~\ref{fig:global_performance_nonlin}, it is clear that these frequencies coincide with small increases in efficiency. Moreover, unlike for $S=2$, the separated flow behavior is not associated with aperiodic system dynamics for $S=20$, nor does it correspond to substantial changes in performance. This demonstrates that leading-edge separation and vorticity generation, while not accounted for in the linear analysis, do not necessarily indicate significant changes in performance from what one would predict through resonance-based arguments.  

\section{Conclusions and connections to other work}

In this article, we investigated the role of resonance in finite-amplitude swimming of a flexible flat plate in a viscous fluid ($Re=240$). First, we unambiguously defined resonance using linear stability modes of the fully coupled fluid-structure interaction system. These modes represent eigenvalues and eigenvectors that account for the viscous fluid, the plate, and the coupling between them. The eigenvalues associated with modes that maximized the energy put into the plate, quantified in terms of an energy ratio ($E_r$), were shown to provide an appropriate measure for system resonance.

We then probed the connections between finite-amplitude swimming and resonant behavior through high-fidelity nonlinear simulations of systematically increased heave amplitude. By comparing the results for different heave amplitudes with one another and with the linear stability modes, connections to and distinctions from resonant behavior were established for finite-amplitude swimming over a wide range of stiffnesses ($S = 0.02, 0.2, 2, 20$). The primary observations of this article (with connections to other work provided where appropriate) are:

\textbf{\textit{Small heave amplitudes}}. For sufficiently small heave amplitudes ($h_0\le0.01$ in our studies) the dynamics were similar across heave amplitude, suggesting primarily linear behavior. Beyond this, peaks in trailing-edge displacement, thrust, and input power all occurred near the resonant frequency corresponding to the mode with largest $E_r$ value, indicating that resonance is a dominant contributor to performance in this small-amplitude regime. 

\textbf{\textit{Large heave amplitudes}}. At larger heave amplitudes more reflective of fish swimming ($h_0=0.1$ in our studies), the results indicated that both resonant and non-resonant mechanisms played a role.

Regarding resonant behavior, different plate shapes were observed depending on the plate stiffness and heave frequency, and these shapes were consistent with the shapes predicted by the global linear analysis. More flexible plates were more likely to exhibit shapes that are reminiscent of higher-modes of an Euler-Bernoulli beam. This phenomenon was also observed by \citet{Quinn2014} through experiments at higher Reynolds numbers.  

In addition, peaks in trailing-edge motion and thrust persisted near the resonant frequency for all stiffnesses considered, with the general trend that the peaks weakened, broadened, and occurred at slightly shifted frequencies from the resonant frequency identified through the linear stability analysis. This distinction from the linear behavior can be due to a number of phenomena, including added mass effects from finite amplitude heaving, leading-edge separation, and nonlinear vortex generation and interaction not present for small-amplitude kinematics. Interestingly, the broader and weaker frequency response is consistent with the response of a spring-mass system subjected to damping. The aforementioned nonlinear mechanisms could therefore be providing an effective damping mechanism, which in turn could be related to the nonlinear fluid damping effect argued by \citet{Ramananarivo2011} (though the authors considered larger mass ratios and heave amplitudes than those of this study that almost certainly led to differences in the detailed system response). Clarifying the specific nonlinear mechanisms that dictate the broader, shorter peaks associated with finite-amplitude heaving is an avenue for future research efforts.

We note that connections to resonance have been argued for at higher Reynolds numbers as well. \citet{Quinn2014} identified peaks in thrust and efficiency at specific frequencies from experimental data at $Re \sim O(10^4)$, and used this to claim that resonance was a contributor to performance. Our results clarify that performance peaks in finite-amplitude swimming are indeed connected to resonant behavior at lower Reynolds numbers. This observation indicates that performance peaks observed at higher Reynolds may also be related to resonance of the FSI system, though a global stability analysis at these higher Reynolds is necessary to make conclusive statements.

Despite these peaks near the resonant frequency, non-resonant behavior was observed as well. For all stiffnesses considered, the input power exhibited a qualitatively different amplitude and phase response from the essentially linear dynamics at lower heave amplitudes. In addition, leading-edge separation (and associated vorticity generation) was present for the larger stiffness values, $S=2,20$. These cases were associated with either marginal improvements (for $S = 20$, $1.8 \le f \le 2.4$) or substantial decreases (for $S = 2, f\ge2.4$) in efficiency. This is consistent with the observations of \citet{Quinn2015} that leading-edge separation was generally deleterious to efficiency. \citet{Quinn2015} used experiments at Reynolds numbers of $O(10^3-10^4)$ to arrive at their conclusions, which suggests potential extensions of our conclusions at lower Reynolds numbers to higher Reynolds numbers.

We also note that leading-edge separation corresponded to an asymmetric wake. \citet{Zhu2014} did an extensive study of the parameters that led to asymmetric wake responses, and arrived at the conclusion that the wake circulation had to be above a threshold value for asymmetry to occur. Our studies suggest that sufficiently large leading-edge vorticity is another prerequisite.

Irrespective of the connections to resonance, the maximal mean thrust in large-amplitude swimming occurs at or very near the frequency for which the wake circulation is maximized. This was observed for all stiffnesses considered in this study, and is consistent with the observations of \citet{Moored2014}, who used a local linear analysis of the wake to demonstrate that performance gains were obtained by actuating at a frequency that maximally destabilized the wake. \citet{Moored2014} considered experimental data of $O(10^4)$, which again indicates that there may be extensions of the conclusions drawn here for the moderate Reynolds number of 240 to higher Reynolds numbers.

\section{Acknowledgments}

The authors gratefully acknowledge funding from the Office of Naval Research (grant~\#~00014-14-1-0533).

\appendix

\section{Simulation parameters and grid convergence study}
\label{app:grid_properties}

The flow equations are treated using a multidomain approach: the finest grid surrounds the body and grids of increasing coarseness are used at progressively larger distances \citep{Colonius2008}. In all computations performed for this article, the domain size of the finest sub-domain is $[-0.2, 1.8] \times [-0.5, 0.5]$ and the total domain size is $[-15.07, 16.67] \times [-7.87, 7.87]$. The grid spacing on the finest domain is $\Delta x = 2/480 \approx 0.0042 $ and the grid spacing for the flag is $\Delta s = 2 \Delta x \approx 0.0083$. The time step is $\Delta t = 0.00015$, which gives a maximum Courant-Friedrichs-Levy number of approximately $0.2$. 

The suitability of these parameters is demonstrated in this appendix for the case when $S=20$, $f=3.2$, and $h_0 = 0.1$. Note that this is a strenuous test case, as this choice of parameters leads to one of the largest vorticity and plate accelerations of the physical parameters considered in this study. Table \ref{tab:grid_conv} contains details of the various grids used in this convergence study. The final column of the table provides the quantity $||\bm{\chi}-\bm{\chi}_f||_\infty/||\bm{\chi}_f||_\infty$, where $\bm{\chi}$ is the plate's position computed on the grid associated with the current row at $t=30$ and $\bm{\chi}_f$ is the plate's position computed on grid 7 at $t=30$. The position of the plate is well-converged for all grids considered, which demonstrates that the chosen grid is adequate for the simulations used in this article.

\begin{table}
\centering
\begin{tabular}{ c c c c c c}
Grid & $\Delta x$ & $\Delta t$ & Smallest sub-domain size & Total domain size & $\frac{||\bm{\chi}-\bm{\chi}_f||_\infty}{||\bm{\chi}_f||_\infty}$ \\ \hline
1 & 0.0083 & 0.0004 & $[-0.2, 1.8] \times [0.5, 0.5]$ & $[-15.07, 16.04] \times [-7.87, 7.87]$ & 0.00337 \\
2 & 0.0083 & 0.0002 & $[-0.2, 1.8] \times [0.5, 0.5]$ & $[-15.07, 16.04] \times [-7.87, 7.87]$ & 0.00474 \\
3 & 0.0056 & 0.0001 & $[-0.2, 1.8] \times [0.5, 0.5]$ & $[-15.07, 16.04] \times [-7.87, 7.87]$ & 0.00182 \\
4* & 0.0042 & 0.00015 & $[-0.2, 1.8] \times [0.5, 0.5]$ & $[-15.07, 16.04] \times [-7.87, 7.87]$ & 0.00070 \\
5 & 0.0042 & 0.00015 & $[-0.2, 1.8] \times [0.5, 0.5]$ & $[-30.98, 32.75] \times [-15.78, 15.78]$ & 0.00019 \\
6 & 0.0042 & 0.000075 & $[-0.2, 1.8] \times [0.5, 0.5]$ & $[-15.07, 16.04] \times [-7.87, 7.87]$ & 0.00058 \\
7 & 0.0033 & 0.0001 & $[-0.2, 1.8] \times [0.5, 0.5]$ & $[-15.07, 16.04] \times [-7.87, 7.87]$ & --- \\
\end{tabular}
\caption{Simulation parameters for the grid convergence study. Grid 4 was used to obtain the data presented in the article.}
\label{tab:grid_conv}
\end{table}

\bibliographystyle{jfm}
\bibliography{swimming_GFR}{}

\begin{thebibliography}{22}
\expandafter\ifx\csname natexlab\endcsname\relax\def\natexlab#1{#1}\fi
\def\au#1{#1} \def\ed#1{#1} \def\yr#1{#1}\def\at#1{#1}\def\jt#1{\textit{#1}}
  \def\bt#1{#1}\def\bvol#1{\textbf{#1}} \def\vol#1{#1} \def\pg#1{#1}
  \def\publ#1{#1}\def\arxiv#1{#1}\def\org#1{#1}\def\st#1{\textit{#1}}

\bibitem[Alben(2008)]{Alben2008}
{\sc \au{Alben, Silas}} \yr{2008}  \at{Optimal flexibility of a flapping
  appendage in an inviscid fluid}.  \jt{Journal of Fluid Mechanics}
  \bvol{614},  \pg{355--380}.

\bibitem[Alben {\em et~al.\/}(2012)Alben, Witt, Baker, Anderson \&
  Lauder]{Alben2012}
{\sc \au{Alben, Silas}, \au{Witt, Charles}, \au{Baker, T~Vernon}, \au{Anderson,
  Erik} \& \au{Lauder, George~V}} \yr{2012}  \at{Dynamics of freely swimming
  flexible foils}.  \jt{Physics of Fluids}  \bvol{24}~(5),  \pg{051901}.

\bibitem[Colonius \& Taira(2008)]{Colonius2008}
{\sc \au{Colonius, Tim} \& \au{Taira, Kunihiko}} \yr{2008}  \at{A fast immersed
  boundary method using a nullspace approach and multi-domain far-field
  boundary conditions}.  \jt{Computer Methods in Applied Mechanics and
  Engineering}  \bvol{197}~(25),  \pg{2131--2146}.

\bibitem[Criesfield(1991)]{Crisfield1991}
{\sc \au{Criesfield, MA}} \yr{1991} {\em Non-linear finite element analysis of
  solids and structures, vol. 1\/}.  \publ{Wiley, New York}.

\bibitem[Dewey {\em et~al.\/}(2013)Dewey, Boschitsch, Moored, Stone \&
  Smits]{Dewey2013}
{\sc \au{Dewey, Peter~A}, \au{Boschitsch, Birgitt~M}, \au{Moored, Keith~W},
  \au{Stone, Howard~A} \& \au{Smits, Alexander~J}} \yr{2013}  \at{Scaling laws
  for the thrust production of flexible pitching panels}.  \jt{Journal of Fluid
  Mechanics}  \bvol{732},  \pg{29--46}.

\bibitem[Ebert {\em et~al.\/}(2003)Ebert, Musgrave, Peachey, Perlin \&
  Worley]{Ebert2003}
{\sc \au{Ebert, David~S}, \au{Musgrave, F~Kenton}, \au{Peachey, Darwyn},
  \au{Perlin, Ken} \& \au{Worley, Steven}} \yr{2003} {\em Texturing \&
  modeling: a procedural approach\/}.  \publ{Morgan Kaufmann}.

\bibitem[Floryan \& Rowley(2018)]{Floryan2018}
{\sc \au{Floryan, Daniel} \& \au{Rowley, Clarence~W}} \yr{2018}  \at{Clarifying
  the relationship between efficiency and resonance for flexible inertial
  swimmers}.  \jt{Journal of Fluid Mechanics}  \bvol{853},  \pg{271--300}.

\bibitem[Goza \& Colonius(2017)]{Goza2017}
{\sc \au{Goza, Andres} \& \au{Colonius, Tim}} \yr{2017}  \at{A strongly-coupled
  immersed-boundary formulation for thin elastic structures}.  \jt{Journal of
  Computational Physics}  \bvol{336},  \pg{401--411}.

\bibitem[Goza {\em et~al.\/}(2018)Goza, Colonius \& Sader]{Goza2018JFM}
{\sc \au{Goza, Andres}, \au{Colonius, Tim} \& \au{Sader, John~Elie}} \yr{2018}
  \at{Global modes and nonlinear analysis of inverted-flag flapping}.
  \jt{Journal of Fluid Mechanics}  \bvol{857},  \pg{312--344}.

\bibitem[Goza {\em et~al.\/}(2016)Goza, Liska, Morley \& Colonius]{Goza2016}
{\sc \au{Goza, Andres}, \au{Liska, Sebastian}, \au{Morley, Benjamin} \&
  \au{Colonius, Tim}} \yr{2016}  \at{Accurate computation of surface stresses
  and forces with immersed boundary methods}.  \jt{Journal of Computational
  Physics}  \bvol{321},  \pg{860--873}.

\bibitem[Hua {\em et~al.\/}(2013)Hua, Zhu \& Lu]{Hua2013}
{\sc \au{Hua, Ru-Nan}, \au{Zhu, Luoding} \& \au{Lu, Xi-Yun}} \yr{2013}
  \at{Locomotion of a flapping flexible plate}.  \jt{Physics of Fluids}
  \bvol{25}~(12),  \pg{121901}.

\bibitem[Lehoucq {\em et~al.\/}(1998)Lehoucq, Sorensen \& Yang]{Lehoucq1998}
{\sc \au{Lehoucq, Richard~B}, \au{Sorensen, Danny~C} \& \au{Yang, Chao}}
  \yr{1998} {\em {ARPACK} users' guide: solution of large-scale eigenvalue
  problems with implicitly restarted Arnoldi methods\/}.  \publ{SIAM}.

\bibitem[Michelin \& Llewellyn~Smith(2009)]{Michelin2009}
{\sc \au{Michelin, S{\'e}bastien} \& \au{Llewellyn~Smith, Stefan~G}} \yr{2009}
  \at{Resonance and propulsion performance of a heaving flexible wing}.
  \jt{Physics of Fluids}  \bvol{21}~(7),  \pg{071902}.

\bibitem[Moored {\em et~al.\/}(2014)Moored, Dewey, Boschitsch, Smits \&
  Haj-Hariri]{Moored2014}
{\sc \au{Moored, KW}, \au{Dewey, PA}, \au{Boschitsch, BM}, \au{Smits, AJ} \&
  \au{Haj-Hariri, H}} \yr{2014}  \at{Linear instability mechanisms leading to
  optimally efficient locomotion with flexible propulsors}.  \jt{Physics of
  Fluids}  \bvol{26}~(4),  \pg{041905}.

\bibitem[Quinn {\em et~al.\/}(2014)Quinn, Lauder \& Smits]{Quinn2014}
{\sc \au{Quinn, Daniel~B}, \au{Lauder, George~V} \& \au{Smits, Alexander~J}}
  \yr{2014}  \at{Scaling the propulsive performance of heaving flexible
  panels}.  \jt{Journal of Fluid Mechanics}  \bvol{738},  \pg{250--267}.

\bibitem[Quinn {\em et~al.\/}(2015)Quinn, Lauder \& Smits]{Quinn2015}
{\sc \au{Quinn, Daniel~B}, \au{Lauder, George~V} \& \au{Smits, Alexander~J}}
  \yr{2015}  \at{Maximizing the efficiency of a flexible propulsor using
  experimental optimization}.  \jt{Journal of Fluid Mechanics}  \bvol{767},
  \pg{430--448}.

\bibitem[Ramananarivo {\em et~al.\/}(2011)Ramananarivo, Godoy-Diana \&
  Thiria]{Ramananarivo2011}
{\sc \au{Ramananarivo, Sophie}, \au{Godoy-Diana, Ramiro} \& \au{Thiria,
  Benjamin}} \yr{2011}  \at{Rather than resonance, flapping wing flyers may
  play on aerodynamics to improve performance}.  \jt{Proceedings of the
  National Academy of Sciences}  \bvol{108}~(15),  \pg{5964--5969}.

\bibitem[Vanella {\em et~al.\/}(2009)Vanella, Fitzgerald, Preidikman, Balaras
  \& Balachandran]{Vanella2009}
{\sc \au{Vanella, Marcos}, \au{Fitzgerald, Timothy}, \au{Preidikman, Sergio},
  \au{Balaras, Elias} \& \au{Balachandran, Balakumar}} \yr{2009}  \at{Influence
  of flexibility on the aerodynamic performance of a hovering wing}.
  \jt{Journal of Experimental Biology}  \bvol{212}~(1),  \pg{95--105}.

\bibitem[Webb(1988)]{Webb1988}
{\sc \au{Webb, Paul~W}} \yr{1988}  \at{Simple physical principles and
  vertebrate aquatic locomotion}.  \jt{American Zoologist}  \bvol{23}~(2),
  \pg{709--725}.

\bibitem[Wu(1961)]{Wu1961}
{\sc \au{Wu, T Yao-Tsu}} \yr{1961}  \at{Swimming of a waving plate}.
  \jt{Journal of Fluid Mechanics}  \bvol{10}~(3),  \pg{321--344}.

\bibitem[Zhang {\em et~al.\/}(2017)Zhang, Zhou \& Luo]{Zhang2017}
{\sc \au{Zhang, Yang}, \au{Zhou, Chunhua} \& \au{Luo, Haoxiang}} \yr{2017}
  \at{Effect of mass ratio on thrust production of an elastic panel pitching or
  heaving near resonance}.  \jt{Journal of Fluids and Structures}  \bvol{74},
  \pg{385--400}.

\bibitem[Zhu {\em et~al.\/}(2014)Zhu, He \& Zhang]{Zhu2014}
{\sc \au{Zhu, Xiaojue}, \au{He, Guowei} \& \au{Zhang, Xing}} \yr{2014}  \at{How
  flexibility affects the wake symmetry properties of a self-propelled plunging
  foil}.  \jt{Journal of Fluid Mechanics}  \bvol{751},  \pg{164--183}.

\end{thebibliography}

\end{document}